\documentclass[12pt,preprint]{aastex}
\usepackage[top=0.8in,bottom=1.8in,right=1in]{geometry}

\begin{document}
\title{Origins of the $H$, $HeI$, and $CaII$ Line Emission in Classical
T Tauri Stars}

\author{John Kwan,$^{1}$ and William Fischer$^{2}$}

\affil{$^{1}$Dept. of Astronomy, University of Massachusetts, Amherst,
MA 01003, kwan@nova.astro.umass.edu\\
$^2$Dept. of Physics and Astronomy, University of Toledo, Toledo, OH 43606,
wfische@utnet.utoledo.edu}

\email{kwan@nova.astro.umass.edu}

\slugcomment{Accepted by MNRAS, 2010 October 12}

\begin{abstract}

We perform local excitation calculations to obtain line opacities and 
emissivity ratios and compare them with observed properties of $H$,
$HeI$, $OI$, $CaII$, and $NaI$ lines to determine the requisite conditions
of density, temperature, and photon ionization rate. We find that $UV$
photoionization is the most probable excitation mechanism for generating
the $HeI\lambda 10830$ opacities that produce all the associated absorption
features. We also calculate the specific line flux at an observed velocity
of $v_{obs}=\pm 150~km~s^{-1}$ for both radial wind and infall models.
All the model results, together with observed correlations between absorption
and emission features and between narrow and broad emission components,
are used to deduce the origins of the strong $H$, $HeI$, and $CaII$ broad line
emission. We conclude that the first two arise primarily in a radial outflow
that is highly clumpy. The bulk of the wind volume is filled by gas at a
density $\sim 10^9~cm^{-3}$ and optically thick to $HeI\lambda 10830$ and
$H\alpha$, but optically thin to $HeI\lambda 5876$, $Pa\gamma$, and the
$CaII$ infrared triplet. The optically thick $HeI\lambda 5876$ emission
occur mostly in regions of density $\ge 10^{11}~cm^{-3}$ and temperature
$\ge 1.5\times 10^4~K$, while the optically thick $H\alpha$ and $Pa\gamma$
emission occur mostly in regions of density
around $10^{11}~cm^{-3}$ and temperature between
$8750$ and $1.25\times 10^4~K$. In producing the observed line fluxes at a
given $v_{obs}$ the covering factor of these emission clumps is sufficiently
small to not incur significant absorption of the stellar and veiling continua
in either $HeI$ or $H$ lines. The strong $CaII$ broad line emission likely
arise in both the magnetospheric accretion flow and the disk boundary 
layer where the gases dissipate part of their rotational energies before
infalling along magnetic field lines. The needed density and temperature
are $\sim 10^{12}~cm^{-3}$ and $\le 7500~K$ respectively. 

\end{abstract}

\section{Introduction}

Spectroscopic observations and analyses have been and will be an essential
tool in discovering the intricate details in the formation of a classical 
T Tauri star, as the spectral lines carry information on the kinematics
and physical conditions of the gases close to the star. Thus, red
absorptions extending to velocities in excess of $100~km~s^{-1}$ 
in Balmer lines and the $NaI$ doublet 
(Appenzeller \& Wolf 1977,
Edwards et al. 1994) indicate active accretion. Blue absorptions are rare
in optical lines, but are seen in Balmer lines, so there is also hint of 
outflows. In addition, the profiles of forbidden lines like $[OI]\lambda
6300$ and $[SII]\lambda 6731$ reveal the often presence of jet-like flows
(Kwan \& Tademaru 1988, Hirth, Mundt, \& Solf 1994). Even though the 
forbidden line emission originate at densities much lower than expected 
values in the vicinity of the star, the high speed of the jet signals that
the outflow likely starts from a deep potential well as either a stellar wind
or an inner disk wind that is subsequently collimated. In that case ejection
of matter is as inherent a characteristic as accretion of matter in the
star formation process.

Ultimately the unravelling of the contributions to the line intensities and
profiles from different kinematic flows and the derivation of quantitative 
measures like mass flow rates require modelling and analysis of the
strong emission lines. Natta et al. (1988) first examined the excitation and
ionization of hydrogen in a stellar wind and found that the hydrogen line 
fluxes calculated for the same mass loss rate can span a wide range owing
to differences in the gas temperature and in the stellar Balmer continuum.
Hartmann et al. (1990) modelled both line fluxes and profiles in a stellar
wind which, driven by Alfven waves, is characterised by large turbulent
velocities in the accelerating portion of the flow. While the calculated
fluxes cover the observed range, the Balmer line profiles are clearly unlike
observed ones in being highly asymmetric with the red side much stronger
than the blue side. The weakness of the blue side occurs because while the
bulk of the hydrogen emission arise from the inner, denser and highly
turbulent region, the outer, less strongly excited expanding envelope blocks
the blue emission to an observer. Mitskevich, Natta, \& Grinin (1993) also
modelled the $H\alpha$ profile by postulating a flow that accelerates to a
peak velocity and then decelerates towards zero. This double-valued
velocity structure will produce an anomaly between the red and blue emission
of an optically thick line at intermediate velocities, as the blue(red)
emission will then arise from the outer(inner) part of the surface of
constant observed velocity ($v_{obs}$), where the excitation temperature is 
lower(higher). Then, to account for the observed ranges in shape and depth of
the apparent blue absorption, Mitskevich et al. (1993) advocated a clumpy flow
for the flexibility in varying the degree of shielding between the outer and
inner parts of the constant $v_{obs}$ surface. They, however, adopted a
parametric function for the dependence of the line excitation temperature
on position, and it is not clear if the observed fluxes and profiles 
of several Balmer lines can be reproduced self-consistently.

The higher Balmer lines, as well as $Pa\beta$ and $Br\gamma$ occasionally
show red absorptions (Edwards et al. 1994, Folha \& Emerson 2001). The 
clear indication of an infalling flow by the broad red absorption, and the
realization that the optical/UV continuum excess can arise from the
impact footpoints on the star (Calvet \& Gullbring 1998) likely contribute
to motivating the studies of hydrogen emission in an accretion flow from
the disk along a dipolar trajectory (Muzerolle, Calvet, \& Hartmann 1998a,
2001, Kurosawa, Romanova, \& Harries 2008). The model hydrogen line fluxes
generally agree with observed values and the model line profiles are
centrally peaked with small blue centroids. The much better comparison of
these calculated profiles with observed ones, in conjunction with findings
of strong magnetic field strengths (Johns-Krull 2007) and theoretical
investigations that probe the initiation of the accretion flow and the
associated angular momentum exchange between disk and star (K\"{o}nigl \&
Pudritz 2000, Mohanty \& Shu 2008, Romanova et al. 2007) spurs the burst of 
activities on the current paradigm of magnetospheric accretion.

While magnetospheric accretion clearly occurs, the origin of the
hydrogen emission in the accretion flow is not without question. On
observational grounds the main issue is the narrow width of the model line
profile (Folha \& Emerson 2001, Kurosawa et al. 2008). Other issues concern the
observed high blue wing velocities, stronger blue emission at the line wings, 
and the sometimes large blue centroids. These characteristics are used as
arguments against the $HeI\lambda 5876$ emission originating in the
accretion flow (Beristain, Edwards, \& Kwan 2001, hereafter BEK01),
but are also present in observed hydrogen profiles
(Folha \& Emerson 2001). With the advance of infrared spectroscopy, it is
natural to follow up the $HeI\lambda 5876$ study by observing the 
$HeI\lambda 10830$ line which, being the transition immediately below
$HeI\lambda 5876$, will have a higher opacity and be more effective in
absorbing the stellar and veiling continua. The ensuing $1\mu m$ 
spectroscopic survey of 38 CTTSs (Edwards et al. 2006, hereafter EFHK06)
indeed produces additional information not conveyed by previously observed
lines. It reveals that $HeI\lambda 10830$ has the strongest propensity of
showing absorption features, including broad blue absorptions indicative
of radial outflows, sharp, narrow, blue absorptions indicative of disk
winds (Kwan, Edwards, \& Fischer 2007), red absorptions some of which are
so broad and deep that challenge conventional assumptions of accretion
flow structure (Fischer et al. 2008), and central absorptions.

The new $1\mu m$ observational results bring forth several new insights
into the problem of helium and hydrogen line formation in CTTSs. First, the 
significant $HeI\lambda 10830$ optical depth ($\sim 1$ or higher) in several
different kinematic flows, together with the high excitation energy
($\sim 20~eV$) of its lower state, suggests that excitation via UV
photoionization needs to be considered. Second, with the establishment of 
frequent presence of radial outflows, the issue regarding the 
comparatively rare occurrence of broad
blue absorptions in Balmer lines now concerns the structures and physical 
conditions of those winds. Third, $HeI\lambda 10830$ emission is common and 
comparable in strength to $Pa\gamma$ emission among the CTTSs with the 
strongest hydrogen lines (EFHK06). Origin of the helium emission in an
accretion flow faces high hurdles because of the arguments put forth earlier
with respect to the $HeI\lambda 5876$ emission, and because the physical
conditions needed for strong $HeI$ emission (e.g., $T>10^4~K$) may be taxing
for a flow in which the gas is primarily in free fall. If the $HeI$ emission
originate in a radial wind, hydrogen emission from the wind may also be
significant, as it is more easily produced, and needs to be re-examined.

We will attempt to address the above issues in this paper. Unlike earlier
investigations which adopt a particular flow structure and evaluate how the
model line fluxes and/or profiles fare with observations, we take a simpler
but hopefully  more general approach. We first calculate the atomic/ionic 
excitation at a local point and determine how the line opacites depend
on the local physical conditions. We will use the observed relative
opacities among the lines to shed light on the requisite physical conditions.
The local excitation calculations also produce line emissivity ratios that
can be compared with observed line flux ratios to further delimit the
physical conditions. These local excitation results should be fairly
independent of the flow structure. Then, to contrast between the outflow
and infall velocity fields, we note that a major difference lies in the
emission area contributing to the observed flux at a high $\vert v_{obs}
\vert$ and calculate the specific fluxes of the more important lines
at $\vert v_{obs} \vert = 150~km~s^{-1}$ for both a radial wind and an
accretion flow.

Crucial to our endeavor is an observational data set that covers simultaneously
both optical and $1\mu m$ spectral regions. $HeI\lambda 10830$ is a key line 
because of its propensity in showing absorption features, thereby indicating
presences of particular kinematic structures. Deriving strong constraints
on the physical conditions giving rise to $HeI\lambda 10830$ formation,
however, needs another helium line, and $HeI\lambda 5876$ is ideal both
for its being a fairly strong line and its position as antecedent of
$HeI\lambda 10830$ in a radiative cascade. Such a data set of CTTSs,
observed at optical and $1\mu m$ wavelengths simultaneously or nearly
simultaneously has been procured by Edwards et al.
(2010). Only a few of the objects are selected for use here to provide
line ratios that are key to unravelling the origins of the
line emission.

The outline of this paper is as follows. We describe the rationale and
methodology of our model calculations in the next section and give details
of the atomic/ionic models employed in \S 3. We present in \S 4 the
relevant observational information. In \S\S 5, 6, 7 we present model results
and compare them with observational data on line opacities, flux ratios, and
specific fluxes respectively to delimit the requisite physical conditions.
We summarize the findings thus deduced in those three sections and utilize
them, together with other observational information, to decide on the
locations of the $H$, $HeI$, and $CaII$ broad emission in \S 8. We discuss the
implications of our proposed origins of the line emission in \S 9, and
review the major conclusions in \S 10.

\section{Line Excitation Model}

The line emission from a gas depends on the local physical conditions and
the radiative transfer. For three of the four kinematic structures revealed 
by the $He I \lambda 10830$ absorptions the high speed of each flow and
the consequent large velocity gradients present isolate the radiative
interaction to a region small in comparison with the overall size of 
the kinematic structure. Then if the physical conditions within that
region are taken to be uniform, both the excitation of the gas and the
consequent line emission depend only on local physical quantities,
namely, density, kinetic temperature, and velocity gradient. The 
spontaneous emission, stimulated absorption and emission of photons
in a line together produce an effective emission rate given by
$A\beta$, with $A$ being the spontaneous emission rate and $\beta$ the
escape probability given by $(1-e^{-\tau})/\tau$, where the line optical depth
\begin{equation}
\tau={g_u\over g_l}{{A\lambda^3}\over 8\pi}(N_l-N_u{g_l\over g_u})
{dl\over {dv}} 
\end{equation}
depends on the local population $N_l(N_u)$ of the lower(upper) level and
the velocity gradient $dv/dl$ (Sobolev 1960). The local emissivity of the
line ($erg~s^{-1}~cm^{-3}$) is then $N_u A \beta h\nu$.

We take advantage of the above reduction of a global radiative transfer
problem to a local one in generating model results for comparison with
observational data. We will be primarily interested in the stellar wind
and the accretion flow in our attempt to ascertain the origin of the broad
line emission. We consider a point roughly in the middle of the
flow, specifically at a distance $r$ from the star where the flow speed
$\vert v \vert$ reaches $150~km~s^{-1}$. The calculated line opacity
at this position is taken to be representative, and the ordering in
magnitude of the opacities of different lines will be compared with
the observational ordering of the lines in propensity of showing an
absorption to delimit the requisite physical conditions. The 
local ratio of emissivities of two lines will also
be compared with the observed line flux ratio. This appears to be,
at first sight, a gross approximation, since each line flux is an 
integration of the emissivity over the entire kinematic
structure and clearly the density at other parts
of the flow, at least, is likely several times  
larger or smaller than the value at $\vert v \vert =150~km~s^{-1}$.
However, the same local excitation calculation applies to all positions,
only for different physical parameters, so when the value of an
emissivity ratio is presented as a function of density and temperature,
one can judge from the dependences how the results will be affected 
when averaged over a range of density and/or temperature. The local
model calculation has the advantage of enabling us to explore a much
broader parameter space, e.g. over four orders of magnitude in density
span, as well as to include many more lines, both from a given atom by
incorporating more energy levels and from different atoms. It will be seen
later that a comparison of different emissivity ratios with observed
line flux ratios does indicate clearly enough the necessary
physical conditions for us to infer the source of the line emission,
and that this deduction is not affected by the mentioned approximation.

For the stellar wind we assume radial streamlines from the star. The
velocity gradient transverse to the radial direction is $v/r$. We expect
the velocity gradient in the radial direction to be larger, for an
acceleration of the gas sufficiently strong to enable the gas to
escape. When the velocity gradient is not isotropic the escape probability
is dependent on direction. We will not be concerned with this nuance,
and simply assume an effective isotropic velocity gradient of $2v/r$.
The error in this estimate is not significant since the velocity gradient
enters only as a factor in the line opacity and is always multiplied with the
density, which varies over a much broader range. We take $r=4R_\ast$
to illustrate the stellar wind model.

We also assume radial streamlines toward the star for the accretion
flow. At $\vert v \vert \geq 150~km~s^{-1}$ the infall trajectory,
even in the dipolar geometry, is approaching radial. The investigation
of $HeI\lambda 10830$ red absorptions also finds that some red
absorptions are so broad and strong that they are accounted for best by
radial infall (Fischer et al. 2008). The same radial geometry also
facilitates comparison of the model results between the two flows.
A gas particle infalling onto a star of $0.5 M_\odot$ and $2R_\odot$
will attain a speed of $150~km~s^{-1}$ at $2.77$ and $2.06 R_\ast$
if it starts from $8$ and $4 R_\ast$ respectively, so
we take $r$ to be $2.5 R_\ast$ for the
accretion flow, and assume an effective isotropic velocity gradient
of $2v/r$. The latter assumption can be quite wrong because, unlike the
stellar wind, the accretion flow at high speeds
fills only a very small solid angle, and the above work on red
absorptions also finds that the infall streamlines may not uniformly fill
the solid angle. In this case the velocity gradient is better given by
$\delta v/\delta l$, where $\delta v$ is the thermal/turbulence line
width and $\delta l$ the transverse size of an infall bundle. We have
no grasp on $\delta l$ and only note that $\delta v/\delta l$ may be
closer to $2v/r$ even if $\delta l$ is very different from $r$. As
mentioned before, the velocity gradient enters into the opacity with the 
density, which is varied over a wide range, and emissivity ratios will 
be presented also as a function of line opacity, so the effect of
a very different velocity gradient can still be gleaned from the results.

The difference between the assumed velocity gradients for the two flows
is not large. It will also be seen that the model results depend much more
strongly on density than the velocity gradient, so a change in the latter
can be compensated by a much smaller change in the former. Thus the
calculated dependences of the local line opacities and emissivity ratios
on physical conditions are applicable to all flow structures with a
comparable velocity gradient to within an order of magnitude.

The specific flux of a line, on the other hand, depends on the volume of
emission, so it is sensitive to $r$, the actual flow geometry, and the
velocity at which it is calculated. In particular, because the correlation 
of velocity with position is vastly different between the stellar wind
and accretion flow, we expect their line fluxes to be quite different, 
even if line emissivity ratios are similar. It is with this consideration
in mind that $\vert v \vert =150~km~s^{-1}$ is specifically chosen. We
think the line flux near this velocity or higher is a truer test of the
accretion flow model than that near zero velocity, as it is much more
constrained. Independent of the accretion flow geometry, the emission at
high speeds must arise from distances quite close to the star. Then the
focussing of the streamlines toward the star confines the solid angle of
the accretion flow at high speeds, which is also constrained by the
small area covering factor, typically $\leq 0.03$ (Calvet \& Gullbring 1998),
of the shocks marking the accretion footpoints. Thus the emission volume
of the gas at a high speed is well constrained, thereby making the
corresponding line flux a more revealing diagnostic.

The observed line emission at $v_{obs}=v$ arise from locations at distances
other than $r$ because emission from positions with higher speeds than $v$
also contribute when their projected velocities along the line of
sight equal $v_{obs}$. To calculate it for the stellar wind we assume that
from $r$ to $1.5r$ the flow speed increases
linearly from $v$ to $2v$. When the
line is optically thick the observed flux at $v_{obs}$ depends on the
excitation temperature and the
projected area with an observed velocity of $v_{obs}$. In Figure 1 the right
dashed curve shows, in the $x-z$ plane, the contour of $v_{obs}=-150~km~s^{-1}$
for a spherical wind. It can be seen that the projected area equals
$\pi(1.5r sin 60^o)^2$ or $27(r/4R_\ast)^2\pi R_\ast^2$. When
the line is optically thick the observed specific flux is then, in the
simple case of a constant excitation temperature,
\begin{equation}
F_{v_{obs}}={2h\nu^4\over c^3}{1\over {e^{h\nu/
kT_{ex}}-1}}{27\pi R_\ast^2\over {d^2}}\ ,
\end{equation}
where $T_{ex}$ is the excitation temperature between the upper and lower
levels, and $d$ is the distance to the star.
When the line 
is optically thin, the observed flux is obtained from integrating the
line emissivity over volume. If the latter varies with distance $p$ 
from the star as $p^{-a}$, the observed specific flux is
\begin{eqnarray}
F_{v_{obs}}&=&{{N_u A\beta h\nu}\over{4\pi d^2}}\int_r^{1.5r} 
{({r\over p})}^a{{4\pi p^2}\over{2v({2p\over r}-1)}}dp\nonumber\\
&=&{{N_u A(1-e^{-\tau})h\nu r^3} \over{2d^2 \tau v}}\int_1 ^{1.5}
{x^{2-a}\over{2x-1}} dx\ .
\end{eqnarray}
Both $N_u$ and $\tau$ are determined at
$r$. For $a$ between $0$ and $4$ the integral ranges from $0.53$ to $0.25$.
Substituting the expression for $\tau$ given earlier in the denominator,
the factor in front of the integral can be rewritten as
\begin{equation}
(1-e^{-\tau}){2h\nu^4\over c^3}
{1\over{e^{h\nu/kT_{ex}}-1}}{4\pi r^2\over {d^2}}\ .
\end{equation}
In order that, when a line transits from optically thin to optically thick, 
the observed specific flux increases smoothly to the earlier limit for
an optically thick line,
we choose the integral to
be $0.422$ or an $a$ of $1$. The error in deriving the flux of an optically
thin line is then about $\pm 40$\%. It turns out that, with the exception 
of $OI\lambda 8446$, the other lines studied are all optically thick under
the physical conditions responsible for the observed line emission. The 
error in the specific flux of an optically thick line arises mostly from the 
assumption of a constant excitation temperature, but one can gauge from
the presented results how a distribution of density or temperature affects
$F_{v_{obs}}$.

For a spherical radial infall the calculation of the specific flux when a line  
is optically thin is analogous to the spherical wind case. It involves 
integration from $R_\ast$ to $r$ with an infall velocity distribution, and is
also made to ensure a smooth transition between the optically thin and thick 
regimes. In the latter regime the specific flux depends on the projected
area, and the contour of $v_{obs}=-150~km~s^{-1}$ is shown by the left
dashed curve in Figure 1. The corresponding projected area is $1.286\pi 
R_\ast^2$.
This much smaller projected area at $\vert v_{obs}
\vert=150~km~s^{-1}$ is a fundamental characteristic of any accretion
flow model. The contrast against the stellar wind value is even larger when
the small solid angle of the accretion flow near the star is taken into
account. When the wind/infall is not spherical, the projected area depends
on viewing angle, but we will not consider this nuance and simply obtain
the observed line flux by multiplying the result for the spherical case
by the filling factor of the flow in solid angle, $F_\Omega$. For the
stellar wind and accretion flow we adopt $F_\Omega=0.5$ and $0.2$
respectively, keeping in mind that the $HeI\lambda 10830$ line sometimes
shows strong emission but only a highly displaced shallow blue absorption
(Kwan et al. 2007), and that $F_\Omega=0.2$ is needed to model the few
very strong red absorptions (Fischer et al. 2008). Occultation of the
line emission by the star and the disk is not taken into account. If it
were, it can be seen from Figure 1 that in the infall case the line flux
at $v_{obs}\leq -150~km~s^{-1}$ would be severely curtailed.

Three physical parameters are important for the local excitation
calculations. They are density, temperature, and ionization flux.
The range of kinetic temperature, $T$, investigated is $0.5\times 10^4$
to $3\times 10^4~K$. We use the hydrogen nucleon number density, $N_H$,
to indicate the number density, and assume a solar composition of the
gas. With the number density of $H:He:O:Ca:Na$ being in the ratio
$1:0.0793:4.9\times 10^{-4}:2.2\times 10^{-6}:2\times 10^{-6}$ (Lodders
2003), the total nucleon number density is then $1.08 N_H$ and the
mass density is $\rho =1.325 N_H m_H$, with $m_H$ being the mass of the
hydrogen atom. The range of $N_H$ explored is $10^8$ to $2\times
10^{12}~cm^{-3}$. For a laminar flow the mass flux is then given by
$4\pi r^2 \rho v F_\Omega$ or $10^{-9}(N_H/10^9~cm^{-3})(r/4R_\ast)^2
(v/150~km~s^{-1})(F_\Omega/0.5)M_\odot~yr^{-1}$.

We include photoionization as a means of excitation. 
As mentioned in $\S 1$ the high opacities of $HeI\lambda 10830$ in
the accretion flow, stellar wind, and disk wind suggest that photoionization
of helium from its ground state is likely. As a
rough estimate of this rate, we note that a luminosity of $10^{-4} L_\odot$
in photon energies above $24.6~eV$, situated at the star, will produce 
at $4R_\ast$ a $HeI$
photoionization rate of 
\begin{equation}
\gamma_{HeI}=2\times 10^{-2}({{\alpha -1}\over {\alpha +2}})({{4R_\ast}
\over r})^2~{L\over 
{10^{-4} L_\odot}}~s^{-1}\ ,
\end{equation}
where $\alpha >1$ is the power law index of the
luminosity energy distribution. The greatest uncertainty in this rate,
however, is the attenuation between the $UV$ luminosity source and
the local point considered, since the mean free path for an optical depth of
unity at the ionization threshold is small, $\sim 1.3\times 10^9~cm$
for a $HeI$ density of $10^8~cm^{-3}$. To circumvent this problem we
take $\gamma_{HeI}$ as
a parameter. The $UV$ source will also ionize
hydrogen from its ground state. The ratio $\gamma_{HI}/\gamma_{HeI}$ ranges
from 1.6 to 4 for $\alpha$ between $1.5$ and $3$ if the luminosity
source is not attenuated. The attenuations at the two thresholds can
differ a lot and it is not clear which one is stronger. In most of the
calculations we simply adopt $\gamma_{HI}=2\gamma_{HeI}$, but will comment
on the effects of differential attenuation. The condition
of $\tau_{HeI\lambda 10830}\geq 1$ and $\tau_{Pa\gamma}\leq 1$, posed by the
much more frequent occurrence of absorption features in $HeI\lambda 10830$
than $Pa\gamma$, will require a minimum $\gamma_{HeI}$ at temperatures low
enough that collisional excitation of $HeI$ is ineffective. It turns out
that this limit is $\sim 10^{-5}~s^{-1}$, so we will present results for
$\gamma_{HeI}=10^{-4}$ and $10^{-5}~s^{-1}$. The latter are much smaller
than the unattenuated value given in the above equation for the hypothetical
$10^{-4}~L_\odot$ UV source.

The stellar and veiling continua, which peak at optical wavelengths,
are effective in ionizing the excited states of hydrogen and helium, as
well as the ground state and excited states of $CaII$ and $NaI$. We
assume a CTTS of temperature $T_\ast = 4000~K$ and radius $R_\ast = 2R_\odot$,
and a veiling continuum given by a blackbody of temperature $8000~K$
covering 3\% of the stellar surface area. At 5000\AA $~$ the veiling continuum
is then about as strong as the stellar continuum.

To summarize, the physical parameters in our local excitation model are
primarily $T$, $N_H$, $\gamma_{HeI}$, and $2v/r$. The first three 
parameters are explored over broad enough ranges to cover all expected
possibilities, such as line optical depths from $10^{-2}$ to $10^3$, and
emissivity ratios from values smaller to values larger than corresponding
observed line flux ratios. From these local excitation calculations 
line opacities and emissivity ratios are compared with observational
data to delimit the requisite physical conditions. These results are fairly
general and not strongly dependent on the kinematic structure. The line flux,
on the other hand, is sensitive to the actual flow geometry. To illustrate
the contrast between the stellar wind and the accretion flow, the
line fluxes at $\vert v_{obs} \vert =150~km~s^{-1}$
are calculated for a stellar wind and an accretion flow
reaching a speed of $v=\vert v_{obs} \vert $ at $r=4R_\ast$
and $2.5R_\ast$ respectively. All these
results aid in deciphering the observed
line emission for their origin.

\section{Atomic Models}

CTTSs show many emission lines. The forbidden lines, such as $[OI]
\lambda 6300$, $[S II]\lambda 6731$, are likely formed at more than ten
stellar radii away and not germane to probing the accretion flow and
inner structures of disk and stellar winds. Among the permitted lines,
their profiles can be narrow ($FWHM\sim 20~km~s^{-1}$), broad ($FWHM
\sim 200~km~s^{-1}$), or composite with a narrow component atop a
broad component (BEK01). 
The narrow lines/components
are most likely formed at the sites where accreting streamlines impact
the star, so their fluxes convey information on the summed area of such
regions and the cooling history of the shocked gas. Here we are concerned
with the broad lines/components, as their widths indicate that they are 
likely formed as the gas accelerates either away from the star in a wind
or towards the star in an accretion flow. In the optical and infrared
domains these broad lines/components are, ordered roughly in decreasing
emission strength, Balmer lines, $CaII~H,~K$, and infrared triplet,
($HeI\lambda 10830$, Paschen and Brackett lines, $FeII$ lines), and
($HeI\lambda 5876$, $NaI~D$, $OI\lambda 8446$, $HeI\lambda 6678$, 
$OI\lambda 7773$, $FeI$ lines).

Several intrigues posed by the observed strengths of the above-mentioned
lines indicate that an understanding of the CTTS spectra must involve
examining the $HI$, $HeI$, and $CaII$ line excitations altogether.
One puzzle is the relative strength between the $HeI\lambda 10830$
and $Pa\gamma$ emission. The observed
$HeI\lambda 10830$ spectra(EFHK06), with prominent absorption features,
clearly demonstrate the high $\lambda 10830$ opacity, a consequence of
the metastability of its lower state, $2s~^3 S$. 
With $\tau_{HeI\lambda
10830}$ at least as large as
$\tau_{H\alpha}$, as inferred from their relative propensity in showing
an absorption, a $HeI~2s~^3 S$ level population comparable to or larger than 
the $HI~n=2$ level population is implied, and while $Pa\gamma$ competes
against $Pf\alpha$, $Br\beta$, and $H\delta$ for the de-excitation of $n=6$
$HeI\lambda 10830$ is the sole permitted radiative decay channel for its
upper level, $2p~^3 P$. Then the $HeI~2s~^3 S\rightarrow 2p~ ^3 P$ collisional 
excitation rate is larger than the hydrogen $n=2\rightarrow 6$ collisional
excitation rate. Thus collisional excitation should highly favor $HeI\lambda 
10830$ over $Pa\gamma$ emission, yet the two observed fluxes in emission
are almost the same.

The second puzzle is the very strong $CaII$ infrared triplet emission
in those CTTSs with strong $HeI\lambda 10830$ and $\lambda 5876$ emission.
The summed flux of the $CaII$ triplet rivals that of $H\alpha$. Collisional
excitation of the triplet has the advantage that their upper state is only
$3.1~eV$ above the ground state, so the triplet can be strong,
relative to $H\alpha$, at low temperatures, despite the low $Ca$ abundance.
However $CaII$ has an ionization potential of $11.9~eV$, much less than
the $HeI$ ionization potential of $24.6~eV$. Furthermore, the lower state
of the $CaII$ infrared triplet is metastable and only $1.7~eV$ above the
ground state, so it is well populated.Ionization from this level takes
$10.2~eV$, very slightly less than the energy of the $Ly\alpha$ photon.
A strong buildup of the $Ly\alpha$ intensity, through radiative trapping,
is required to sustain population in levels $n\geq 2$ of hydrogen 
for strong Balmer, Paschen, and Brackett line emission. It will
at the same time ionize $CaII$ from its metastable state and reduce the
$CaII$ fraction. Thus the strong $CaII$ triplet emission appear incongruous
with not only the $HeI\lambda 10830$ and $\lambda 5876$ emission but also
the hydrogen Paschen and Brackett line emission. In addition, the infrared
triplet are very optically thick, as their fluxes are nearly equal despite
a factor of $10$ difference among their oscillator strengths. Yet they
rarely show any absorption feature, even in CTTSs where their peak fluxes
are only comparable to the continuum flux.

In addition to calculating the $HI$, $HeI$, and $CaII$ line excitations,
we will also include those of $OI$ and $NaI$.
The $OI\lambda 11287$ and $\lambda 8446$ lines are produced primarily 
through the $Ly\beta$ fluorescence process (Bowen 1947), in which the
$OI~3d~^3D$ state is excited upon absorption of a $Ly\beta$ photon by the ground
state, and decays via emission of a $\lambda 11287$, a $\lambda 8446$,
and a $\lambda 1303$ photon in succession. Thus the $OI\lambda 8446$
flux is expected to correlate with the $Pa\gamma$ flux. The $OI\lambda
7773$ triplet, on the other hand, are formed via recombination and cascade and
collisional excitation, in a fashion similar to the $CaII$ infrared
triplet. So inclusion of the above $OI$ lines will further check
on the $HI$ and $CaII$ excitation conditions. The $NaI$ atom, with an
ionization potential of $5.14~eV$, is easily photoionized by the stellar
and veiling continua, and requires a much lower temperature for collisional
ionization than helium. The often appearance of absorption in the
$NaI$ doublet, in conjunction with absorption in $HeI\lambda
10830$, then sheds light on the pertinent physical conditions.

We will not examine the $FeII$ and $FeI$ lines here. There are many of 
them, from infrared to UV wavelengths. Their summed emission strength
may even exceed that of the hydrogen lines, so they are an important heat
sink, and must be counted in deriving the total energy generation rate.
With ionization potentials of $7.9$ and $16.2~eV$ for $FeI$ and $FeII$
respectively, the $FeI$ to $FeII$ flux ratio will also provide a constraint
on the ionization condition. A proper study of the $FeI$ and $FeII$
line excitations, however, needs good observational data on the fluxes
of the many $UV$ multiplets, which are not presently available. The large
number of lines and multiplets also makes it more suitable to study
them in a separate paper. We will also not consider the lines of highly
ionized metals, such as $CIV\lambda 1549$ and $OVI\lambda 1034$, partly
because of the lack of simultaneous $UV$ spectra and partly because 
understanding the excitations of these lines will likely benefit from
understanding the $HI$, $HeI$ and $CaII$ excitations first, rather
than vice versa.

In the following subsections we describe our model atom/ion for $HI$,
$HeI$, $OI$, $CaII$, and $NaI$, but leave the references for the atomic
parameters to an appendix.

\subsection{HI}

In modelling the hydrogen atom we assume that the level population among
degenerate energy states are thermally distributed, and use 15 distinct
energy levels, labelled $n=1-15$. Limiting the hydrogen atom to a fixed
number of levels can be questionable. This is because the
$n\rightarrow n+1$ collisional rate coefficient, $C_{n,n+1}$, increases
while the spontaneous emission rate of level $n$, $A_n$, decreases as
$n$ increases, so once the electron density $N_e$ exceeds the value
$A_n/C_{n,n+1}$, $\sim 5\times 10^7~cm^{-3}$ for $n=15$,
or an even lesser value if some of the radiative transitions
are optically thick, successive collisional excitations to higher levels,
effectively leading to ionization, provide the quickest route of depopulating
level $n$. Ignoring this population transfer to higher levels
produces error in the population of not
only level $n$, but also lower levels, since there is less population
return to the lower levels from collisional and radiative de-excitations of
upper levels. The proper number of levels to use is clearly dependent on
$N_e$, which in turn is dependent on $N_H$, $T$, and $\gamma_{HI}$.
To ensure that $15$ levels are adequate over our explored density range,
we have performed the following test. The net rate of population transfer 
from level $15$ to $16$ via collisions is given by $R_{15\rightarrow 16}
=(N_{15}N_e C_{15,16} - N_{16}N_e C_{16,15})$, where $N_{15}$ and
$N_{16}$ are the population in level $15$ and $16$ respectively. The
local excitation calculations with $15$ levels show that the
$n=13\rightarrow 14$ and $n=14\rightarrow 15$ excitation
temperatures are higher than $6000~K$ for $T$ between $7500$
and $1.5\times 10^4~K$, so, if the $n=15\rightarrow 16$ excitation 
temperature is $\geq 6000~K$, $R_{15\rightarrow 16}$ has a top value of
$\sim 0.01~N_{15}N_e C_{15,16}$, which is about one-half the direct
collisional ionization rate, $N_e C_{15,\infty}$, from level 15. To estimate
the effect of including this additional route of population transfer for
level $15$, we have artificially doubled $C_{15,\infty}$ and repeated
the calculations. The differences in the hydrogen level population
and line fluxes are less than a few percent between the two sets of
calculations, and we are confident that the use of $15$ levels is adequate to
obtain reliable Paschen line fluxes up to the transition from $n=13$
to 3.

\subsection{HeI}

Our HeI model atom consists of the 19 lowest energy states. The spin of
the two electrons in each state can add up to $1$ or $0$, so the 19
states can be separated, according to the degeneracies of the total spin,
into a ladder of triplets and one of singlets, since radiative transitions
across the ladders are forbidden by the electric-dipole selection rules.
For the same energy quantum number $n$ the angular momentum states
$l=1-n$, unlike those in the case of $HI$, are separated in energy by
many thermal Doppler widths, so we treat them as distinct energy levels.
Thus, our 19 level atom consists of 10 singlets with energy quantum number
$n=1-4$, and 9 triplets with energy quantum number $n=2-4$. Figure 2 shows
a schematic of the $11~n=1-3$ levels, and indicates several important 
transitions.

The lower level of the $\lambda 10830$ transition is highly metastable. 
Its radiative decay rate is only $1.7\times 10^{-4}~s^{-1}$. For our explored
density and temperature ranges, collisions with electrons provide the 
speediest way of returning population to the ground state. A more important
route than direct collisional de-excitation is collisional excitation to
the singlet $2s~^1 S$, followed by another collisional excitation to
$2p~^1 P$ and then emission of a $\lambda 584$ photon, which can either
escape or photoionize hydrogen. The latter means of depleting $\lambda 584$
photons is particularly important once $\tau_{HeI\lambda584}$
becomes large. To determine its rate of depopulating $2p~^1 P$, we note
that the mean free path for hydrogen ionization is $l_{Hiz}=1/(N_{HI}
\sigma_{584A})$, where $\sigma_{584A}$ is the photoionization cross-section
at $584$\AA, while the absorption mean free path, averaged over a thermally
broadened profile, is 
\begin{equation}
l_{abs}\approx {{8\pi \Delta v_D[ln(\tau_{HeI\lambda 584} +2.72)]^{0.5}}
\over {B_{lu} hcN_{HeI}}}\ ,
\end{equation}
where $\Delta v_D=(2kT/m_{He})^{0.5}$ is the helium thermal Doppler velocity
width, and $B_{lu}$ is the Einstein stimulated absorption coefficient. The
rate ($s^{-1}$) of depopulating $2p~^1 P$ due to $\lambda 584$ ionizing
hydrogen is then given by the product of the spontaneous emission rate,
$A_{HeI\lambda 584}$, and the ratio $l_{abs}/l_{Hiz}$. As an illustration, in
the case when hydrogen and helium are mainly neutral, this rate is about
$2\times 10^{-3} A_{HeI\lambda 584}$, but almost $2A_{HeI\lambda 20581}$. This
process is important in determining not only the population in
excited singlets, but also, through the reduced population flow from
singlets to triplets via collisions, the triplet population.

The ionization of hydrogen by $\lambda 584$ photons also affects the hydrogen
ionization structure, but only slightly. If helium excitation is produced 
primarily by $UV$ continuum photoionization, the 
production rate ($s^{-1}~cm^{-3}$) of $\lambda 584$ photons must be less
than the rate of continuum photoionization or $N_{HeI}\gamma_{HeI}$.
The rate of hydrogen ionization by $\lambda 584$ photons is then less
than $N_{HeI}\gamma_{HeI}=0.079 N_H \gamma_{HeI}$, which is less than the
rate of continuum photoionization of hydrogen, $N_{HI}\gamma_{HI}$. If
thermal motion is the energy source for helium excitation, then collisional
excitation and ionization are much more efficient for hydrogen than for
helium. Therefore, we have not implemented this coupling between the two
ionization structures, which would require an iterative procedure.

In analogy to $\lambda 584$ photons ionizing hydrogen, $Ly\alpha$ photons
can ionize helium from its two metastable states, $2s~^3S$ and $2s~^1S$,
which will have the bulk of the excited state population. This ionization
rate ($s^{-1}$) is readily deduced from the earlier derivation. It
equals $N_2 A_{Ly\alpha} l_{abs} \sigma_{1216A}$, where $N_2$ is the
hydrogen population in $n=2$, $l_{abs}$ the analogous $Ly\alpha$ 
absorption mean free path, and $\sigma_{1216A}$ the $2s~^3S$ or $2s~^1S$
photoionization cross-section at $1216$\AA. We have included this process
in addition to the usual photoionizations by the stellar and veiling
continua. The corollary effect of depopulating the hydrogen $n=2$ level
is insignificant, as we have verified from the results of
the calculations.

\subsection{OI}

Figure 3 shows a schematic of the energy levels of $OI$ pertinent to
our calculations and the important radiative transitions. The ground state
$2p^4~^3P$ has three fine-structure levels, with energy separations such that
the forbidden transitions $^3P_1\rightarrow ^3P_2$ and $^3P_0\rightarrow ^3P_1$
have wavelengths of $63~\mu m$ and $146~\mu m$ respectively. The upper
state $3d~^3D$ likewise has three fine-structure levels, but with much
smaller energy separations. The $2p^4~^3P_2$ transitions to 
$3d~^3D_1$, $^3D_2$, and $^3D_3$ have wavelengths almost the same as
$Ly\beta$, to within a hydrogen thermal Doppler width. With charge exchanges
maintaining $OI/OII$ close to $HI/HII$, absorption of $Ly\beta$
photons is a fortuitous enhancement of $3d~^3D$ excitation, thereby 
leading to strong $\lambda 11287$ and $\lambda 8446$ emission.

The $\lambda 11287$ photon emissivity ($cm^{-3}~s^{-1}$) resulting
from this $Ly\beta$ fluorescence process (Bowen 1947) depends on the
$Ly\beta$ intensity and the probability that $3d~^3D$ decays via emission
of a $\lambda 11287$ photon. Because $Ly\beta$ is expected to be very
optically thick, we assume its specific intensity to be given by the blackbody
value at the hydrogen $n=1\rightarrow 3$ excitation temperature, i. e.,
$I_{Ly\beta}=(2h\nu^3_{Ly\beta} /c^2)(9N_1/N_3 -1)^{-1}$, where $N_3$
and $N_1$ are the hydrogen $n=3$ and $1$ level population respectively.
For the probability of $\lambda 11287$ emission, we consider first the
upper level $3d~^3D_1$. It can 
decay to all three fine-structure levels of the ground state as well
as to $3p~^3P$, so the probability of $\lambda 11287$ emission upon
absorption of a $Ly\beta$ photon through the $2p^4~^3P_2
\rightarrow 3d~^3D_1$ transition is
\begin{equation}
P={A_{\lambda 11287}\beta_{\lambda 11287}\over{A_{\lambda 11287}
\beta_{\lambda 11287}+A_{^3D_1\rightarrow ^3P_0}\beta_{^3D_1\rightarrow ^3P_0}+
A_{^3D_1\rightarrow ^3P_1}\beta_{^3D_1\rightarrow ^3P_1}
+A_{^3D_1\rightarrow ^3P_2}}}\ ,
\end{equation}
where $A$ signifies the spontaneous emission rate and $\beta$ the 
escape probability. The $^3D_1\rightarrow ^3P_2$ escape probability equals 1,
even though the transition is most likely very optically thick, because
the $Ly\beta$ opacity is even larger, so the emitted $\lambda 1025.77$
photon will not be absorbed by $OI$. The probabilities of
$\lambda 11287$ emission via $Ly\beta$ absorptions through the $^3P_2
\rightarrow ^3D_2$ and $^3P_2\rightarrow ^3D_3$ transitions
can be written out analogously. The most common situation is that
$\lambda 11287$ is optically thin and all the $2p^4~^3P\rightarrow 3d~^3D$
transitions are sufficiently optically thick that $3d~^3D$ decays to
the ground state via emission of $\lambda 1026$ photons that are absorbed by
hydrogen. In this case the above expression reduces to $P=A_{\lambda 11287}
/(A_{\lambda 11287} + A_{^3D_1\rightarrow ^3P_2})$, and the
$\lambda 11287$ photon emissivity ($s^{-1}~cm^{-3}$)
via $Ly\beta$ pumping is simply
\begin{eqnarray}
R_{Ly\beta}&=&{5\over 9}N_{OI}{N_3\over{9N_1 - N_3}}2.077
A_{\lambda 11287} \nonumber\\
&=&1.41\times 10^6~N_{OI}{N_3\over N_1}\ , 
\end{eqnarray}
with the approximation that $5/9$ of the $OI$ population is in the
$2p^4~^3P_2$ level. Even if $N_3/N_1$ is as low as $10^{-9}$, $R_{Ly\beta}$
is still larger than what can be brought about through $UV$ continuum
photoionization or direct collisional excitation at $T\leq 10^4~K$.

The absorption of $Ly\beta$ photons by $OI$ has the corollary effect
of depopulating the $HI~n=3$ population. We have included this process
in the hydrogen excitation calculations, but only with the simple
expression enumerated above for the most common situation in order to
avoid an iterative procedure involving $HI$ and $OI$ calculations.
With $N_{OI}/N_1 = 4.9\times 10^{-4}$ it is readily seen from the above
expression for $R_{Ly\beta}$ that the corresponding $n=3$ de-excitation
rate is $690~s^{-1}$. While it can rival $Ly\beta$ escape, it is much
weaker than $H\alpha$ escape at low densities and collisional de-excitation
at high densities, so the actual effect on the hydrogen level population
is insignificant.

The $Ly\beta$ fluorescence process enhances not only $\lambda 11287$
emission but, through the subsequent radiative cascade back to the ground
state, also $\lambda 8446$ and $\lambda 1303$ emission. The resulting
$\lambda 11287/\lambda 8446$ flux ratio is then just the ratio of the
photon energies or $0.75$. However, collisional excitation favors 
$\lambda 8446$ over $\lambda 11287$ emission, so a smaller observed flux
ratio is a measure of the relative contribution between the two
processes. Absorption in $\lambda 8446$ is occasionally seen. It is
facilitated by the population built up in $3s~^3S$ through $Ly\beta$
fluorescence and sustained via radiative trapping of $\lambda 1303$
photons.

The $\lambda\lambda 7772,7774,7775$ triplet emission is produced from
recombination and cascade, and collisional excitation. The metastable
nature of the lower level $3s~^5S$, whose spontaneous emission rate is
only $5\times 10^3~s^{-1}$, sustains a comparatively large population
in that level and enhances the collisional pathway of triplet emission.
It likewise helps to bring about absorption of the stellar and veiling
continua through the larger line opacity.

In performing the $OI$ excitation calculations we assume that the 
fine-structure states are populated in proportion to their degeneracies,
and use a single level to represent them. This simplification is appropriate
when the transitions between those
fine-structure states and another level are separated in energy by about
one thermal Doppler width or less. When this is not the case, as for the
$3s~^3S\rightarrow 2p^4~^3P$ or $3p~^5P\rightarrow 3s~^5S$ transitions, 
the simplified procedure works fine when the transitions are optically thin  
but, when they are optically thick, does not take into account the 
availability of several radiative channels for de-excitation. Also, when
collisional de-excitation dominates over radiative de-excitation, it
underestimates the total emitted flux by a factor equal to the number of
distinct (Doppler-width separated) lines. To remedy this situation, we
re-define the line optical depth as the usual definition divided by the above
factor. It represents a sort of average of the optical depths of the separate
lines. For example, in our procedure this optical depth between $2p^4~^3P$
and $3s~^3S$, $\tau_{\lambda 1303}$, equals 0.6, 1.0, and 3.0 of 
$\tau_{\lambda 1302.2}$, $\tau_{\lambda 1304.9}$, and
$\tau_{\lambda 1306}$ respectively. This 
modification also produces the correct total emitted flux when the distinct
lines are all optically thin and when they all have optical depths greater
than $\sim 2$. There may be a small error in the level population when
some of the line optical depths lie between 0.5 and 2, but this regime occupies
a very narrow strip of our explored density range, and we are not overly
concerned.

\subsection{CaII, NaI}

$CaII$, with an ionization potential of $11.87~eV$, can be ionized quite
readily by even the veiling continuum if the latter has an energy
distribution that extends towards the far $UV$ like a blackbody. For
example, with our adopted veiling continuum of temperature $8000~K$ and
area covering factor of $0.03$, the photoionization rate of the 
$4s~^2S$ ground state at $4R_\ast$ is $3.3\times 10^{-4}~s^{-1}$, higher
than our assumed $UV$ continuum photoionization rate, which is
$\gamma_{CaII}\sim (0.5/6.3)\gamma_{HI}$. Moreover, the excited state
$3d~^2D$, the lower state of the $\lambda\lambda 8498, 8542, 8662$ triplet,
is metastable, with an Einstein A rate of only $1~s^{-1}$, so it is most
likely populated in thermal equilibrium with the ground state. Its lower
ionization potential and larger photoionization cross-section produce an
even stronger ionization rate, $3.4\times10^{-2}~s^{-1}$ with our adopted 
veiling continuum. As an example, if the $4s~^2S\rightarrow 3d~^2D$
excitation temperature is $7000~K$, ionization of $CaII$ from $3d~^2D$ is 30
times that from $4s~^2S$. Also, as noted in the beginning of \S 3, the $3d~^2D$
ionization potential is very slightly less than the $Ly\alpha$ photon
energy. This ionization rate by $Ly\alpha$ photons ($s^{-1}$) is (cf. \S 3.2)
\begin{eqnarray}
\gamma_{Ly\alpha}&=&N_2 A_{Ly\alpha} l_{abs} \sigma_{3d~^2D}\nonumber\\
&=&5.54\times 10^4~{N_2\over N_1}({T\over{10^4~K}})^{0.5}
[ln(\tau_{Ly\alpha} + 2.72)]^{0.5}\ ,
\end{eqnarray}
where $\sigma_{3d~^2D}=6.15\times 10^{-18}~cm^2$ is the $3d~^2D$ 
photoionization cross-section at threshold. In the excitation calculations
the hydrogen $n=2$ to $1$ population ratio, $N_2/N_1$, has a value often
larger than $2\times 10^{-7}$, that needed for $\gamma_{Ly\alpha}$ to
equal $3.4\times 10^{-2}~s^{-1}$, the $3d~^2D$ photoionization rate by the
veiling continuum. Thus, $Ly\alpha$ photons constitute an even more 
potent source of $CaII$ ionization.

Our model $CaII$ atom consists of just the three levels $4s~^2S$, $3d~^2D$,
and $4p~^2P$. Both $3d~^2D$ and $4p~^2P$ have two fine-structure states.
The two $4p~^2P\rightarrow 4s~^2S$ transitions are the well known $H$ and
$K$ lines at $\lambda\lambda 3968$ and $3934$ respectively, while
the three $4p~^2P\rightarrow 3d~^2D$ transitions constitute the infrared
triplet. Like $OI$, we assume the fine-structure states to be populated
in proportion to their degeneracies and re-define the line optical depth to
reflect the number of distinct lines. The observed line fluxes indicate
that all five lines are very optically thick, so our simplified procedure
of lumping into one single level the population in all associated 
fine-structure states is fine.

Our model $NaI$ atom consists of just the ground level $3s~^2S$ and an
excited level comprising the two fine-structure states of $3p~^2P$,
whose radiative decays give rise to the well known $\lambda\lambda 5896,
5890$ doublet. The lower state of the doublet being the ground state
helps to bring about strong absorption and/or emission in the doublet.
On the other hand $NaI$, with an ionization potential of only $5.14~eV$,
is easily photoionized or collisionally ionized from the ground level,
and even more so from the excited level, so the sodium ionization 
structure is key. 

\section{Observational Information}

The local excitation calculations can produce line opacities and 
emissivities over an extensive parameter space of physical conditions,
but need information from observational data to demarcate the pertinent
regions. Here we summarize the observational input on line ratios, line
specific fluxes, and line opacities that will be utilized to compare
with model results.

For the information on hydrogen line ratios we make use directly of
Bary et al.'s(2008) collection of
$Pa~n_u/Pa\beta$ and $Br\gamma/Pa~n_u$ ratios, where
$n_u$, from 5 to 14, is the energy quantum number of the upper level of the
Paschen transition. For line ratios involving
the $CaII$ infrared triplet and $OI\lambda\lambda 8446, 7773$ we utilize
the data available in Muzerolle, Hartmann, \& Calvet (1998b). Three 
objects (DL, DG, \& BP Tau)
are selected because they are the only ones whose $OI\lambda 7773$
is not dominated by a strong red absorption. Because the lines considered
have quite close wavelengths, we simply use the listed equivalent
widths to obtain ratios among them. For the $HeI$ lines we
obtain the $HeI\lambda 10830$ information from EFHK06 and the 
$HeI\lambda 5876$ information from BEK01. To avoid the issue of estimating
the true emission when an absorption feature is present in $HeI\lambda 10830$,
only those objects among the reference sample (cf. Fig. 4 of EFHK06) with
emission much stronger than absorption are used. They (CW, DL, DG, HN, BP,
GG, \& DG Tau, and RW Aur), totalling 8, are then looked up in the 
reference sample of BEK01 for the $HeI\lambda 5876$ data. The $HeI\lambda 5876/
\lambda 10830$ ratio is equal to $(1+r_R)EW_{\lambda 5876}F_{\lambda 5876}/
[(1+r_Y)EW_{\lambda 10830}F_{\lambda 10830}]$, where $r_{\lambda}$ 
denotes the veiling, $EW_{\lambda}$ the equivalent width and $F_{\lambda}$
the photospheric specific flux, assumed to be given by that of a $4000~K$
blackbody. For $HeI\lambda 5876$ the $EW$ of the broad component is
being used. In the same Echelle order of $HeI\lambda 10830$ $Pa\gamma$ is
observed, so the data set of EFHK06 conveniently provides the $Pa\gamma/
HeI\lambda 10830$ ratios for the same 8 objects. Unfortunately the
$HeI\lambda\lambda 10830, 5876$ data sets are procured at very different
times, so the $HeI\lambda 5876/\lambda 10830$ ratios derived from them
may have uncertainties associated with time variation of the line emission.
Partly to remedy this situation, we will also utilize the preliminary
information on three objects (DL, DG, \& HN Tau) in the data set of
Edwards et al. (2010). 
The $HeI\lambda 5876/\lambda 10830$ and
$CaII\lambda 8498/Pa\gamma$ ratios of those three objects 
are particularly helpful because the
optical and $1\mu m$ spectra are procured at the same time.

For the observed specific fluxes of lines we make use of the $HeI\lambda
10830$ and $Pa\gamma$ profiles in EFHK06, the $HeI\lambda 5876$ profiles
in BEK01, the $Pa\beta$ and $Br\gamma$ profiles in Folha \& Emerson (2001),
and the $CaII\lambda 8542$ profiles in Muzerolle et al. (1998b).
In the model calculations the specific flux of a line at $\vert v_{obs}\vert
=150~km~s^{-1}$ is measured relative to the continuum level that is given
by a $4000~K$ photosphere of radius $2R_\odot$ veiled by a $8000~K$
blackbody over 3\% of its surface. This is done to facilitate comparison
with observed spectra which are usually plotted with velocity as abscissa
and strength relative to the continuum as ordinate. Normally the
veiling and photospheric temperature of the star are determined, so it
is straightforward to take account of the differences in veiling and
photospheric temperature between model and observed star. The remaining
uncertainty lies in the unknown surface area of the stellar photosphere,
which depends on both the stellar radius and, because of the presence
of an opaque disk, the viewing angle. 

The ratio of two emission lines conveys information on the line opacities
in addition to the physical conditions of density and temperature, but
usually detailed excitation calculations are needed to disentangle the
effects of opacity, density, and temperature. An absorption feature,
on the other hand, directly reveals that the line optical depth is $\sim 1$ or
more, while an absence of a similar feature in another line indicates
its optical depth is much smaller, except when its emission is
sufficiently strong to fill in the absorption. Thus we can use the
relative propensity among the observed lines in showing an absorption to
constrain the physical conditions of the absorbing gas. 

Clearly $HeI\lambda
10830$ is most proficient in showing absorption features. EFHK06 find that
47\% of 38 CTTSs observed show red absorptions at $HeI\lambda 10830$
compared with 24\% at $Pa\gamma$, and 71\% show blue absorptions 
at $HeI\lambda 10830$ compared
with 0\% at $Pa\gamma$. If the blue absorptions are separated into broad
and narrow ones that are likely formed in a stellar wind and disk wind
respectively (Kwan et al. 2007), then the two kinds are present in 39\%
and 29\%, respectively, of the 38 objects. Among the 15 CTTSs observed
by Edwards et al. (1994) red absorptions are seen in $Na~D$ and $H\delta$
in 9 and 8 objects respectively. The rarity of $H\alpha$ red absorptions
(2/15) is due to the strong $H\alpha$ emission. Blue absorptions are present
in $H\alpha$ and $Na~D$ for about 50\% of the stars, but except for one
star(As 353A), are narrow and sharp, indicative of a disk wind origin.
Among the 8 $OI\lambda\lambda 7773, 8446$ profiles shown in Muzerolle et al.
(1998b) 2 have red absorptions in both lines, and another 4 have red 
absorptions in $OI\lambda 7773$ only, while none has a blue absorption.
Among the 11 $CaII\lambda 8542$ profiles shown 1(DS Tau) has a red
absorption, and 1 (RW Aur) has a narrow blue absorption. BEK01 shows 31
$HeI\lambda 5876$ profiles, of which 3 have red absorptions and none has a
blue absorption.

From the above observational input on absorption features we gather that a
broad blue absorption indicative of a stellar wind is present
quite often in $HeI\lambda 10830$, only rarely in $H\alpha$ and
$Na~D$ and thus far, not seen in the Paschen lines, $OI\lambda\lambda 7773$,
8446, $HeI\lambda 5876$, or the $CaII$ infrared triplet. In decreasing 
probability of showing a red absorption, the order of lines is roughly
$HeI\lambda 10830$, Balmer lines, $Na~D$, $OI\lambda 7773$, Paschen lines,
$OI\lambda 8446$, $HeI\lambda 5876$, and $CaII$ infrared triplet. The
rarity of red absorptions in the $CaII$ infrared triplet, however, does
not indicate the lines are optically thin. They are known to be optically
thick from their nearly equal emission strengths despite a factor of 5
between $\tau_{CaII\lambda 8662}$ and $\tau_{CaII\lambda 8498}$. Often
their emission is very strong, as measured in relation to the continuum
level, so it is likely the emission will fill in the red absorption
produced from scattering the stellar and veiling continua. It also
happens that $Pa~13$, $Pa~15$, and $Pa~16$ lie on the red side (at 
$\sim 120~km~s^{-1}$) of $CaII\lambda \lambda 8662, 8542$, and $8498$
respectively, and their contribution will further obliterate a red 
absorption. 

\section{Line Opacities and Importance of UV Photoionization}

\subsection{Model Results}

Figure 4 shows the optical depth contours of four important lines in the 
($N_H, T$) plane for $r=4R_\ast$, and $\gamma_{HeI}=10^{-4}$(top panel)
and $10^{-5}~s^{-1}$(bottom panel).
They are $\tau_{H\alpha}=1$ and $3.16$, 
$\tau_{HeI\lambda 10830}=1$ and $3.16$, 
$\tau_{Pa\gamma}=0.1$ and $0.316$, and 
$\tau_{HeI\lambda 5876}=0.1$ and $0.316$, with the right
contour of each pair having the higher value. The different contour levels
between ($H\alpha$, $HeI\lambda 10830$) on the one hand and ($Pa\gamma$,
$HeI\lambda 5876$) on the other reflect their relative tendencies to
show an absorption. They help to locate the appropriate region in ($N_H, T$)
space causing absorption in one line but not in another. For example, to
produce a discernible blue absorption in $HeI\lambda 10830$
but not in $Pa\gamma$, the bulk of the gas need to have an $N_H$ between
the $\tau_{HeI\lambda 10830}=1$ and $\tau_{Pa\gamma} =0.1$ contours. For 
$\gamma_{HeI}=10^{-4}~s^{-1}$ the leeway in $N_H$ is a factor of between
$10$ and $50$, depending on $T$. 

The $HeI$ and $HI$ optical depth contours demonstrate the role of 
photoionization in $HeI$ and $HI$ excitation. Figure 4 shows that
at $T$ above $2.25\times 10^4~K$($10^4~K$) the $HeI$($HI$) 
contours are almost identical between the two cases of $\gamma_{HeI}$
($\gamma_{HI}$). This occurs because for
$\gamma_{HeI}=10^{-4}~s^{-1}$
($\gamma_{HI}=2\times 10^{-4}~s^{-1}$) and an $N_e$ of $10^9~cm^{-3}$ the
rate of collisional excitations to the $HeI\lambda 10830$($H\alpha$) lower
level will equal the rate of population via photoionization at 
$T=2.25\times 10^4~K$($10^4~K$). Thus, above this temperature $HeI$($HI$)
photoionization at the assumed strength is less effective than collisional
excitation.
The fairly large separation
between the $\tau_{HeI\lambda 10830}=1$ and $\tau_{Pa\gamma}=0.1$ contours
in $N_H$ space at $T>2.25\times 10^4~K$ indicates that the condition of
$\tau_{HeI\lambda 10830}\gg \tau_{Pa\gamma}$ can be readily produced by 
collisional excitations alone. As the effectiveness of $HeI$ collisional
excitation declines with $T$ decreasing towards $10^4~K$, photoionization 
becomes increasingly more important and the locations of the $\tau_{HeI
\lambda 10830}$ contours depend on $\gamma_{HeI}$. Clearly a lower
$\gamma_{HeI}$ requires a higher $N_H$ to produce the same line optical depth
and, as seen in Figure 4, the $\tau_{HeI\lambda 10830}$ contours for
$\gamma_{HeI}=10^{-5}~s^{-1}$ shift to the right from those for
$\gamma_{HeI}=10^{-4}~s^{-1}$. A comparison between the $\tau_{HeI
\lambda 10830}=1$ and $\tau_{Pa\gamma}=0.1$ contours will then delimit
the ($\gamma_{HeI}$, $N_H$) parameter space available to meet the
observational constraints. For example, at $9000~K\leq T\leq 1.5\times
10^4~K$ a minimum $\gamma_{HeI}$ of $\sim 10^{-4}~s^{-1}$ is needed. At
$T<10^4~K$ photoionization of $HI$ becomes important and the $\tau_{H\alpha}$
and, more noticeably, $\tau_{Pa\gamma}$ contours also depend on 
$\gamma_{HI}$. It is interesting that in this temperature range comparing
between the $\tau_{HeI\lambda 10830}=1$ and $\tau_{Pa\gamma}=0.1$
contours in the $\gamma_{HeI}=10^{-5}~s^{-1}$ case shows that a larger
($\gamma_{HeI}$, $N_H$) parameter space is again available to meet the
$\tau_{HeI\lambda 10830}\gg \tau_{Pa\gamma}$ condition. Thus, at $T<8750~K$
a $\gamma_{HeI}$ of $10^{-5}~s^{-1}$ is more than sufficient.

The above rather surprising result is caused by the 
$HeI\lambda 10830$ lower level being extremely metastable, 
making spontaneous decay
ineffective in comparison with depopulation via collisions with electrons 
for $N_e>10^4~cm^{-3}$. As mentioned in \S 3.2, collisional excitation to
the singlet $2s~^1S$ is more important than direct collisional 
deexcitation to the ground state. The rate of this depopulation route is 
then $\propto N_e e^{-9240~K/T}$ and the $HeI\lambda 10830$ optical depth,
in the limit when photoionization dominates over collisional excitation,
can be expressed as
\begin{equation}
\tau_{HeI\lambda 10830}\propto {N_{HeI}\over N_{He}}{N_H\over N_e}
\gamma_{HeI}~e^{9240~K/T}\ . 
\end{equation}
As $T$ decreases, both $N_{HeI}/N_{He}$ and $e^{9240~K/T}$ increase while
$N_e/N_H$ decreases, and they work in concert to make $\tau_{HeI\lambda 10830}$
larger, thereby enabling a given optical depth to be obtained at a lower $N_H$,
hence the reversal of the $\tau_{HeI\lambda 10830}=1$ contour direction seen
distinctly in the bottom panel of Figure 4.

An illustration of how the electron fraction $N_e/N_H$ depends on the 
physical parameters will help understand the behaviors of the optical depth
contours as well as results presented later.
This fraction is contributed primarily by hydrogen ionization, but its
dependences on $T$ and $N_H$ are not straightforward. 
In Figure 5 we plot $N_e/N_H$ versus $log(N_H)$ for seven temperatures
and $\gamma_{HI}=2\gamma_{HeI}=2\times 10^{-4}$ (top panel)  
and $2\times 10^{-5}~s^{-1}$ (bottom panel).
With the expectation that $N_e/N_H$ is lower when $\gamma_{HI}$
is lower, the dependences of $N_e/N_H$ on $N_H$ and $T$ in the two panels
are qualitatively the same. At $8750~K\leq T \leq 
10^4~K$, $N_e/N_H$ initially falls with $N_H$ increasing from $10^8~cm^{-3}$
as hydrogen ionization through $UV$ photoionization dominates. It levels
off and rises slowly when photoionization from $n=2$, bolstered by $N_2/N_1$
increasing with increasing $N_H$ and $Ly\alpha$ trapping, grows stronger,
and rises faster once sufficient population is built up into the high
$n$ levels for collisional ionization to take over. At $T\leq 7500~K$ the
contributions from excited state ionizations only slow down the rate at
which $N_e/N_H$ decreases with increasing $N_H$.

The quantitative values of $N_e/N_H$  
make it easy to see the behavior of the
$\tau_{HeI\lambda 10830}$ contours. At $\gamma_{HeI}=10^{-5}~s^{-1}$
the $\tau_{HeI\lambda 10830}=1$  contour near $10^4~K$ lies at 
$N_H\sim 10^{10}~cm^{-3}$. From Figure 5(bottom panel) 
it is seen that there $N_e/N_H$
decreases rapidly with $T$ decreasing from $10^4$ to $7500~K$. This
sharp decrease is largely responsible for the switchback of the
$\tau_{HeI\lambda 10830}=1$ contour toward a lower $N_H$. The same figure
also helps to explain the large
separation between the $\tau_{HeI\lambda 10830}=1$ and $3.16$
contours seen at $\gamma_{HeI}=10^{-5}~s^{-1}$ 
and $8750~K\leq T \leq 10^4~K$. 
This arises because in that temperature range increasing $N_H$
above $\sim 10^{10}~cm^{-3}$ leads to
$N_e/N_H$ increasing also, which counteracts the
effect of $N_{HeI}/N_{He}$ increasing in the expression for 
$\tau_{HeI\lambda 10830}$. As a result $\tau_{HeI\lambda 10830}$ barely 
changes for $3\times 10^9~cm^{-3}\leq N_H \leq 10^{12}~cm^{-3}$. This
large separation between the $\tau_{HeI\lambda 10830}=1$ and 3.16 contours
is not seen in the $\gamma_{HeI}=10^{-4}~s^{-1}$ case because in the same
temperature range the $\tau_{HeI\lambda 10830}=3.16$ contour there lies at
$N_H< 10^9~cm^{-3}$.

At $T\leq 7500~K$ the $\tau_{HeI\lambda 10830}$ contours for the two
$\gamma_{HeI}$s are not far apart. This insensitivity of $\tau_{HeI
\lambda 10830}$ to decreasing $\gamma_{HeI}$ at low temperatures may
explain why $HeI\lambda 10830$ shows such a penchant for absorption. Seen in 
$HeI\lambda 10830$ are not only the broad blue and red absorptions that arise
from kinematic regions close to the star, but also the narrow, sharp blue
absorption indicative of a disk wind, and the central absorption that has been
suggested as arising from a disk corona even farther away from the
star (Kwan 1997). Presumably the ionization sources are located close to
the star and, even if intervening attenuation is ignored, $\gamma_{HeI}$
will decrease with distance. It turns out that
the density required to produce a $\tau_{HeI\lambda 10830}$ of unity is not
prohibitive. To see
this, we find that if $N_e$ is supplied through $UV$ photoionization of
hydrogen, as likely at $T< 7500~K$,
then $N_e\propto (N_H \gamma_{HI})^{0.5}$ and
\begin{equation}
\tau_{HeI\lambda 10830}\propto (N_H \gamma_{HeI})^{0.5}
({\gamma_{HeI}\over \gamma_{HI}})^{0.5}~e^{9240~K/T}\ . 
\end{equation}
Thus $\tau_{HeI\lambda 10830}$ is more sensitive to $T$ for $T\leq 7500~K$
than to $\gamma_{HeI}$. Also $\gamma_{HeI}/\gamma_{HI}$ can be quite large
if the $UV$ continuum near the hydrogen ionization threshold
is attenuated more severely by intervening hydrogen.
In the limit that both helium and hydrogen are ionized by photons with
energies $\ge 24.6~eV$ $\gamma_{HeI}/\gamma_{HI}$ is $\sim 6$, which is
a factor of 12 larger than the earlier assumed value of 0.5.
This means that $\gamma_{HeI}$ can be
smaller than $10^{-5}~s^{-1}$ by a factor of 12 and still produce at $T=5000~K$
a $\tau_{HeI\lambda 10830}$ of one with $N_H=5\times 10^8~cm^{-3}$.
Even lower values of $\gamma_{HeI}$ are possible if $T$ is lower. The
temperature at which direct collisional de-excitation of $2s~^3S$ occurs as 
rapidly as the postulated de-population route is $2700~K$, so $\gamma_{HeI}$  
can be smaller by another factor of $\sim 20$ when direct collisional 
de-excitation dominates.

In Figure 4 the pair of contours for each line purports to illustrate its
sensitivity to varying $N_H$. It is seen that the $H\alpha$, $Pa\gamma$,
and $HeI\lambda 5876$ optical depths increase faster than linearly with
increasing $N_H$. This is because the population in each of their lower 
level is sustained via radiative trapping, so for $H\alpha$, for example,
$\tau_{H\alpha}\propto N_2\propto N_1(\gamma_{HI} + N_e C_{12})
\tau_{Ly\alpha}/A_{Ly\alpha}\propto N_1^2(\gamma_{HI} + N_e C_{12})$.
The $HeI\lambda 10830$ optical depth, by comparison, is less responsive to
increasing $N_H$, particularly at $T$ between $8750$ and $10^4~K$
when $\gamma_{HeI}=10^{-5}~s^{-1}$, as discussed earlier.

The $H\alpha$ contours are dependent on $\gamma_{HI}$ over its explored range
only at $T<10^4~K$, as mentioned before,
and then only weakly so. The $Pa\gamma$ contours are
more sensitive because the $n=3$ population relies on the $n=2$ population
being built up first. Above $10^4~K$ the hydrogen contours move towards 
lower densities with increasing temperature, owing to the stronger collisional
excitations from the ground level, but then reverse direction at
$T\sim 1.5\times 10^4~K$. This reversal is caused by the increase in collisional
ionizations which rapidly reduces the hydrogen neutral fraction.

Figure 6 shows the contours $\tau_{NaI\lambda
5892}=1$, $\tau_{CaII\lambda 3945}=1$,
$\tau^{\ast}_{CaII\lambda 8498}=1$, 
$\tau_{OI\lambda 8446}=0.1$, and $\tau_{OI\lambda
7773}=0.1$. The $\tau_{Pa\gamma}=0.1$ contour
is also re-plotted for comparison. For the line optical depth that
represents a multiplet with well separated components (in Doppler widths),
it needs be mentioned that, with our single level stand-in, the plotted
optical depth is a rough average of the individual component optical depths. 
To be specific,
\begin{eqnarray}
(\tau_{NaI\lambda 5890}, \tau_{NaI\lambda 5896})/\tau_{NaI\lambda 5892}
&=&(1.33,~0.67)\nonumber\\
(\tau_{CaII\lambda 3934}, \tau_{CaII\lambda 3968})/\tau_{CaII\lambda 3945}
&=&(1.33,~0.67)\nonumber\\
(\tau_{CaII\lambda 8498}, \tau_{CaII\lambda 8542}, \tau_{CaII\lambda 8662})
/\tau^{\ast}_{CaII\lambda 8498}&=&(0.2,~1.8,~1.0)\nonumber\\
(\tau_{OI\lambda 7772}, \tau_{OI\lambda 7774}, \tau_{OI\lambda 7775})
/\tau_{OI\lambda 7773}&=&(1.4,~1.0,~0.6)\ .
\end{eqnarray}
The wavelength used to denote the $CaII$ infrared triplet as a group is the
same as that of one member because that particular member is used to produce
observed line ratios for comparison with model results.
A similar denotation is not adopted for the $OI\lambda 7773$ triplet
because its three members are separated by only $\sim 200~km~s^{-1}$
and all are seen within one line profile, so their total flux is used.

It is seen that the $CaII$ and $NaI$ optical depth 
contours exhibit little difference
between the two $UV$ ionization fluxes, as the $CaII$ and $NaI$ fractions
depend only indirectly on the postulated $UV$ continuum through the
resulting $N_e$ that affects the recombination rate and, in the case of $CaII$,
also through the hydrogen $N_2/N_1$ ratio that affects the 
$Ly\alpha$ photoionization rate. The $CaII$ and $NaI$ contours 
also respond similarly to $T$ increasing from $5000~K$, owing
to their common susceptibility to collisional ionization. The $OI$
ionization structure is tied to hydrogen's via charge exchange, so the $OI$
optical depth contours at $T\leq 10^4~K$ respond to varying $\gamma_{HI}$ in
much the same way as the hydrogen optical depth contours. Above $T=10^4~K$ the
$\tau_{OI\lambda 7773}$ contour stays at nearly the same $N_H$ while
the $\tau_{OI\lambda 8446}$ contour moves towards much higher densities.
This is due to the lower state of $OI\lambda 7773$ being metastable 
while that of $OI\lambda 8446$ decaying rapidly with
$\tau_{OI\lambda 1303}$ decreasing with increasing temperature.

Both Figures 4 and 6 refer to local excitation calculations at $r=4R_\ast$,
a location appropriate for either a stellar wind or the farther portion
of an accretion flow. Figure 7 shows similar optical depth contours for
$\gamma_{HeI}=10^{-4}~s^{-1}$, and $r=2.5R_\ast$ to evaluate the dependence
on position. Comparing it with Figures 4 and 6(top panels)
indicates that the differences are
quite small. For fixed values of $N_H$ and $T$, the 
$HI$, $HeI$, and $OI$ optical depths are somewhat
smaller at $r=2.5R_\ast$ because of the factor $r/v$ in the opacity expression, 
hence their contours are displaced towards higher densities, with the
amounts dependent on the responses of the line optical depths to increasing
$N_H$. For the $NaI$ and $CaII$ optical depths, there is the additional effect
of a stronger stellar and veiling continuum flux at $r=2.5R_\ast$, which 
increases the photoionization rate and leads also to a displacement of
their contours towards higher densities. From Figures 4, 6 and 7 we can infer
that the line optical depths and, by analogy, line emissivity ratios, which
will be discussed in the next section, are much more dependent on $N_H$,
$T$, and $\gamma_{HeI}$ than on $r$.

\subsection{Comparison with Observations}

With in mind the observed propensities of the various lines in showing an
absorption, we are in a position to draw the following conclusions from
the model results presented in this section.

1. $UV$ photoionization is necessary to produce the broad 
red and blue absorptions seen in $HeI\lambda 10830$ because both gases have
a temperature $\sim 10^4~K$ or less. The most direct
argument comes from the observed red absorptions. They are seen in
$HeI\lambda 10830$ and $NaI\lambda 5892$ frequently, and $Pa\gamma$
occasionally. For $NaI$ absorption
to be more prevalent than $Pa\gamma$ absorption the temperature
needs to be less than $10^4~K$, as a comparison between the $Pa\gamma$
and $NaI$ optical depth contours indicates. Then for $HeI\lambda 10830$
absorption to be as prevalent as $NaI$ absorption at $T<10^4~K$, $UV$
continuum photoionization is paramount. In the next two sections one 
result drawn from the analyses will be that
the gas producing the broad blue absorption has a temperature not higher than
$\sim 10^4~K$, which is also the condition needed for a broad blue absorption
to be present in $NaI\lambda 5892$ but not $Pa\gamma$.  

2. In the temperature range of $T<10^4~K$ we can compare the observed order
of lines in their propensity of showing a red absorption to the order 
gleaned from the model results by judging the available
volume in ($N_H, T, \gamma_{HeI}$) space where the line optical depth
exceeds 0.1 as reflecting the propensity. The latter decreases roughly in the
order of $HeI\lambda 10830$, Balmer lines, $CaII\lambda 3945$, $NaI\lambda 
5892$, Paschen lines, $CaII\lambda 8498$, $OI\lambda 8446$, $OI\lambda 7773$,
and $HeI\lambda 5876$. It needs to be noted that the range in $\gamma_{HeI}$
is specifically chosen to place $HeI\lambda 10830$ close to the top of the
order, so a $\gamma_{HeI}\sim 10^{-5}~s^{-1}$ or higher is needed.
With in mind the explanation given in \S 4 regarding the lack
of red absorptions in $CaII\lambda 8498$, this order is close to that 
observed except for the position of $OI\lambda 7773$. This line, however,
is a triplet and, with its members separated by only $\sim 200~km~s^{-1}$,
is more effective than a single line in producing an absorption because
of its greater velocity coverage, and this interesting property of $OI\lambda
7773$ more than compensates for its smaller optical depth in relation to 
$OI\lambda 8446$. The occasional presence of red absorptions in $OI\lambda
8446$ and $HeI\lambda 5876$ also indicates a density $N_H\sim 10^{11}~cm^{-3}$
or higher for the red absorption gas.

3. The density of the gas producing the broad blue absorption is much 
lower than that of the accretion flow. This is obvious if the two flows have
similar temperatures over most of their volumes, since $Pa\beta$, $Pa\gamma$,
and $OI\lambda\lambda 8446,7773$ show red absorptions but almost never
blue absorptions. But even if that were not the case, the absence of
$Pa\beta$ and $Pa\gamma$ blue absorptions would by themselves require,
at $T>10^4~K$, $N_H < 3\times 10^9~cm^{-3}$. A rough estimate, based on
Figures 4, 6, and 7, of the typical gas density occupying the bulk of
the stellar wind and accretion flow is  $10^9$ and $10^{11}~cm^{-3}$
respectively. The corresponding mass flux, however, can be quite similar,
as the product of the factors $r^2$ and $F_\Omega$ is likely $30$ to
$100$ times larger for the stellar wind.

4. When $\tau_{HeI\lambda 10830}$ is $\sim 1$ the $H\alpha$ optical depth is 
close to or higher than 1 for $T\le 2\times 10^4~K$ and $\gamma_{HeI} \sim 
10^{-4}~s^{-1}$ or less (cf. Fig. 4). Thus we expect $H\alpha$ to be
as effective as $HeI\lambda 10830$ in absorbing the stellar and veiling
continua. The rarer occurrence of broad blue or red 
absorption in $H\alpha$ is attributed to the strength of the $H\alpha$ 
emission, which is considerably greater than the continuum level.

\section{Line Ratios and Different Sets of Physical Conditions}

Line ratios are very good diagnostics of physical conditions because while
the observed flux of a line depends on the local line emissivity and
the volume of emission, the flux ratio of two lines bypasses to a large
extent the effect of emission volume and probes directly the local
physical conditions. Here we examine several pairs of line ratios that are
particularly illuminating in this regard. 

The first pair is $HeI\lambda
5876/\lambda 10830$ and $Pa\gamma/HeI\lambda 10830$. The nature 
of $HeI\lambda 10830$ being the only allowed radiative transition following 
the $\lambda 5876$ transition means that emission of a $\lambda 5876$ 
photon is usually followed by emission of a $\lambda 10830$ photon, but
not vice versa, while collisional excitation from $2s~^3S$ strongly favors
$\lambda 10830$ emission because of both the lower excitation energy and
the larger collisional cross-section. The $HeI\lambda 5876/\lambda 10830$
ratio is then expected to be small at low densities. It will rise only
when the $\lambda 10830$ transition becomes sufficiently optically thick
that at the same time collisional de-excitation becomes competitive with
$\lambda 10830$ escape, population is built up into $2p~^3P$ from where
collisional excitation of $\lambda 5876$ is more effective. It is therefore
very sensitive to density and, with $\lambda\lambda 10830, 5876$ being
the strongest observed helium lines, clearly the most important diagnostic
of physical conditions giving rise to helium emission. To find out how
the hydrogen emission fare in the same conditions, we want to contrast
$HeI\lambda 5876/\lambda 10830$ against a ratio involving a hydrogen line
and a helium line. The $Pa\gamma/HeI\lambda 10830$ ratio is the most 
logical choice since, as mentioned in \S 3, its observed value of close
to unity appears incongruous with their perceived opacities. It also has
the advantage of being quite easily and accurately measured because
both lines occur in the same Echelle order.

It is possible and highly probable that hydrogen emission arise from
more than one kinematic region, because the physical conditions conducive
to strong hydrogen emission are likely to be more wide-ranging than those
conducive to either strong helium or $CaII$ emission. Ratios involving only
hydrogen lines are therefore useful to ascertain the relevant region in
the ($N_H,~T,~\gamma_{HI}$) parameter space. We will examine first $Pa\gamma
/Pa\beta$ and $Pa\gamma/H\alpha$ in tandem to see, for a given $\gamma_{HI}$,
how they depend on $N_H$ and $T$, and correlate with each other. The
$Pa\gamma/Pa\beta$ ratio is interesting because there is a substantial
amount of recent data pointing to a fairly uniform but somewhat surprising
value (Bary et al. 2008), while $Pa\gamma/H\alpha$ involves both the
strongest optical line observed and the line selected earlier to compare
with $HeI\lambda 10830$. These two ratios involve hydrogen levels up to
$n=6$. We will also compare $Pa~n_u/Pa\beta$ with observed
ratios for $n_u$ up to 12 so the observed hydrogen emission from high
$n$ levels are also brought into play.

The next two pairs of ratios focus on the $CaII$ infrared triplet emission in
relation to hydrogen and oxygen emission. We examine first $CaII\lambda 8498
/Pa\gamma$ and $OI\lambda 8446/Pa\gamma$, and then $CaII\lambda
8498/OI\lambda 8446$ and $OI\lambda 7773/\lambda 8446$. The latter pair has the
added bonus that the three wavelengths involved are in close proximity, so
the observed ratios have less uncertainties.

In the following three subsections we study in order the above
mentioned line ratios. In each subsection we present first the results
of the model calculations and then compare them with observations to
demarcate the prevailing physical conditions.

\subsection{$HeI\lambda 5876/\lambda 10830$ and $Pa\gamma/HeI\lambda 10830$}

\subsubsection{Model Results}

Figure 8 shows how $HeI\lambda 5876/\lambda 10830$ and $Pa\gamma/HeI\lambda
10830$ depend on $N_H$ (top panel) or line optical depth (bottom panel) for
$r=4R_\ast$, $\gamma_{HeI}=10^{-4}~s^{-1}$, and various temperatures. In
this and subsequent figures we use different line types to denote 
different temperatures. Each of six line types signifies the same temperature
in all the relevant figures, but a seventh line type, dot-long dash, can
signify a different temperature in a different figure in order to allow
for more flexibility in the selection of appropriate temperatures to
illustrate. Table 1 lists the designations between line types and temperatures,
and serves as a reference. When a model track in a figure has space nearby
its temperature (in unit of $10^4~K$) is marked for ready identification.

The most important physical process influencing the $HeI\lambda 5876/
\lambda 10830$ ratio is collisions with electrons. The summed rate of
$2s~^3S\rightarrow 3d~^3D$ and $2s~^3S\rightarrow 4f~^3F$ collisions,
which lead to $HeI\lambda 5876$ emission, is much weaker than the summed
rate of collisions from $2s~^3S$ that lead to $HeI\lambda 10830$
emission. In the regime when level de-excitation is dominated by radiative
decay, the $HeI\lambda 5876/\lambda 10830$ ratio resulting from
collisional excitations ranges from $1.3\times 10^{-3}$ at $T=5000~K$ to
$4\times 10^{-2}$ at $T=2\times 10^4~K$. The ratio resulting from
recombination and cascade is higher. When the cascade ends at $2s~^3S$
roughly two $\lambda 10830$ photons are produced for each $\lambda 5876$
photon. However, $2s~^3S$ de-excites to the ground level primarily through
collisions, and the ratio of $2s~^3S\rightarrow triplets$ to $2s~^3S
\rightarrow singlets$ collisional rate gives the number of additional 
$\lambda 10830$ photons emitted before each de-excitation of $2s~^3S$.
This number is $2.4$, $7.1$, and $12.2$ at $T=5000$, $10^4$, and 
$2\times 10^4~K$ respectively. Thus, even if helium excitation were all
caused by $UV$ photoionization of the ground state, the $HeI\lambda 5876
/\lambda 10830$ ratio would only be $0.27$, $0.1$, and $0.07$ at
$T=5000$, $10^4$, and $2\times 10^4~K$ respectively. A larger ratio can only
be obtained in the regime when the $HeI\lambda 10830$ optical depth is
sufficiently high that collisional de-excitation begins to rival
$\lambda 10830$ escape and effectively retards the growth of the 
$HeI\lambda 10830$ flux.

The detailed calculations bear out the above analysis. In Figure 8 
results for $T\leq 2\times 10^4~K$ are shown because those for 
$T=3\times 10^4$ and $2\times 10^4~K$ are almost the same. At low densities
$HeI\lambda 5876/\lambda 10830$ is very small because of the prominent
contribution to the $HeI\lambda 10830$ emissivity from stimulated absorption of 
the stellar and veiling continua. With increasing $N_H$ $HeI\lambda 5876/
\lambda 10830$ roughly levels off at a value determined by the relative
contribution between collisional excitation and recombination and
cascade to the line emissivity. Then, depending on $T$,
$HeI\lambda 5876/\lambda 10830$ begins to rise sharply when 
$\tau_{HeI\lambda 5876}$ exceeds between $1$ and $10$ 
while $\tau_{HeI\lambda 10830}$ is between $30$ and $350$. The actual
criterion for collisional quenching of $HeI\lambda 10830$ emission 
is $N_e C_{2p~^3P\rightarrow 2s~^3S}\sim A_{HeI\lambda 10830}
\beta_{HeI\lambda 10830}$, and it is seen from Figure 8 that the density $N_H$
needed to meet this criterion drops from $\sim 10^{12}~cm^{-3}$ at $T=5000~K$
to $\sim 3\times 10^{10}~cm^{-3}$ at $T=2\times 10^4~K$.

Turning to $Pa\gamma/HeI\lambda 10830$, it is seen that its dependences on
$N_H$ and $T$ are drastically different. At a given temperature 
$Pa\gamma/HeI\lambda 10830$ peaks at a lower $N_H$ than that marking
the sharp rise of $HeI\lambda 5876/\lambda 10830$. Over the range of
$N_H$ explored, $Pa\gamma/HeI\lambda 10830$ is strongest at $T$ between
$8750$ and $10^4~K$ and weakest at $T\geq 2\times 10^4~K$, while it is the 
exact opposite for $HeI\lambda 5876/\lambda 10830$. Thus these two ratios
anti-correlate. This relationship is readily understood when examined with
the temperature range separated into two regimes.

At $T\geq 10^4~K$ the $HeI\lambda 10830$ emissivity is clearly very sensitive
to temperature since the $1s~^1S\rightarrow 2s~^3S$ collisional excitation
rate, which supplements $UV$ photoionization in populating $2s~^3S$, and
collisional excitation of $HeI\lambda 10830$ from $2s~^3S$ are both
strongly dependent on temperature. The $HeI\lambda 5876$ emissivity is even
more responsive to increasing temperature because of both a higher
excitation energy from $2s~^3S$ and its reliance also on the build up of the
$2s~^3S$ population. Consequently the $HeI\lambda 5876/\lambda 10830$
ratio increases with increasing $T$ from $10^4$ to $2\times 10^4~K$. It
does not change much with $T$ between $2\times 10^4~$ and $3\times 10^4~K$ 
because the reduction of $N_{HeI}/N_{He}$ due to ionization begins to
take effect. The $Pa\gamma$ emissivity also benefits from the higher
collisional excitation rate when $T$ increases from $10^4~K$, but the
concurrent decreases in $N_{HI}/N_H$ (due to ionization) and $Ly\alpha$
opacity retard the build up of population into the excited levels. A
confirmation of this very different behavior of the excited state population
between $HI$ and $HeI$ can be seen from Figure 4, 
since the $\tau_{HeI\lambda 5876}$
and $\tau_{Pa\gamma}$ contours are essentially contours of the population
in $HeI~2p~^3P$ (lower state of the $HeI\lambda 5876$ transition),
$N_{2p~^3P}$, and $HI~n=3$, $N_3$, respectively. As $T$ increases above
$10^4~K$, the $N_{2p~^3P}$ contour shifts towards decreasing $N_H$,
indicating an increasing $N_{2p~^3P}/N_H$ ratio at a fixed $N_H$, and then
hardly shifts between $T=2.25\times 10^4$ and $3\times 10^4~K$, while the
$N_3$ contour shifts very little between $T=10^4$ and $1.5\times 10^4~K$,
but then shifts towards increasing $N_H$ with further increase in $T$. 
This contrasting behavior accounts for $Pa\gamma/HeI\lambda 10830$ decreasing
while $HeI\lambda 5876/\lambda 10830$ increases with $T$ increasing above
$10^4~K$.

As $T$ decreases from $10^4~K$ the ionization of hydrogen begins to rely
more on $UV$ photoionization and the electron fraction $N_e/N_H$ decreases
with decreasing $T$ (cf. Fig. 5). Both decreases in $N_e/N_H$ and $T$ 
reduce the $HeI~2s~^3S$ de-excitation rate, as alluded to earlier in
\S 5.1, and the consequent higher $2s~^3S$ population counteract the decrease 
in the $2s~^3S\rightarrow 2p~^3P$ collisional excitation rate. As a
result the $HeI\lambda 10830$ emissivity does not decrease
as rapidly with decreasing temperature as the $Pa\gamma$ emissivity, and
$Pa\gamma/HeI\lambda 10830$ drops.

At a given temperature $Pa\gamma/HeI\lambda 10830$ rises rapidly as $N_H$
increases from $10^9~cm^{-3}$, peaks at an $N_H$ between $10^{10}$ and
$10^{11}~cm^{-3}$, and then falls. The rapid rise is related to the high
sensitivity to $N_H$ of the hydrogen population build up into upper
levels, relying on radiative trapping of successive $\Delta n=1$
transitions. The rise to the peak corresponds to $\tau_{Pa\gamma}$
increasing from less than $1$ to a value between $2$ and $20$. Further
increase in $\tau_{Pa\gamma}$ reduces the photon escape probability and slows
the growth of $Pa\gamma$ emissivity, leading to a falling $Pa\gamma/HeI
\lambda 10830$ ratio. Curbing of $Pa\gamma$ emissivity occurs before
that of $HeI\lambda 10830$ emissivity
because $C_{6\rightarrow 5}\sim 35C_{HeI~2p~^3P\rightarrow 2s~^3S}$.

The earlier discussion in \S 5.1 on how $\gamma_{HeI}$ affects the $HeI$ and
$HI$ optical depth contours also hints at the corresponding responses of
the $HeI\lambda 5876/\lambda 10830$ and $Pa\gamma/HeI\lambda 10830$ ratios,
which are shown in Figure 9 for $\gamma_{HeI}=10^{-5}~s^{-1}$. As
expected, $HeI\lambda 5876/\lambda 10830$ is affected little and then only
at $T< 10^4~K$, since both line emissivities, like the line optical depths,
respond similarly to varying $\gamma_{HeI}$, while $Pa\gamma/HeI\lambda 10830$
clearly peaks higher at $8750~K\leq T\leq 1.25\times 10^4~K$, the 
temperature range over which the $HeI~2s~^3S$ population is most
susceptible to $\gamma_{HeI}$. The basic contrast between the two ratios in
their dependences on $N_H$ and $T$, however, remains the same.

\subsubsection{Comparison with Observations}

The anti-correlation between $HeI\lambda 5876/\lambda 10830$ and
$Pa\gamma/HeI\lambda 10830$ is more conspicuous when the two ratios are
plotted against each other, as shown in Figure 10. The logarithmic scale
is used for greater clarity of the model tracks which will be bunched 
towards the $x$ or $y$ axis in a linear plot. 
$HeI\lambda 5876/\lambda 10830$ is clearly high when $Pa\gamma/HeI\lambda
10830$ is low, and vice versa. Many of the observed values 
that are marked lie
at positions very much off the tracks that model results produce as
$N_H$ increases at a fixed $T$. The track closest to them is
generated at $T=5000~K$, and it is more easily seen from Figure 8 that
$HeI\lambda 5876/\lambda 10830$ is slightly higher than $0.2$ over the
range $5\times 10^{10}~cm^{-3}\leq N_H\leq 10^{12}~cm^{-3}$ when 
$Pa\gamma/HeI\lambda 10830$ drops from $\sim 1.0$ to $0.4$. Several
arguments, however, point to a temperature of the emitting gas much higher
than $5000~K$. They are based on ratios of hydrogen lines and on line
fluxes, and will be propounded when the pertinent results are presented.
The remaining way to explain the observed ratios is for $HeI\lambda 5876$
and $Pa\gamma$ emission to arise from two different locales, e. g., a
region of $T=2\times 10^4~K$ for the former and one of $T=10^4~K$ for
the latter. The $HeI\lambda 5876/\lambda 10830 $ and $Pa\gamma/
HeI\lambda 10830$ ratios in their respective locales are then actually
higher than the observed ratios since both locales will produce 
$HeI\lambda 10830$ emission efficiently. For example, Figure 8 shows that
at $T=2\times 10^4~K$ $Pa\gamma/HeI\lambda 10830$ is less than $0.125$
for $N_H>3\times 10^{10}~cm^{-3}$, while at $T=10^4~K$ $HeI\lambda 5876/
\lambda 10830$ is less than $0.06$ for $N_H<3\times 10^{11}~cm^{-3}$. Then,
if the observed $HeI\lambda 10830$ emission is contributed equally by each
locale, to produce observed $HeI\lambda 5876/\lambda 10830$ and
$Pa\gamma/\lambda 10830$ ratios of $0.23$ and $0.6$ respectively, the $2\times 
10^4~K$ ($10^4~K$) locale will need to produce a $HeI\lambda 5876/
\lambda 10830$ ($Pa\gamma/HeI\lambda 10830$) ratio of $0.4$ ($1.08$).

In a way the model results confirm the perceived enigma mentioned in
\S 3 concerning $HeI\lambda 10830$ and $Pa\gamma$. They
demonstrate that a common set of physical conditions, a natural and
reasonable premise implicit in formulating the enigma, cannot produce
at the same time the $HeI\lambda 10830$ and $Pa\gamma$
emission strengths and their relative opacities as suggested by their
propensities in showing an absorption. 

We draw the following conclusions
concerning the 
observed strong $HeI\lambda\lambda 5876,10830$ and $Pa\gamma$ emission.

1. $HeI\lambda 5876$ and $Pa\gamma$ emission do not arise from the same
locale. In particular, the gas emitting $HeI\lambda 5876$ has a
temperature close to $2\times 10^4~K$ while that emitting $Pa\gamma$ has
a temperature close to $10^4~K$.

2. The gas producing the $HeI\lambda\lambda 5876, 10830$ emission is
very optically thick in those lines. Thus $\tau_{HeI\lambda 5876}$ needs
to exceed $30$ for $HeI\lambda 5876/\lambda 10830$ to exceed 0.3 (cf.
Figs. 8 \& 9 left bottom panel).
The gas producing the $HeI\lambda 10830$ blue absorption, which needs
to be optically thin in $HeI\lambda 5876$ to avoid a $HeI\lambda 5876$
blue absorption,
is  then distinct from the gas producing the $HeI$
emission. This does not exclude the emission to
arise from a stellar wind. It only stipulates that the emitting regions
fill a sufficiently small volume of the wind that their summed surface
area with a particular projected $v_{obs}$ subtends a small solid angle.

3. The gas producing the red absorption is not producing the $HeI\lambda
5876$ emission. Even though $HeI\lambda 5876$ shows a red absorption in
a few CTTSs, indicating in those objects an optical thickness that fulfills
one requirement for strong emission, the deduced temperature of less than
$10^4~K$ for that gas (cf. \S 5.2) does not meet the temperature 
requirement.

\subsection{$Pa\gamma/Pa\beta$, $Pa\gamma/H\alpha$, $Pa~n_u/Pa\beta$,
and $Br\gamma/Pa~n_u$}

\subsubsection{Model Results}

The hydrogen line ratios are expected to correlate with one another,
owing to the ladder-like energy level structure and very similar
de-excitation pathways for each excited level. Figure 11 shows how
$Pa\gamma/Pa\beta$ and $Pa\gamma/H\alpha$ relate to $N_H$ and $T$ for 
$r=4R_\ast$ and $\gamma_{HI} =2\gamma_{HeI} =2\times 10^{-4}~s^{-1}$. We
use the case of $T=10^4~K$ to illustrate the $Pa\gamma/H\alpha$
dependence on $N_H$. As $N_H$ increases from $10^8~cm^{-3}$, $Pa\gamma/
H\alpha$ falls because of the strong contribution to the $H\alpha$
emissivity from stimulated absorption of the stellar and veiling
continua. This contribution diminishes when $\tau_{H\alpha}$ begins to
exceed unity beyond $N_H\sim 6\times 10^8~cm^{-3}$, whereupon $Pa\gamma/
H\alpha$ increases with increasing $N_H$. The initial rise from $N_H
=6.3\times 10^8$ to $3\times 10^9~cm^{-3}$ is due to the continual 
contribution of $Pa\gamma$ stimulated absorption. At $N_H\sim 3\times
10^9~cm^{-3}$ collisional excitation of $H\alpha$ also becomes important,
and population is rapidly built up into $n>2$ levels. The steeper response of 
the $n=6$ population to $N_H$ increasing from $6.3\times 10^9$ to
$2.5\times 10^{10}~cm^{-3}$ produces the rise of $Pa\gamma/H\alpha$ from $\sim 
0.025$ to $0.06$. The further rise of $Pa\gamma/H\alpha$ as $N_H$ exceeds
$\sim 6.3\times 10^{10}~cm^{-3}$ is caused by collisional de-excitation
rivalling $H\alpha$ escape, effectively stunting $H\alpha$ emissivity
growth. As seen from the accompanying plot with $\tau_{Pa\gamma}$ as
the abscissa the important rise of $Pa\gamma/H\alpha$ from $0.025$ begins
at $\tau_{Pa\gamma}\sim 1$. The $Pa\gamma/Pa\beta$ ratio also increases from
$\sim 0.45$ at about the same $\tau_{Pa\gamma}$. Once the lines involved are
optically thick their emissivity ratios are solely functions of the 
excitation temperatures. Thus, at $N_H=1.6\times 10^{10}$, $10^{11}$,
$6.3\times 10^{11}~cm^{-3}$, when $\tau_{Pa\gamma}=5.6$, 33.4, 210, the
($H\alpha$, $Pa\beta$, $Pa\gamma$) excitation temperatures are ($4710$,
$2570$, $2350$), ($5930$, $3240$, $3020$), and ($6290$, $3920$, $3940$)
respectively, giving rise to $Pa\gamma/H\alpha$ and $Pa\gamma/Pa\beta$
ratios of ($0.05$, $0.54$), ($0.065$, $0.77$), and ($0.15$, $1.14$).
Just like the case of $HeI\lambda 5876/\lambda 10830$, both $Pa\gamma/
H\alpha$ and $Pa\gamma/Pa\beta$ can, by means of collisional excitation,
rise significantly above values produced from recombination and cascade
only when the upper transition ($Pa\gamma$ here) is optically thick, and
higher ratios require higher opacities and consequently higher excitation
temperatures.

At a given $N_H$ $Pa\gamma/Pa\beta$ is almost the same for $8750~K\leq T
\leq 3\times 10^4~K$, but much smaller for $T\leq 7500~K$. As a function
of $\tau_{Pa\gamma}$ the behavior of $Pa\gamma/Pa\beta$ at $T\leq 7500~K$
is actually more similar to those at higher temperatures, only that the
highest $\tau_{Pa\gamma}$ reached is much smaller. This stronger similarity is
because, at the same $\tau_{Pa\gamma}$, 
the comparison between the rates of photon escape,
$A_{Pa\gamma}/\tau_{Pa\gamma}$, and collisional de-excitation is almost 
independent of temperature. 

The dependences of $Pa\gamma/Pa\beta$ and $Pa\gamma/H\alpha$ on $N_H$ are only
weakly sensitive to $\gamma_{HI}$. At $T\geq 10^4~K$ there is no discernible
difference between the cases of $\gamma_{HI}=2\times 10^{-5}$ and 
$2\times 10^{-4}~s^{-1}$. At $T=(7500, 5000)K$, $Pa\gamma/Pa\beta$ is
lower by ($0-10$, $0-25$)\% and $Pa\gamma/H\alpha$ lower by ($0-30$,
$20-40$)\% over the density range $3\times 10^{10}~cm^{-3}\leq N_H \leq
10^{12}~cm^{-3}$ for $\gamma_{HI}=2\times 10^{-5}~s^{-1}$.

\subsubsection{Comparison with Observations}

In Figure 12 $Pa\gamma/Pa\beta$ is plotted against $Pa\gamma/H\alpha$ for
$\gamma_{HI}=2\times 10^{-4}~s^{-1}$.
As expected, $Pa\gamma/Pa\beta$ correlates with $Pa\gamma
/H\alpha$. The figure 
shows that $Pa\gamma/Pa\beta$ and $Pa\gamma/H\alpha$ do not delineate
temperature well since the model tracks for
different temperatures between 5000 and $3\times 10^4~K$ run similarly.
They, however, are sensitive to density.
From Figure 11 it is seen that the dependence of $Pa\gamma/Pa\beta$ on $N_H$
is strong and quite similar for $8750\leq T\leq 3\times 10^4~K$, but changes
rapidly as $T$ decreases from $8750~K$. For example, to obtain $Pa\gamma
/Pa\beta > 0.7$ the density needed at $T\leq 7500~K$ is higher than that
needed at $T\geq 8750~K$ by a factor $\geq 4$. 

The data sample of Bary et al. (2008), including observations
of the same object at multiple epochs, has $73$ values of $Pa\gamma/Pa\beta$
that cluster closely about $0.86$ with an estimated variance of $\pm 0.11$.
Among case B models the best fit to $Pa\gamma/Pa\beta$ as well as $Pa~n_u
/Pa\beta$ for $n_u$ up to $14$ has $T=1000~K$ and $N_e=10^{10}~cm^{-3}$.
While this best fit to $8$ data points has a reduced $\chi^2$ surpassing
$99.9$\% confidence, there is a glaring discrepancy at $Pa\gamma/Pa\beta$
where the fit value is only $0.73$, and none of the other models explored
has a higher value.

To see how the other $Pa~n_u/Pa\beta$ values from our calculations fare
with the data, we show them in Figure 13 for five densities at each of
four temperatures. In general, as $N_H$ increases above $2.5\times 
10^{10}~cm^{-3}$, $Pa~n_u/Pa\beta$ ($6\leq n_u \leq 12$) increases, with
the higher $n_u$ increasing more. Although there is no single $N_H$ that
produces an excellent match to data, the range in $N_H$ generating $Pa~n_u/
Pa\beta$ ratios that bracket the data is quite small, of order less than
a factor of $10$. This range is almost the same for $T\geq 8750~K$, 
being $5\times 10^{10}~cm^{-3}\leq N_H \leq 2\times 10^{11}~cm^{-3}$,
while that for $T=7500~K$ is $2\times 10^{11}~cm^{-3}\leq N_H \leq 
10^{12}~cm^{-3}$.
Figure 14 shows an analogous plot
for $Br\gamma/Pa~n_u$, and the comparison between model results and data
is similar.

Bary et al. (2008) has also compared their data with line ratios
calculated with a constant excitation temperature (LTE), but find that
in neither the optically thick nor thin case is there an acceptable fit.
Our calculations are more similar to this set of models, only with the
level excitations and line optical depths calculated self-consistently for a
given $N_H$ and $T$. Thus the $Pa~n_u$ excitation temperatures are not the
same and the lines are not all completely optically thick. For example,
at $T=10^4~K$ and $N_H=10^{11}~cm^{-3}$, the excitation temperatures and
line optical depths of ($Pa\beta$, $Pa\gamma$, $Pa~10$, $Pa~12$) are
($3240$, $3020$, $2910$, $2940$)$K$ and ($104$, $33.4$, $3.44$, $1.78$)
respectively. The fair agreement between model results and observed
values makes understanding the hydrogen emission as resulting from
collisional excitation an attractive alternative to the proposal by
Bary et al.(2008) of a recombining gas at $T=1000~K$.

We draw the following conclusions from the comparison between model and
observational results on the hydrogen line ratios.

1. $Pa\gamma$ is optically 
thick in the hydrogen emission region. Thus $\tau_{Pa\gamma}$ needs to 
exceed $20$ for $Pa\gamma/Pa\beta$ to exceed $0.7$ at $T\leq 10^4~K$
(Fig. 11 left bottom panel).
Even many of the
higher order Paschen lines have optical depths exceeding unity.

2. The temperature in the hydrogen emission region is higher than $5000~K$.
At $5000~K$ $Pa\gamma/Pa\beta$ reaches a value of only $0.81$ even at the
highest $N_H$ explored, $2.5\times 10^{12}~cm^{-3}$. The more limited
and extreme range on $N_H$ needed to produce $Pa\gamma/Pa\beta \geq 0.86$
at $T\leq 5000~K$ is one argument. Another is the difficulty in matching
the observed $Pa\gamma/HeI\lambda 10830$ ratio. As seen from Figure 8,
$Pa\gamma/HeI\lambda 10830$ is only $0.38$ at $1.6\times 10^{12}~cm^{-3}$
and falls with increasing $N_H$. This match-up is better in the case of
$\gamma_{HeI}=10^{-5}~s^{-1}$ (cf. Fig. 9), but then $Pa\gamma/Pa\beta$
is only $0.6$ at $N_H=1.6\times 10^{12}~cm^{-3}$.

3. The density in the hydrogen emission region is centered around
$N_H\sim 10^{11}~cm^{-3}$ if $8750~K\leq T \leq 3\times 10^4~K$, and 
around $N_H\geq 5\times 10^{11}~cm^{-3}$ if $T\leq 7500~K$,
in order to produce the observed
$Pa~n_u/Pa\beta$, $n_u\geq 6$, ratios.

\subsection{$CaII\lambda 8498/Pa\gamma$, $OI\lambda 8446/Pa\gamma$,
$CaII\lambda 8498/OI\lambda 8446$, and $OI\lambda 7773/\lambda 8446$}

\subsubsection{Model Results}

We compare the $CaII$ infrared triplet to $Pa\gamma$ for its being a
fairly strong hydrogen line not too far away in wavelength, and to 
$OI\lambda 8446$ for its tie to the hydrogen excitation through $Ly\beta$
fluorescence and its close proximity in wavelength. Figure 15 shows the
dependences of $CaII\lambda 8498/Pa\gamma$ and $OI\lambda 8446/Pa\gamma$
on $N_H$ and $\tau_{Pa\gamma}$ for $r=4R_\ast$, $\gamma_{HI}=2\times 10^{-4}
~s^{-1}$ and seven temperatures. Figure 16 is analogous, but with $CaII\lambda 
8498/OI\lambda 8446$ versus $N_H$ and $\tau^\ast_{CaII\lambda 8498}$, and
$OI\lambda 7773/\lambda 8446$ versus $N_H$ and $\tau_{OI\lambda 7773}$.
Because of our single level representation of multiple fine-structure
levels, we divide the emissivity from our model $CaII$ ion by three, thereby
implicitly assuming that all three components of the triplet are optically
thick, a reasonable assertion as all three components have nearly identical
fluxes and profiles. Our model $\tau^\ast_{CaII\lambda 8498}$
is five times the optical depth of the component with the weakest oscillator
strength (cf. \S 5.1), so the $CaII$ infrared emission requires $\tau^\ast
_{CaII\lambda 8498}$ to exceed ten. Our $OI\lambda 7773$ line also 
represents a triplet, but the whole emissivity is employed here since the
observed flux also comprises contributions from all three components.

Figure 15 shows that $OI\lambda 8446/Pa\gamma$ rises continually to beyond
unity as $N_H$ increases. This is largely because the much lower $OI\lambda
8446$ optical depth (cf. Fig. 6) keeps $\lambda 8446$ escape more effective 
than collisional de-excitation. The dependence of $OI\lambda 8446/Pa\gamma$ on
temperature depends slightly on $N_H$. At $2\times 10^{10}~cm^{-3}\leq~N_H\leq
2\times 10^{11}~cm^{-3}$ $OI\lambda 8446/Pa\gamma$ increases as $T$ increases
from $5000~K$, peaks at $T\sim 8750~K$ and then drops with further increase
in $T$. This behavior can be seen from the dependences of the $OI\lambda 8446$
emissivity, $\propto N_{OI} N_3/N_1$, and the $Pa\gamma$ emissivity,
$\propto N_6 A_{Pa\gamma}\beta_{Pa\gamma}~\propto N_6/N_3$, on the hydrogen
level population. Thus $OI\lambda 8446/Pa\gamma$ is $\propto (N_{OI}/N_1)
(N^2_3/N_6)$. As $T$ increases from $5000~K$ at a fixed $N_H$, $N_3/N_6$
always decreases, while $N_3$ first increases but then decreases when
hydrogen becomes more ionized. The countering effects of $N_3/N_6$ and
$N_3$ on $OI\lambda 8446/Pa\gamma$ in the beginning of the temperature
rise, and their concerted effects subsequently produce the mentioned
behavior. At $N_H > 2\times 10^{11}~cm^{-3}$ the same behavior holds
except that $OI\lambda 8446/Pa\gamma$ increases again as $T$ increases
beyond $1.5\times 10^4~K$. This occurs because, with the rapid decrease
of the hydrogen neutral fraction, radiative recombination of $OII$ becomes
important and $N_{OI}/N_1$ increases from the constant value of $1.1N_O/N_H$
maintained by charge-exchange reactions.

The two ratios $CaII\lambda 8498/Pa\gamma$ and $CaII\lambda 8498/OI\lambda
8446$ have very similar dependences on $N_H$ and $T$, as to be expected. 
The requirement of $\tau^\ast_{CaII\lambda 8498} > 10$ confines the
pertinent densities to $N_H\geq 10^{11}~cm^{-3}$ for $T\leq 8750~K$ and even
higher limits for $T > 8750~K$. Thus the large $CaII\lambda 8498/Pa\gamma$
and $CaII\lambda 8498/OI\lambda 8446$ model values at $N_H < 10^{10}~
cm^{-3}$ are not relevant. They actually involve very small fluxes, but are
shown for uniformity in the density range covered. As $N_H$ increases
from $\sim 10^{11}~cm^{-3}$ at a fixed temperature, both $CaII\lambda 8498
/Pa\gamma$ and $CaII\lambda 8498/OI\lambda 8446$ rise rapidly. Two
factors contribute to this. First, while $N_{HI}/N_H$ is fairly constant
in that density range (cf. Fig. 5), the fraction of calcium in $CaII$ 
increases with $N_H$. Second, the onset of collisional de-excitation of
the $CaII$ population occurs, by comparison with hydrogen, at a higher $N_H$.
When the latter is reached, the consequent collisional quenching of the
$CaII$ triplet emissivity causes the decline of the ratio from its peak.

Both $CaII\lambda 8498/Pa\gamma$ and $CaII\lambda 8498/OI\lambda 8446$
reach their highest value at $T=7500~K$. This can be understood as follows.
The hydrogen line emissivities are fairly insensitive to temperature for
$8750~K\leq T \leq 3\times 10^4~K$ (cf. \S 6.2). The $CaII\lambda 8498$
emissivity, on the other hand, decreases with increasing $T$ above 
$8750~K$, owing to the rapidly decreasing $CaII$ fraction. Thus the two
ratios fall beyond $T=8750~K$. At $5000~K \leq T \leq 8750~K$ collisional
excitations of both $CaII$ and $HI$ are sensitive functions of temperature
because of the strong dependence of $N_e/N_H$ on $T$ (cf. Fig.5), and
the higher ratio of excitation potential to thermal energy. Above $N_H=
10^{11}~cm^{-3}$, however, collisional de-excitation comes into play for
hydrogen, and the $Pa\gamma$ emissivity increases only modestly with
increasing temperature in comparison with the $CaII$ triplet emissivity
which, until $N_H\geq 6\times 10^{11}~cm^{-3}$, is still proportional
to ($N_{CaII}/N_{Ca}$)($N_e/N_H$), and the steeper rise of $Ne/N_H$ with
increasing $T$ from $5000~K$ more than overcomes the counter effect of
a lower $N_{CaII}/N_{Ca}$ ratio. As a result, $CaII\lambda 8498/Pa\gamma$ 
increases as $T$ increases from $5000~K$ and, in conjunction with its fall
at $T > 8750~K$, peaks at an intermediate temperature.

Regarding the $OI\lambda 7773/\lambda 8446$ ratio, both recombination and
cascade and collisional excitation strongly favor $OI\lambda 7773$ over
$OI\lambda 8446$ emission. This ratio is less than unity only because of
the $Ly\beta$ fluorescence process which, as seen from Figure 16, is most
effective at $N_H$ between $10^{10}$ and $3\times 10^{10}~cm^{-3}$.
At $N_H < 10^{10}~cm^{-3}$, even though $Ly\beta$ fluorescence remains
more effective than the other two processes, $OI\lambda 7773/\lambda 8446$
increases as $N_H$ decreases. This is because stimulated absorption of
the stellar and veiling continua contributes prominently to the
$OI\lambda 7773$ emissivity. But the line fluxes involved are very small.
$OI\lambda 7773/\lambda 8446$ also increases as $N_H$ increases above $\sim
3\times 10^{10}~cm^{-3}$. This is because collisional de-excitation begins to 
affect the hydrogen level population, causing $N_3/N_1$, the important
factor in the $Ly\beta$ fluorescence rate, to be nearly constant, while
collisional excitation of $OI\lambda 7773$ continues to grow roughly as
$N^2_H$. Thus, in this regime $OI\lambda 7773/\lambda 8446$ responds to
increasing density in a fashion similar to $CaII\lambda 8498/OI\lambda 8446$,
but it has not reached a peak even at $N_H=1.6\times 10^{12}~cm^{-3}$
because collisional de-excitation of $OI\lambda 7773$ remains weaker than
$\lambda 7773$ escape.

The dependences of the four emissivity ratios on $\gamma_{HI}$ between
$2\times 10^{-4}$ and $2\times 10^{-5}~s^{-1}$ are not strong, with $CaII\lambda8498/Pa\gamma$ and $CaII\lambda 8498/OI\lambda 8446$ reaching somewhat
higher values for $5000~K\leq T\leq 7500~K$ when $\gamma_{HI}=2\times
10^{-5}~s^{-1}$. There is also a weak dependence on $r$. Among the four
cases of $\gamma_{HI}$ and $r$, the ($\gamma_{HI}=2\times 10^{-4}~s^{-1}$,
$r=4R_\ast$) and ($\gamma_{HI}=2\times 10^{-5}~s^{-1}$, $r=2.5R_\ast$)
cases show the largest difference. 

\subsubsection{Comparison with Observations}

Figure 17 shows the relations between $CaII\lambda 8498/Pa\gamma$ and
$OI\lambda 8446/Pa\gamma$ and between $CaII\lambda 8498/OI\lambda 8446$ and
$OI\lambda 7773/\lambda 8446$ for the case ($\gamma_{HI}=2\times 10^{-4}~
s^{-1}$, $r=4R_\ast$), and Figure 18 shows those for the case
($\gamma_{HI}=2\times 10^{-5}~s^{-1}$, $r=2.5R_\ast$).
In addition to the data
points marked, we mention that $CaII\lambda 8498/OI\lambda 8446$ can
be determined for the remaining 5 objects in the data set of
Muzerolle et al.(1998b). It is 4.5, 12.7, 9.8, 0.81, and 1.31 for DR, CW,
RW, DS, and UY Tau respectively. Thus it appears that the observed values
of $CaII\lambda 8498/OI\lambda 8446$ separate into two groups. The full
data set of Edwards et al.(2010) also indicates this dichotomy, namely,
one group with $CaII\lambda 8498/Pa\gamma$
or $CaII\lambda 8498/OI\lambda 8446$ close to or less than 1, 
the other with $CaII\lambda 8498/Pa\gamma$ or
$CaII\lambda 8498/OI\lambda 8446$ greater than $\sim 5$. The physical 
conditions that will produce the first group of values are $N_H$ around
$10^{11}~cm^{-3}$ and $T$ between $5000$ and $1.25\times 10^4~K$ (cf.
Figs. 15 \& 16), but the temperature is narrowed to $8750~K\leq T\leq
1.25\times 10^4~K$ when the constraints imposed by the observed 
$Pa~n_u/Pa\beta$ ratios are also implemented (cf. \S6.2.2).
To produce the second group of values a considerably
higher $N_H$, $\sim 10^{12}~cm^{-3}$, is required, while $T$ is restricted
to $\leq 7500~K$. 

The CTTSs with $CaII\lambda 8498$ emission much stronger than $Pa\gamma$ 
or $OI\lambda 8446$ emission also tend to be the ones with the most
prominent $HeI\lambda 5876$ emission. The physical conditions needed to 
produce the $CaII\lambda 8498/Pa\gamma$ ratio cannot produce the 
$HeI\lambda 5876/\lambda 10830$ ratio, so a separate region producing the
bulk of the $HeI\lambda 5876$ emission is additionally required, as
already inferred in \S 6.1. The 
$CaII$ emission region itself appears to be capable of generating
the observed ratios among the hydrogen lines (cf. Figs 13 \& 14 
$T=7500~K$ panel). 

The CTTSs with $CaII\lambda 8498$ emission comparable
to or less than $Pa\gamma$ emission have a broad spread in helium 
emission, from $HeI\lambda 5876/\lambda 10830$ ratios close to those
of the previous group to little $HeI\lambda 5876$
emission and $HeI\lambda 10830$ primarily in absorption (Edwards et al. 2010).
When $HeI\lambda 5876$ emission is weak, the physical conditions needed
to produce the hydrogen line ratios(cf. \S 6.2) can also 
generate the requisite $CaII$ emission. When $HeI\lambda 5876$ emission
is strong, a separate region producing it is again called upon.

Based on the above understanding of the ratios involving $HeI$, $H$,
and $CaII$ lines, we draw the following deduction.

1. Three sets of physical conditions are identified. The first, characterized
by a comparatively high temperature ($T$ close to $2\times 10^4~K$) is
necessary to produce strong $HeI\lambda 5876$ emission. The second, 
characterized by a comparatively high density ($N_H\sim 10^{12}~cm^{-3}$)
and low temperature ($T\leq 7500~K$), is necessary to produce strong $CaII$
infrared triplet emission. The third, with less stringent constraints
($8750~K\leq T\leq 1.25\times 10^4~K$, $N_H$ around $10^{11}~cm^{-3}$),
produce primarily hydrogen emission. When the second set is present,
the first set is often present concurrently. The third set is present
either by itself or concurrently with the first set. It is also
likely to be present when both the first and second sets are.

\section{Line Fluxes and Sizes of Emission Regions}

As mentioned in \S 2, the specific flux $F_{v_{obs}}$ of an optically thick  
line depends on the line excitation temperature and 
the emission area with a projected velocity of $v_{obs}$, 
$A_{v_{obs}}$. The excitation temperature, in turn, depends on the
local physical conditions ($N_H$, $T$, etc.) and the emission area depends
on the specific kinematic model. In particular, for $\vert v_{obs} \vert
=150~km~s^{-1}$ a spherical wind model with $r=4R_\ast$ has $A_{v_{obs}}
=27(r/4R_\ast)^2~\pi R_\ast^2$ and some leeway in its adjustment by varying 
$r$, while a spherical infall model with $r=2.5R_\ast$ has $A_{v_{obs}}
=1.286\pi R_\ast^2$ and little room for maneuver. In a CTTS either wind or
accretion flow likely fills only a fraction, $f_\Omega$, of the $4\pi$
solid angle. This dilution affects the actual observed area in a way that
depends on viewing angle, whose variation is not included here to keep
the paper manageable. We simply assume an effective observed area given
by $f_\Omega A_{v_{obs}}$. Thus, for the same excitation temperature the
line flux in the infall model will be $(0.2/0.5)(1.286/27)$ or $1/52.5$
of that in the wind model. We think this is a conservative estimate in that
the true ratio is likely smaller.

For ease in comparison with an observed line, we scale the model 
$F_{v_{obs}}$ to the continuum flux $F_c$, and present
the ratio $y_{v_{obs}}=F_{v_{obs}}/F_c$ as a function of density or line
optical depth. Our model continuum is represented by a $4000~K$ blackbody
veiled by a $8000~K$ blackbody over 3\% of its surface. Thus our model
continuum has a veiling contribution at $HeI\lambda 5876$ of $r_V=0.67$
and one at $HeI\lambda 10830$ or $Pa\gamma$ of $r_Y=0.19$. When a CTTS
has a different $T_\ast$, $r_V$, or $r_Y$ the model $y_{v_{obs}}$ can
be easily rescaled to compare directly with the observed line profile.

We shall make use of the information deduced from line ratios to identify the
appropriate physical conditions. Since we are calculating the specific flux
at $\vert v_{obs}\vert = 150~km~s^{-1}$ it is the ratio of $F_{v_{obs}}$s
that is pertinent here. If the two lines have similar profiles, this ratio is
close to the ratio of their integrated fluxes, which was used in \S 6 to
identify the physical conditions responsible for the bulk of the emission. This
appears to be the case among $HeI\lambda\lambda 10830, 5876$, and $Pa\gamma$.
However, $CaII\lambda 8498$ is distinctly narrower than $Pa\gamma$. The
contrast is obvious in DR Tau , noticeable in DG Tau, and measurable in
DL Tau (Edwards et. al 2010). When we come to discussing the $CaII\lambda 8498$ 
specific flux we will take note of the distinction.

\subsection{$H\alpha$, $Pa\gamma$, $Pa\beta$, $Br\gamma$}

In this subsection we select among the hydrogen lines for presentation
$H\alpha$, the strongest optical member, and $Pa\gamma$, $Pa\beta$, and
$Br\gamma$ for the availability of large data sets (EFHK06,
Folha \& Emerson 2001).

Figures 19 and 20 show, for the wind and infall 
model respectively, how the $H\alpha$ and $Pa\gamma$ specific
fluxes depend on $N_H$ and line optical depth. The y-axis limits in Figure 20
are $1/52.48$ (-1.72 in log) of those in Figure 19, so chosen that, if the
specific flux in the infall model is $1/52.48$ of that in the wind model, the
two figures will be visually identical. They do look very much alike. When
one figure is overlaid on the other, the largest difference in the density
dependence plots occurs for $y_{v_{obs}}(H\alpha)$ at $N_H\sim 10^9~cm^{-3}$,
where the line emissivity is contributed primarily by stimulated absorption
of the incident continuum flux which is a factor of $2.56$ larger at
$r=2.5R_\ast$. The optical depth dependence plots, when overlaid together, show
larger differences. This is because at the same $\tau$ ($\propto N_l~ 
r/v$) the density in the $r=2.5R_\ast$ case is higher and the line specific
flux is more sensitive to density than to the velocity gradient, as
will be propounded next.

Figures 19 and 20 show not only that the primary distinction in the specific
flux produced between the wind and infall models arises from the emission
area, but also that variation in the velocity gradient (a factor of $1.6$
larger in the $r=2.5R_\ast$ model) generates only a small change. As a
specific example, we consider the case of $N_H=10^{11}~cm^{-3}$ and 
$T=10^4~K$, conditions suitable to bringing about the observed $Pa\gamma
/Pa\beta$ ratio. The value of $y_{v_{obs}}(Pa\gamma)$ in the wind model is
$3.81$, while that in the infall model is $0.7242=3.802/52.5$. Their
nearly identical emissivity, $y_{v_{obs}}/(f_\Omega A_{v_{obs}})$, is
not surprising since $Pa\gamma$ is very optically thick ($\tau_{Pa\gamma}
=33.44$ and $20.46$ in the wind and infall model respectively,
cf. Figs. 19 \& 20 right bottom panel). Even when
$Pa\gamma$ is optically thin, the difference in emissivity is small. For
example, at $N_H=10^{10}~cm^{-3}$, $T=10^4~K$, $y_{v_{obs}}=0.356$
and $\tau_{Pa\gamma}=1.63$ in the wind model, while $y_{v_{obs}}
=4.78\times 10^{-3}=0.251/52.5$ and $\tau_{Pa\gamma}=0.355$ in the infall
model. The $Pa\gamma$ emissivity is a factor of 1.41 smaller with the
larger velocity gradient. However, at the next density step of $1.585\times 
10^{10}~cm^{-3}$, $y_{v_{obs}}=0.0154$ in the infall model or a
factor of $3.22$ larger than the earlier value. From the slope of the 
log $y_{v_{obs}}$ dependence on log $N_H$ at $10^{10}~cm^{-3}$ it can be 
estimated that the larger velocity gradient by a factor of $1.6$ can be
compensated by an increase in $N_H$ by a factor of only $1.147$ to produce
the same line emissivity. For this reason, even though we emphasize the 
location where $\vert v \vert =150~km~s^{-1}$, the local excitation
calculations of line emissivities and ratios are applicable elsewhere because
a large variation in the local velocity gradient is tantamount to only a 
modest variation in the density.

The two figures also show that over the density range $2\times 10^{10}~cm^{-3}
<N_H < 2\times 10^{12}~cm^{-3}$ the dependences of
$y_{v_{obs}}(H\alpha)$ and $y_{v_{obs}}(Pa\gamma)$ on temperature are
strong for $5000~K\leq T\leq 8750~K$, but weak for $8750~K\leq T\leq
3\times 10^4~K$. This is because, despite the stronger collisional
excitation with $T$ increasing above $8750~K$, the concomitant more
rapid buildup of population into high $n$ levels produces a stronger
collisional ionization and reduces the neutral fraction.

The 38 $Pa\gamma$ profiles in the reference sample of EFHK06
provide a convenient data set of $y_{v_{obs}}$s for comparison with our
model results. To avoid the difficulty of extracting the underlying red 
emission when a red absorption is obviously or probably present, we concentrate
on the blue wing emission and determine $y_{v_{obs}}$ at $v_{obs}=-150~km~
s^{-1}$. Of the 38 objects, 9 (DR, CW, DL, HL, DG, DK \& HN Tau, AS 353 A, 
RW Aur A) have high $1\mu m$ veiling ($r_Y \ge 0.5$), of whom 8 have
$y_{v_{obs}}>0.25$ and 1 (DK Tau) has $y_{v_{obs}}\sim 0.15$;
11 (DS, HK, BP, DF, DO, GG, \& GK Tau, YY, \& GW Ori,
UY Aur, UZ Tau E) have intermediate $1\mu m$ veiling ($0.3\le r_Y \le 0.4$),
of whom 10 (excepting GK Tau) have $0.1\le y_{v_{obs}} \le 0.25$. Of
the remaining 18 objects with low $1\mu m$ veiling ($r_Y\le 0.2$), 2
(SU Aur, TW Hya) have $y_{v_{obs}}$ between $0.1$ and $0.25$, and the rest
have $y_{v_{obs}}<0.1$. 

For members in the first veiling group, our model
$y_{v_{obs}}$s, when scaled to the CTTS's stellar plus veiling continuum,
need to be reduced by a factor of $\sim 2$ or more. Then, at the physical
conditions needed to produce the $Pa~n_u/Pa\beta$ ratios ($N_H$ centered
around $10^{11}~cm^{-3}$, $T\ge 8750~K$, or $N_H$ centered around $5\times
10^{11}~cm^{-3}$, $T=7500~K$, cf. \S 6.2.2 \& Fig. 13) and assuming that
these ratios of integrated fluxes are the same as the ratios of specific
fluxes at $v_{obs}=-150~km~s^{-1}$, the $y_{v_{obs}}$ values in the infall
model will be less than $0.05$. They fail to produce the values of those
9 objects by factors ranging from $\sim 3 $ to $\sim 15$. 
We therefore conclude that a kinematic flow other 
than accretion infall is responsible for those hydrogen emission. We
also draw the same conclusion for the majority of the members in the
second $1\mu m$ veiling group. The infall model has difficulty not only
accounting for their specific fluxes at $v_{obs}\le -150~km~s^{-1}$ but also
their observed line wing asymmetry (say, at $\vert v_{obs} \vert\ge 150~
km~s^{-1}$). Thus, whereas in the infall model the blue wing emission is
occulted by the star and therefore expected to be weaker than the red
wing emission, only GK Tau shows this characteristic, 2 (DF \& DO Tau) have 
comparable blue and red wing emission, and the remaining 8 have stronger
blue wing emission. To be sure, DS Tau and YY Ori have obvious red 
absorptions, and DK Tau and UY Aur may have weak red absorptions that make
it difficult to discern the underlying red wing emission; nevertheless, the
lines of sight that favor seeing a red absorption are also the ones that
incur stellar occultation of the blue wing emission. Then the earlier
conclusion with regard to the high veiling group also implies that whatever
kinematic region responsible for their hydrogen emission can also produce
the moderate hydrogen emission seen in the second $1\mu m$ veiling group.

The wind model poses the opposite conundrum in that the model $y_{v_{obs}}$s
are much larger than those observed in the high $1\mu m$ veiling group,
by factors of about 4-10. It can be responsible for the observed hydrogen
emission provided the optically-thick emitting gas ($\tau_{Pa\gamma}>10$,
cf. Fig. 19 right bottom panel) fills only a portion of the postulated
flow, and covers but a fraction of the contour surface shown
in Figure 1. This also happens to be the necessary proviso to avoid the
occurrence of a blue absorption. Thus the wind flow needs to be highly
inhomogeneous, with the density of the emitting gas considerably higher
than that pervading the bulk of the flow. The temperature of the emitting
gas is probably $\le 10^4~K$, since the line emissivity at a higher
temperature is not higher and the filling factor likely decreases rapidly
with increasing temperature above $10^4~K$, owing to the stronger cooling
at a higher temperature. It is also probably $>7500~K$, otherwise the density
needed to produce the $Pa~n_u/Pa\beta$ ratios would be around $6\times 10^{11}~
cm^{-3}$, an order of magnitude greater than that needed at $T\ge 8750~K$,
yet the $Pa\gamma$ emissivity is about the same, so the filling factor
required of the gas at $N_H\sim 6\times 10^{11}~cm^{-3}$ is as high and
therefore more restrictive.

Among the optical hydrogen lines $H\alpha$ has the highest optical depth,
$\sim 3\times 10^3$. This can be seen from Figure 19 which shows that at
$N_H=10^{11}~cm^{-3}$ and $T=10^4~K$ $\tau_{H\alpha}$ exceeds $10^3$, the
upper bound on the abscissa. The high $\tau_{H\alpha}$ leads to a fair
amount of emission in the damping wings and likely accounts for the
conspicuous extension of the observed $H\alpha$ profile at the base. Thus,
assuming just natural broadening of the emission profile, the rate of
emitting $H\alpha$ photons that are displaced from line center by 
$\geq x$ thermal widths is $n_u A_{H\alpha} 2a/(\pi x)$, where
$a=\Gamma_{R}/(4\pi\Delta\nu_D)$, $\Gamma_R$ is the summed radiative decay
rate from the upper and lower states of $H\alpha$, and $\Delta\nu_D$ the
thermal Doppler frequency width. It is assumed that $x$ is sufficiently
large that the emission is not re-absorbed. With the total $H\alpha$ 
emission rate being $n_u A_{H\alpha}\beta_{H\alpha}$, the fraction of
photons emitted beyond $x$ thermal widths is then $2a/(\pi x \beta_{H\alpha})$.
At $10^4~K$ the thermal Doppler velocity width is $\sim 13~km~s^{-1}$,
so for $x=27$, the fraction of $H\alpha$ photons emitted at
$\vert v_{obs}\vert\ge 350~km~s^{-1}$ is $\sim 0.16$.

Figure 21 shows the analogous plots of $Pa\beta$ and $Br\gamma$ for
the infall model. The stellar plus veiling 
continuum at $1.28\mu m$ of the CTTSs in the sample of Folha \& Emerson
(2001) is not listed but most likely significantly stronger than our model
value because of the typically high $1\mu m$ veiling observed and 
probable contribution from dust emission (Muzerolle et al. 2003). Even 
assuming that our model continuum is representative, the model
$y_{v_{obs}}(Pa\beta)$, at the appropriate conditions indicated by
the $Pa~n_u/Pa\beta$ ratios, is $\le 0.15$ and falls short of
quite a few observed values. Thus, among the 49 objects with $Pa\beta$
profiles 8 (DG, DL, DR, HL, \& CW Tau, GM, \& RW Aur, YY Ori) have,
at $v_{obs}=-150~km~s^{-1}$, $y_{v_{obs}}\ge 0.5$, 8 (GG, RY, DO, DS, \&
FS Tau, UY, \& SU Aur, V1331 Cyg) have $0.2\le y_{v_{obs}}< 0.5$, and 
10 (BM And, DE, DF, DK, HK, BP, GI, HP, \& T Tau, GW Ori) have
$0.1\le y_{v_{obs}}< 0.2$. Again then, a comparison between model results
and $Pa\beta$ data leads to the same conclusions deduced earlier. It is 
interesting that the objects with strong (moderate) $Pa\beta$ flux at
$\vert v_{obs} \vert \ge 150~km~s^{-1}$ also have strong (moderate)
$Pa\gamma$ flux. This may simply be a confirmation of both the finding of
Bary et al. (2008) that the $Pa\gamma/Pa\beta$ ratio does not vary 
strongly among CTTSs and the assumption of similar $Pa\gamma$ and $Pa\beta$
profiles.

In summary, based on the hydrogen line specific fluxes at $\vert v_{obs} \vert 
\ge 150~km~s^{-1}$, we arrive at the following conclusions regarding the
infall accretion and wind flows.

1. The infall accretion flow is not responsible for the strong or moderate
hydrogen emission at $\vert v_{obs} \vert \ge 150~km~s^{-1}$. We can extend 
this inference to the bulk of the observed emission for the following reasons.
First, the emission generated by gas with $\vert v\vert \ge 150~km~s^{-1}$
is a substantial fraction of the total, since the same particles also
contribute to observed emission at $\vert v_{obs}\vert
< 150~km~s^{-1}$ when their
trajectories do not parallel the line of sight. A rough estimate of the line
profile produced by them in an azimuthally symmetric distribution is a 
flat-top shape in the regime $-150~km~s^{-1}\le v_{obs} \le 150~km~s^{-1}$.
Using this guideline to determine the fraction of the blueward emission
(to avoid the red absorption), we find from Figure 3 of EFHK06 that 19
(excepting GK Tau) of 20 CTTSs in the high and intermediate veiling groups
($r_Y\ge 0.3$) have this fraction $\ge 0.5$. Second, with the velocity
distribution expected to be a smooth function, the infall model will also
have difficulty accounting for the observed specific fluxes at $\vert 
v_{obs} \vert$ somewhat less than $150~km~s^{-1}$. Third, whatever kinematic
structure called for to produce the observed emission at $\vert v_{obs}\vert
\ge 150~km~s^{-1}$ will likely have gas particles moving at $<150~km~s^{-1}$.
All these considerations lead us to conclude that the bulk of the $Pa\gamma$
emission in these CTTSs arise from a kinematic structure other than
the accretion flow.

2. For the wind flow to be responsible for the moderate or strong hydrogen 
emission the emitting gas, whose density is
considerably higher than that of the gas occupying the bulk of the flow,
can only fill a small fraction of the volume.

These conclusions remain true, of course, when the two flows are compared
at the lower ionization rate of $\gamma_{HI}=2\times 10^{-5}~s^{-1}$. 
For either flow the specific fluxes are almost identical to those in the
$\gamma_{HI}=2\times 10^{-4}$ case at $T\ge 8750~K$, and are weaker at
$T\le 7500~K$, as can be deduced from the contrast in the $\tau_{Pa\gamma}$ 
contours between the two ionization cases (Fig. 4). Thus, for
($H\alpha$, $Pa\gamma$) the $y_{v_{obs}}$s at $N_H=10^{11}$ and
$10^{12}~cm^{-3}$ with $\gamma_{HI}=2\times 10^{-5}~s^{-1}$ are smaller
by a factor of (2.55, 4.14) and (1.06, 1.16) respectively at $T=7500~K$,
and (3.67, 6.64), and (1.92, 2.69) respectively at $T=5000~K$. This
dependence of $y_{v_{obs}}$ on $\gamma_{HI}$ when $T$ is $\le 7500~K$, as well
as $y_{v_{obs}}$'s rapid rise with $T$ increasing above $7500~K$ at a given 
$\gamma_{HI}$ lead us to favor the temperature in the hydrogen emission
region being $\ge 8750~K$.

\subsection{$HeI\lambda\lambda 10830, 5876$}

Figures 22 and 23 show the $HeI\lambda \lambda 10830, 5876$ $y_{v_{obs}}$s 
as a function of density and line optical depth for the wind and infall model
respectively. Like the $HeI\lambda 5876/\lambda 10830$ ratio, the model
results for the $HeI$ specific fluxes at
$T=3\times 10^4~K$ are very close to those at $T=2\times 10^4~K$ and are
not shown.

The 22 $HeI\lambda 5876$ broad-component profiles in Figure
7 of BEK01 present a convenient data set for comparison
with our model results. We again look at the specific flux at $v_{obs}=
-150~km~s^{-1}$ to avoid the red absorption which appears to affect
several profiles. The 22 optical veilings range from 0.1 to 20, and the
observed $y_{v_{obs}}$s range from 0.01 to 0.5, but the correlation 
between $y_{v_{obs}}$ and $r_V$ does not appear to be strong. Eight 
objects (CW, HN, DG, DL, \& DR Tau, AS 353 A, RW \& UY Aur) have
$0.2\le y_{v_{obs}} \le 0.5$ and $r_V$ between 0.8 and 20, 4 (GG, DQ, \& DF Tau,
GM Aur) have $0.1\le y_{v_{obs}} <0.2$ and $r_V$ between 0.2 and 0.7,
and the rest have $y_{v_{obs}}<0.1$ and $r_V$ between 0.1 and 4.7. With
our model $r_V$ of 0.67, the model $y_{v_{obs}}$s likely need to be
lowered (raised) for comparison with the first (second) veiling group.
Even for observed $y_{v_{obs}}$s as high as 0.5, it appears that the infall
model can accommodate them. As a specific example, we assume that the
$HeI\lambda 5876/\lambda 10830$ ratio is 0.4 and determine from Figure
8 that the required physical conditions can be ($T=2\times 10^4~K$,
$N_H = 1.1\times 10^{11}~cm^{-3}$), ($T=1.5\times 10^4~K$, $N_H=3.2\times
10^{11}~cm^{-3}$), or ($T=1.25\times 10^4~K$, $N_H=1.5\times 10^{12}~cm^{-3}$)
which produce, as seen from Figure 23, a $y_{v_{obs}}\sim 0.6$ in each case.
The $HeI\lambda 5876$ optical depth is greater than 50 (cf. Fig. 23 right 
bottom panel), so a red absorption will be seen for certain lines of sight. For
smaller observed $y_{v_{obs}}$s, however, a small filling factor of the
emitting gas can be invoked to reduce the incidence of red absorption.

The $1\mu m$ spectroscopic survey of CTTSs by EFHK06
also procures a set of $HeI\lambda 10830$ profiles. Most of them have strong 
absorption features, so we consider solely the high $1\mu m$ veiling group
($r_Y\ge 0.5$) whose members are among the objects with the strongest
$HeI\lambda 10830$ emission. Excluding DK Tau which has only absorption
features, we determine the $y_{v_{obs}}$s at $v_{obs}=150~km~s^{-1}$ to
avoid the blue absorptions. They range from 0.3 to 1.3. Our model $r_Y$ is
0.19, but even lowering the model $y_{v_{obs}}$s considerably for
comparison with the CTTSs with particularly high $r_Y$s, it appears that
the infall model can produce the highest observed $HeI\lambda 10830$
specific fluxes. 

For both $HeI\lambda 10830$ and $\lambda 5876$ it is
also clear that their emission are far too strong in a laminar wind flow,
and that the gas producing the helium emission can only occupy a fraction
$\le 0.01$ of the contour area shown in Figure 1 in order to
match the observed specific fluxes.

While the above analysis does not directly reject either model, we 
favor the wind region as the production
site of strong helium emission for the following reasons.

1. Except near the accretion shock where $UV$ photons are produced, 
the required temperature of $T\ge 1.25\times 10^4~K$ is difficult to
generate and maintain in an accretion flow, which is primarily in free
fall. On the other hand, it is reasonable to expect that acceleration of
the gas in a wind would produce heating. Helium emission in the wind
is also compatible with the earlier deduced condition for hydrogen 
emission in the wind in that the decreasing filling factor of the gas with
increasing temperature from $10^4$ to $\sim 2\times 10^4~K$ is
consistent with the cooling and expansion of gas at a higher temperature.

2. An infalling flow is hard pressed to explain the strong blue asymmetry
in the $HeI\lambda 5876$ line wings. Judging the emission at $\vert 
v_{obs} \vert \ge 150~km~s^{-1}$ (cf. Fig. 7 of BEK01),
we find that 18 of the 22 $HeI\lambda 5876$ broad components are 
stronger on the blue side, although 3 of them (AS 353 A, RW \& GM Aur)
have red absorptions that clearly accentuate the asymmetry, 3 (DD, DF, \&
DO Tau) are stronger on the red side, and 1 (DE Tau) is too weak to
discern. While lack of azimuthal symmetry or strong local inhomogeneities
can generate asymmetries in the wing emission, their numbers of blue and
red asymmetries should, for random lines of sight, statistically balance
out. On the other hand, stellar occultation of the approaching accretion
flow and disk occultation of the receding wind flow produce always a red
and blue asymmetry respectively. As seen from Figure 1, the projected area
of the $v_{obs}=-150~km~s^{-1}$ contour in the infall model is 
only slightly larger than the projected stellar surface area. There is
little leeway in adjusting the accretion flow geometry to avoid 
diminishing the blue wing emission, let alone enhancing it above the red
wing emission.

3. The accretion flow model has greater difficulty producing the
$HeI\lambda 10830$ profile morphologies of those objects with strong
emission (cf. Fig. 4 of EFHK06). First, while most of the 38
objects have blue and/or red absorptions that make it difficult to
decipher the wing emission at $\vert v_{obs} \vert\ge 150~km~s^{-1}$,
4 have mostly emission, of whom BP Tau's wing emission is too weak to 
compare its two sides and the remaining 3(CW \& HN Tau, RW Aur A) all
have stronger blue wing than red wing emission, and will pose a challenge
to the infall model. Second, when emission above continuum is observed
along with absorption, the objects with strong emission tend to associate
with blue absorptions indicative of a stellar wind (cf. Fig. 10 of 
Kwan et al. 2007), whereas the objects with strong absorptions on the
red side only do not have comparably strong emission (e.g., AA Tau, BM And,
RW Aur B, LkCa 8 in Fig. 4 of EFHK06 or Fig. 2 of
Fischer et al. 2008). The latter appear to have both a lower $r_Y$ and a
smaller equivalent width in the emission above the continuum. Bearing in
mind that the observed emission is contributed in part already by 
scattering of continuum photons, it is surprising, if the accretion flow
is the site of helium emission, that the in-situ emission are relatively
weak in those accretion flows with particularly large widths and sizes 
that are needed to produce the broad and deep red absorptions (cf. \S 4.2
\& 4.3 of Fischer et al. 2008). On the other hand, the above-mentioned
correlation naturally follows if the wind flow is the site of helium 
emission. Also, in producing the needed in-situ emission there is more
leeway afforded the wind model in the choices of $r$ and filling factor
of the high temperature regions. The Monte Carlo simulations by 
Kwan et al. (2007) already indicate that the $HeI\lambda 10830$ profile
morphologies can be accounted for by a stellar wind, depending on the
opening angle of the wind and the viewing angle. However, in light of
the present finding that the emission regions do not occupy the bulk
of the volume, there is the constraint that these regions share the
velocity distribution of the gas occupying the bulk of the volume and
responsible for the $HeI\lambda 10830$ absorption.

In \S 6.1 where the $HeI\lambda 5876/\lambda 10830$ and $Pa\gamma/HeI
\lambda 10830$ ratios are discussed it was mentioned that, even though
their model values generated at $T=5000~K$, $5\times 10^{10}~cm^{-3}
\le N_H \le 10^{12}~cm^{-3}$ are closest to the observed values (cf. Fig 10),
several reasons argue for a higher temperature. One, given in \S 6.2.2,
is based on the difficulty of simultaneously matching $Pa\gamma/Pa\beta$
at $T=5000~K$. We add a second reason here, based on the weakness of the
$HeI\lambda 5876$ flux at $T=5000~K$. As seen from Figure 22, its model
$y_{v_{obs}}$ in the wind model increases from 0.1 to 1.3 as $N_H$ increases
from $5\times 10^{10}$ to $10^{12}~cm^{-3}$. Reducing it by a factor of
2 upon rescaling it to the continua of the 8 objects with $r_V$s between
0.8 and 20 and observed $y_{v_{obs}}$s between 0.2 and 0.5, it is clear
that the emitting gas need to fill a substantial portion of the contour
surface shown in Figure 1 in order to match the observed values. This
would, however, produce a blue $HeI\lambda 5876$ absorption which is
never seen.

As $T$ increases from $5000~K$, the $HeI\lambda 5876/\lambda 10830$ ratio
first decreases (cf. Fig. 8) and then rises again when $T$ exceeds $10^4~K$,
so we expect the helium emission to arise from regions of temperature above
$10^4~K$. With $T$ increasing from $10^4$ to $2\times 10^4~K$, the specific
fluxes of both $HeI\lambda 5876$ and $\lambda 10830$ increase rapidly (cf. 
Fig. 22), but then stay nearly constant for $T$ between $2\times 10^4$
and $3\times 10^4~K$. If regions in the wind flow are raised to temperatures
above $2\times 10^4~K$ before cooling and expansion come into play, we
anticipate them to be most effective in contributing to the observed helium
emission when they are at temperatures near $2\times 10^4~K$.

When $\gamma_{HeI}$ is reduced from $10^{-4}$ to $10^{-5}~s^{-1}$ the
specific fluxes of both $HeI\lambda 10830$ and $\lambda 5876$ remain
almost the same for $T\ge 1.5\times 10^4~K$, but fall rapidly for
$T< 1.5\times 10^4~K$. This result can be anticipated from the large
difference in the $\tau_{HeI\lambda 10830}$ or $\tau_{HeI\lambda 5876}$
contour between the two cases of $\gamma_{HeI}$ in the temperature range
$10^4~K<T<1.5\times 10^4~K$ (cf. Fig. 4). This strong dependence of 
$HeI$ emission on $\gamma_{HeI}$ in that temperature regime, and the need
for a higher density to obtain the same $HeI\lambda 5876/\lambda 10830$
ratio as $T$ decreases from $2\times 10^4~K$ lead us to favor a temperature
range from $1.5\times 10^4$ to $2\times 10^4~K$ for the $HeI\lambda
5876$ emission.

\subsection{$CaII\lambda 8498$, $OI\lambda 8446$}

It was pointed out in \S 6.3.2 that the observed $CaII\lambda 8498/Pa\gamma$
ratios of integrated fluxes appear to fall into two groups and, earlier
in \S 7, that the strong $CaII\lambda 8498$ lines are narrower than
$Pa\gamma$. For those objects with $CaII\lambda 8498/Pa\gamma\sim 1$ or 
less, the $CaII$ emission can be produced by the set of physical conditions
($8750~K\leq T\leq 1.25\times 10^4~K$, $N_H$ around $10^{11}~cm^{-3}$)
producing primarily hydrogen emission. The larger issue, of course, concerns
the strong emitters with $CaII\lambda 8498/Pa\gamma$ greater than $\sim 5$.
Their $CaII$ profiles appear narrower than the $Pa\gamma$ profiles in the
data set of Edwards et al. (2010). There may already be a hint of this
property from a comparison of non-simultaneous $CaII$ and $Pa\gamma$ 
profiles. Thus, the ratio of specific flux at $v_{obs}=-150~km~s^{-1}$
to that at $v_{obs}=0$ is smaller for $CaII\lambda 8542$, being 0.27, 0.1,
and 0.4 for DG, DR, and DL Tau respectively (Muzerolle et al. 1998b), than
for $Pa\gamma$, being 0.46, 0.33, and 0.54 (EFHK06).

In view of the possibly different $CaII\lambda 8498$ and 
$Pa\gamma$ profiles, we will, in comparing between model and observed
values of the $CaII\lambda 8498$ specific flux at $v_{obs}=-150~km~s^{-1}$,
not utilize the constraint imposed on the physical conditions by the
$CaII\lambda 8498/Pa\gamma$ ratio of integrated fluxes, and consider
all densities in the regime $5000\le T \le 1.25\times 10^4~K$
as possible. Figure 24 shows the dependences of $y_{v_{obs}}$
on $N_H$, $T$, and line optical depth for the infall model.
The observed values of DG, DR, DL Tau and RW Aur are 1.4, 0.6, 2.5 
and 3.7 (Muzerolle et al. 1998b).
Anticipating that the model $y_{v_{obs}}$s have to be reduced by a 
factor of $\ge 2$, owing to the strong veilings at $\lambda 8498$ 
(between 0.7 and 5.1) of 
those objects, we find that, with the exception of RW Aur, the
infall model can produce the wing emission. We do not consider this
strong enough evidence to argue against the infall model. 

The issue concerning the strong $CaII$ emission is, of course, its  
emission near line center. While our specific flux calculation provides no aid 
in discriminating between the wind and infall models,
the earlier derivation of requisite physical conditions for strong $CaII$
emission (i.e., $T\le 7500~K$, $N_H$ close to $10^{12}~cm^{-3}$)
remains true, because of the much stronger dependence of the specific 
fluxes and flux ratios on $N_H$ or $T$ than on the velocity gradient.
With this in mind, we do not think the wind region is the site of strong
$CaII$ emission for the following reasons.

1. The physical condition of high density and low temperature is incongruous
with that inferred from the helium and hydrogen emission, and with the
constraint imposed by the lack of blue absorption in $Pa\gamma$. With the
bulk of the wind region occupied by gas at a density of $<10^{10}~cm^{-3}$
(cf. Fig. 4), the hydrogen emitting regions at $T\le 1.25\times10^4~K$ and
$N_H\sim 10^{11}~cm^{-3}$, and the indication that lower temperature is
associated with lower density and larger filling factor, it does not  
appear the wind region has the physical conditions needed for
strong $CaII$ emission.

2. The strong $CaII$ emission has a linewidth narrower than the $Pa\gamma$
width or the typical velocity extent seen in a broad, blue $HeI\lambda
10830$ absorption. If $CaII$ emission originates from the wind region, it
is surprising , particularly in the case of DR Tau, that, 
while the $Pa\gamma$ and $HeI\lambda 5876$ line
shapes are quite similar, the $CaII$ line shape is so different.

3. The strong $CaII$ line appears symmetrical about line center. It does
not have the characteristic feature of a stronger blue emission as
exhibited often by $Pa\gamma$ and $HeI\lambda 5876$, and attributable
to disk occultation of the receding wind.

The strong $CaII$ emission near line center can be produced in an accretion
flow. Its narrower width and different line shape from $Pa\gamma$ will
not be issues. The condition of $T\le 7500~K$ is not unreasonable in
light of Martin's (1996) calculation of the thermal structure of the
infalling gas, and the density criterion of close to $10^{12}~cm^{-3}$
may not be too restrictive a constraint on the mass accretion rate if the
flow is dilutely filled. There is a second possible site, the disk
boundary layer where the gas dissipates part of its rotational energy before
infalling along a stellar magnetic field line. We will elaborate on our
preference in the next section when all the deductions and arguments put
forth in \S\S 5, 6, 7 are synthesized.

We will not comment much on the $OI\lambda 8446$ specific flux. 
Our calculations indicate that
the specific flux of $OI\lambda 8446$ rises more rapidly than
$Pa\gamma$ with $N_H$ increasing from $10^{11}$ to $10^{12}~cm^{-3}$,
causing $OI\lambda 8446/Pa\gamma$ to be higher at the conditions
suitable for strong $CaII$ emission. Thus there is a larger contribution
to the $OI\lambda 8446$ line from the $CaII$ emission region than is the
case for $Pa\gamma$. 

\section{Origins of the Strong $H$, $HeI$, and $CaII$ Line Emission}

In \S\S 5, 6 we have presented calculations that shed light on the physical
conditions giving rise to the absorption and emission features of most
of the prominent lines observed in CTTSs. These are excitation calculations
that include all the important physical processes affecting the atomic/ionic
level population. They are facilitated by the presence of large velocity
gradients expected in a wind or infalling flow in that the level population
depend only on the local density ($N_H$), temperature ($T$),
photon ionization rate ($\gamma_{HeI}$), and velocity gradient ($2v/r$).
Thus the resulting line emissivity ($erg~s^{-1}~cm^{-3}$) can be calculated
in the ($N_H$, $T$, $\gamma_{HeI}$, $2v/r$) parameter space and the ratio
of two line emissivities will demarcate the requisite physical conditions.
One of our findings is that the $H$ and $HeI$ emission regions can occupy
only a very small fraction of the wind flow. Our use of $2v/r$ for the velocity
gradient is then inaccurate, and a more suitable choice would be $\delta v
/\delta l$, where $\delta l$ is the linear dimension of an emission region
and $\delta v$ the thermal/turbulence velocity width. However, as 
demonstrated in \S 7.1, the line emissivity is much more sensitive to $N_H$
than to $2v/r$ or $\delta v/\delta l$, particularly when the line is
optically thick, as is true for the $H$, $HeI$, and $CaII$ lines studied,
so a large change in $2v/r$ or $\delta v/ \delta l$ occasions only a small
change in $N_H$. Because the requisite physical conditions deduced for the
gases responsible for the $HeI\lambda 10830$ absorption, $H$ emission,
$HeI$ emission, and $CaII$ emission are so disparate, we are confident our
conclusions are not fundamentally altered by the uncertainty in this
parameter. For the same reason, even though we emphasize the two locations of 
$r=4$ and $2.5R_\ast$ and the velocity $v=150~km~s^{-1}$, the results on
the ordering of line opacities and the responses of line emissivity ratios
to $N_H$, $T$, and $\gamma_{HeI}$ are applicable to other locations and
not sensitive to the kinematic structure. Then in \S 7 we distinguish between
the wind and accretion flows by evaluating the specific fluxes of the more
important lines at $\vert v_{obs} \vert =150~km~s^{-1}$. Here we recapitulate
and synthesize the deductions reached separately in 
\S\S 5, 6, 7 and, in conjunction
with other arguments, decide on the locations of the $H$, $HeI$, and $CaII$
line emission.

Our conclusions drawn with regard to the
$H$, $HeI$, and $CaII$ broad line emission in the following subsections
apply only to the strong emitters,
specific examples of which have been mentioned in \S\S 6, 7. This qualification
arises for the following reasons. First, the determination of observed
line ratios, particularly $HeI\lambda 5876/\lambda 10830$, is not as
reliable for weak emitters because of the presence of absorption features
and the contribution to emission from scattering of continuum photons.
Second, a couple of our arguments concern the limitations of an infalling
flow on the production of line photons, and therefore hinge upon the
observational bar placed by the strong emitters.

\subsection{Physical Conditions of the Gases Producing Emission and
Absorption Lines}

Below we summarize and discuss the findings obtained from a comparison between
model results and observational data on line opacities and line ratios. They
shed light on the properties of the gases responsible for the emission 
and absorption lines.

1. Optically thick $H$ and $HeI$ emission lines. The gases producing the
$HeI\lambda\lambda 10830, 5876$ and $Pa\gamma$, $Pa\beta$ emission are very
optically thick in those lines, in order to produce the observed
$HeI\lambda 5876/\lambda 10830$, $Pa\gamma/Pa\beta$, and $Pa\gamma/HeI
\lambda 10830$ ratios (cf. \S\S 6.1.2., 6.2.2.).

2. Separate physical conditions for $H$, $HeI$, and $CaII$ emission. The
need for distinct physical conditions conducive to $H$ and $HeI$ emission
is demonstrated clearly in Figure 10 where it is seen that many of the
marked CTTSs, which are among the strong $HeI\lambda 5876$ emitters, have
$HeI\lambda 5876/\lambda 10830$ and $Pa\gamma/HeI\lambda 10830$ ratios
that cannot be simultaneously produced with a common temperature range, but
must require that the bulk of the $HeI\lambda 5876$ emission be produced
at temperatures higher than those producing the bulk of the $Pa\gamma$
emission (cf. \S 6.1.2.). 

For the strong $HeI\lambda 5876$ emission we favor the temperature range
$1.5\times 10^4~K\leq T\leq 2\times 10^4~K$. The upper bound is adopted
because the line emissivity at a higher temperature 
is not higher (\S 7.2), while presumably the 
emission area is smaller. The lower bound is adopted because at a
lower temperature the line emissivity is much more dependent on
$\gamma_{HeI}$ and a higher density is needed to produce the same 
$HeI\lambda 5876/\lambda 10830$ ratio, as seen in Figures 8 and 9 (cf. \S 7.2).
The same figures also show that within the favored temperature range
$HeI\lambda 5876/\lambda 10830$ rises steeply with increasing density,
and the density needed to produce $HeI\lambda 5876
/\lambda 10830 \geq 0.3$ is $\geq 10^{11}~cm^{-3}$.

The physical conditions responsible for $H$ emission, as deciphered
by examining the ratios among the hydrogen lines themselves, 
namely $Pa~n_u/Pa\beta$ and $Br\gamma/Pa~n_u$, indicate that
$N_H$ is centered around $10^{11}~cm^{-3}$ for $T\geq 8750~K$, and
$\geq 5\times 10^{11}~cm^{-3}$ for $T\leq 7500~K$ (cf. \S 6.2.2.).
In conjunction with the observed $Pa\gamma/HeI\lambda 10830$ ratio, however,
the temperature can be narrowed to the range $5000~K<T\leq 1.25\times
10^4~K$. The upper bound comes from the small $Pa\gamma/HeI\lambda 10830$
ratio ($\leq 0.2$) that can only be generated at $N_H$ around $10^{11}~cm^{-3}$
at $T\geq 1.5\times 10^4~K$ (cf. Figs. 8 \& 9 right top panel). The lower
bound is due to the need of very high densities ($>2\times 10^{12}~cm^{-3}$)
to produce $Pa\gamma/Pa\beta>0.8$ (cf. Fig. 11 left top panel) and the
concomitant result of $Pa\gamma/HeI\lambda 10830$ lower than observed (cf. 
Figs. 8 \& 9 right top panel, \S 6.2.2). We further favor the temperature range 
conducive to $H$ emission being $8750~K\leq T\leq 1.25\times 10^4~K$, because  
the line emissivities, particularly $Pa\gamma$ and $Br\gamma$, 
decrease rapidly with $T$ decreasing below $8750~K$ (cf. Figs. 19, 20, 21
right top panel, \S 7.1). In this temperature range and with $N_H$ around
$10^{11}~cm^{-3}$ the $CaII\lambda 8498/Pa\gamma$ ratio is also in line
with the observed values among the weak $CaII$ broad emitters (cf. \S 6.3.2).

The observed $HeI\lambda 10830/Pa\gamma$ ratio, typically $\sim 1.5$ among
the strong emitters, also indicates that the $HeI$ emission area is smaller
than the $H$ emission area. Thus, if the emission area contributing to $HeI$
emission, primarily with $T$ around $2\times 10^4~K$, were the same as that
with $T$ around $10^4~K$, $HeI\lambda 10830/Pa\gamma$ would be $\geq 10$,
because while the $Pa\gamma$ emission from each area is about the same 
(cf. Fig. 19 right top panel) the
$HeI\lambda 10830$ emission at $2\times 10^4~K$ and $N_H$ around 
$10^{11}~cm^{-3}$ is $\geq 20$ times the $Pa\gamma$ emission (cf.
Figs. 8 \& 9 right top panel).

The physical conditions responsible for strong broad $CaII$ emission are most
evident in Figures 17 and 18 where the observed $CaII\lambda 8498/OI\lambda
8446$ values greater than $\sim 7$ point to a temperature
$T\leq 7500~K$ and then from Figure 16 (left top panel) a density $N_H$
close to $10^{12}~cm^{-3}$ (cf.\S 6.3.2). They can also be inferred from the 
large $CaII\lambda 8498/Pa\gamma$ values.
There are, however, also CTTSs observed to have $CaII\lambda 8498/
OI\lambda 8446$ and $CaII\lambda 8498/Pa\gamma$ of $~\sim 1$
or less (e.g. BP Tau in Figs. 17 \& 18), and $Pa\gamma$ emission strengths not
much weaker than those among the strong $CaII$ emitters. Their $H$ and
$CaII$ emission indicate $8750~K\leq T\leq 1.25\times 10^4~K$ and $N_H$ around
$10^{11}~cm^{-3}$ (cf. \S 6.3.2), as mentioned just earlier.
Such physical conditions
also naturally arise through cooling of the
regions responsible for $HeI$ emission, which is often observed to be strong
when $CaII$ emission is strong. Thus we conclude that this set of physical 
conditions producing primarily $H$ emission 
is also present. It may indeed be the most prevalent since it is
needed when both $HeI\lambda 5876$ and $CaII$ emission are weak. When the
latter emission are strong, the other two sets, namely ($1.5\times 10^4~K
\leq T\leq 2\times 10^4~K$, $N_H\geq 10^{11}~cm^{-3}$) for the bulk of
the $HeI$ emission and ($T\leq 7500~K$, $N_H\sim 10^{12}~cm^{-3}$) for the
bulk of the $CaII$ emission, are also present.

The fundamental cause of the distinct sets of physical conditions 
conducive to $H$, $HeI$, and $CaII$ emission is the different
$H$, $HeI$, and $CaII$ atomic/ionic structures, making their efficacies
in photon production sensitive to different temperature ranges. Thus,
while $HeI\lambda 10830$ and $\lambda 5876$ emission rise
rapidly with increasing $T$ up to $2\times 10^4~K$, hydrogen emission in
Balmer, Paschen, and higher order lines are not more efficient at $T>10^4~K$
because the stronger collisional ionization reduces the neutral hydrogen
fraction, and a temperature decreasing below $10^4~K$ clearly favors $CaII$ 
over $H$ emission. The separate physical conditions we identify for the
$H$, $HeI$, and $CaII$ emission regions
are therefore understandable and not
particular to our excitation model despite its simplifications and
approximations.

3. Higher densities for the gases producing the $H$ and $HeI$ emission 
lines than the gas producing the broad, blue absorption.
Broad, blue absorptions indicative of a stellar wind are often seen in
$HeI\lambda 10830$ but almost never in
$HeI\lambda 5876$, $Pa\beta$, or $Pa\gamma$. This observational constraint,
in conjunction with finding 1 above, means that a laminar wind will not be
able to produce simultaneously the observed absorption and emission features
seen in either $HeI$ or $H$ lines. This dilemma is the same as that realized 
in earlier wind models of hydrogen line emission (Hartmann et al. 1990).
Anticipating that the temperature of the gas occupying the bulk of the wind
volume is $\le 10^4~K$, the constraint placed by the
observed absorption features
limits the density of this gas to be no more than $\sim 10^{10}~cm^{-3}$
(cf. Fig. 4), which is considerably lower than the density of the
$H$ or $HeI$ emission gas. Thus a highly clumpy flow is called for if the
$H$ and $HeI$ emission regions also reside in the wind (cf. \S\S 7.1, 7.2).

4. In order to produce a $\tau_{HeI\lambda 10830}\geq 1$ in the radial
wind, the minimum density needed at the location where the wind reaches 
a speed of $150~km~s^{-1}$ is $\sim 5\times 10^8~cm^{-3}$ (cf. Fig. 4),
and the corresponding mass loss rate is $\sim 0.5\times 10^9~M_\odot~yr^{-1}$.

5. The red absorption gas has a temperature $T< 10^4~K$ and a density
$N_H$ greater than or about $10^{11}~cm^{-3}$. The former condition is 
needed to ensure a more prevalent occurrence of red absorptions in the
$NaI$ doublet than in $Pa\gamma$ (cf. Fig. 6, \S 5.2). The latter condition is 
inferred from the observed ordering of the lines in propensity of showing a
red absorption.

6. $UV$ photoionization is necessary to produce the broad red and
blue absorptions seen in $HeI\lambda 10830$. This 
is demonstrated forcefully by the more prevalent occurrence of
red absorptions in $HeI\lambda 10830$ than in $Pa\gamma$, given that the
low temperature of the absorption gas ($T<10^4~K$ as noted above) renders
collisional excitation futile (cf. \S 5.2). $UV$ photoionization of $HeI$ in 
the gas occupying the bulk of the wind flow is also needed to produce $HeI
\lambda 10830$ blue absorptions since it is most likely that the temperature
of that gas is also $<10^4~K$ if the hydrogen emission regions occupy only
a small fraction of the volume and have a temperature $8750~K\leq T\leq
1.25\times 10^4~K$. With the main depopulation path of the highly metastable 
$HeI\lambda 10830$ lower level being collisional excitation to the $2s~^1S$
state, whose rate decreases with decreasing $T$ and $N_e$, a 
$\tau_{HeI\lambda 10830}\sim 1$ is quite realizable at $T\leq 6250~K$ even
for low ionization fluxes (cf. \S 5.1). We therefore think $UV$ photoionization
is also the excitation mechanism responsible for the sharp blue absorptions
and central absorptions seen in $HeI\lambda 10830$.

$UV$ photoionization of $HI$ in the bulk of either the wind or accretion
flow is not as crucial, but it likely contributes, particularly if the
gas temperature is $<7500~K$, given that $UV$ photons are already present 
for $HeI$ ionization. In the regions producing either $H$ or $HeI$ optical
and infrared line emission collisional excitation is the primary
mechanism, and $UV$ photoionization plays a less significant role.

We will make use of these findings and additional ones deduced in \S 7, as
well as arguments based on correlations among absorption and emission
features, to support our decisions on the emission sites of the strong
$HeI$, $H$, and $CaII$ lines in the following two subsections.

\subsection{Wind Region as the Site of the Strong $HeI$ and $H$ Emission}

We first summarize the arguments against the accretion flow as the 
appropriate site.

1. Difficulty for an infalling flow to account for the stronger blue
wing emission. In \S\S 7.1, 7.2 we have listed CTTSs to show that both
$Pa\gamma$ and $HeI\lambda 5876$ have distinctly a stronger blue wing. In
an infalling flow the emission area at $v_{obs}\le -150~km~s^{-1}$ is not
much larger than $\pi R_{\ast}^2$ (cf. Fig. 1), so it is very difficult
to avoid stellar occultation of the infalling flow approaching an observer.
This preferential attenuation of the blue wing emission is intrinsic to
the infall flow geometry, but is counter to the observed trend.

2. Difficulty for an infalling flow to produce the observed fluxes in the
line wings. The finding that the $H$ and $HeI$ emission lines are optically
thick means that their specific fluxes depend on the emission area and the
line excitation temperatures. For an infalling flow its geometry severely 
confines the projected areas with large $\vert v_{obs}\vert$s. The hydrogen
lines face, in addition, a strong limitation on their excitation 
temperatures through the $n\rightarrow n+1$ collisional excitations that
lead to rapid ionization at $T>10^4~K$. As a result we find that the observed
$Pa\beta$ and $Pa\gamma$ fluxes at $\vert v_{obs}\vert\ge 150~km~s^{-1}$
cannot be produced by our infall model over the broad ranges of physical
parameters explored. This result is not surprising, as it is simply an
alternative, albeit more explicit, way of phrasing previous findings that
the hydrogen profiles calculated from accretion flow models are narrower
than observed (Folha \& Emerson 2001). Interestingly, the steep
rise of the helium line emissivity with increasing $T$ from $10^4$ to
$2\times 10^4~K$ (cf. Fig. 23) enables the infall model to reproduce the
observed $HeI$ line fluxes. However, a temperature $\ge 1.25\times 10^4~K$
is required, and we think it is difficult to heat the gas, which is 
primarily in free fall, to those temperatures, except possibly near the
impact sites where photoionization heating by photons generated in the
shocks can be important. It has also been deduced earlier that the gas
occupying the bulk of the infalling flow has a temperature $<10^4~K$, so we do
not think the accretion flow generally has the physical conditions conducive 
to strong $HeI$ emission.

3. Weak or no correlation between strong $HeI\lambda 10830$ red absorptions
and strong $Pa\gamma$ or $HeI\lambda 5876$ emission. If $Pa\gamma$ or
$HeI\lambda 5876$ is produced in the accretion flow, it is reasonable to
expect that presence of a strong  
$HeI\lambda 10830$ red absorption should be accompanied by a strong $Pa\gamma$
or $HeI\lambda 5876$ line. From Figures 3 and 4 of EFHK06 it is seen that
among the CTTSs with the strongest $HeI\lambda 10830$ red absorptions,
2 (DK Tau, YY Ori) have moderate $Pa\gamma$ emission, but 7 (AA, GI, DN,
\& V830 Tau, BM And, RW Aur B, LkCa 8) have only weak $Pa\gamma$ emission.
In the simultaneous $1\mu m$ and optical data sample of Edwards et al.
(2010), of the 6 objects with strong $HeI\lambda 10830$ red
absorptions, only 1 (DR Tau) has strong $Pa\gamma$ and $HeI\lambda 5876$
emission, while the rest (DK, GK, AA, \& GI Tau, BM And) have weak $Pa\gamma$
and $HeI\lambda 5876$ (broad component) emission.

4. It is noted in \S 5.2 that when $\tau_{HeI\lambda 10830}$ is $\sim 1$ the
$H\alpha$ opacity is close to or higher than $\tau_{HeI\lambda 10830}$.
Then if the $H\alpha$ emission originate in the accretion flow, which is
probably within the radial wind, it will be scattered by the wind, and the
resulting observed profile will be highly asymmetric with the red side
much strong than the blue side. The collection of 31 $H\alpha$ profiles
in Figure 15 of BEK01 shows that such profiles are rare, with AS 353A,
DR Tau, and DO Tau being the lone examples.

5. Absence of correlation between $HeI\lambda 5876$ broad component and
narrow component emission. A narrow line, either on top of
a broad one or by itself, is seen in $Pa\gamma$ (EFHK06), the $CaII$
infrared triplet (Muzerolle et al. 1998b), and most frequently and
prominently in $HeI\lambda 5876$ (BEK01). This observational result is
understandable if the narrow line emission is formed in the post-shock
regions at the impact sites of the infalling flow on the star, since the
$HeI$ lines form at higher temperatures and will likely reach higher 
excitation temperatures than the $H$ and $CaII$ lines. The $HeI\lambda
5876$ narrow component (NC) strength, being sensitive to the surface
area of impact and the pre-shock density, is then a good indicator of
the accretion flow magnitude. Then, if the $HeI\lambda 5876$ broad component 
(BC) arises from the accretion flow, one would expect a correlation between
the BC and NC strengths. 
Even though, with the available information of BC and NC equivalent
widths ($W_\lambda$) and red veilings ($r_R$), a direct plot of the BC
flux versus NC flux for the sample of 31 CTTSs cannot be made, we can
obtain a sense of this correlation from Figure 1 of BEK01. There it is
seen that among those CTTSs with the highest values of $(1+r_R)W_\lambda$
in the BC 4(CW, HN, \& DG Tau, AS 353A) have no or comparatively much weaker
NC, 3(DL Tau, RW \& UY Aur) have a comparatively weaker NC (but typical,
in terms of $(1+r_R)W_\lambda$ in the NC, relative to the whole sample), and
1 (DR Tau) has a comparatively weaker NC (but strong relative to the whole
sample). Then among those CTTSs with the highest values of $(1+r_R)W_\lambda$
in the NC 2(DD \& DO Tau) have a comparable BC, and 2(FM \& HK Tau) have
no BC. We judge from this comparison between the BC and NC values of
$(1+r_R)W_\lambda$ that there is no correlation between their strengths, and
conclude that the broad helium line emission is generally not related to the
accretion flow. We can extend this conclusion to the hydrogen line emission.
Even though the 38 $Pa\gamma$ profiles in EFHK06 and the 31 $HeI\lambda 5876$ 
profiles in BEK01 are not simultaneous, we see that the group of CTTSs with
strong (weak) $HeI\lambda 5876$ BC is almost the same as the group of 
CTTSs with strong (weak) $Pa\gamma$ emission.

While the above-mentioned characteristics of the $H$ and $HeI$ line profiles
pose severe difficulties for an infalling flow, they favor a wind flow for 
the line origin. These include the stronger blue wing emission and the
usually blue centroid (both owing to disk occultation of the wind flow 
receding from an observer), high blue wing velocities (as there is no
limitation on the terminal wind speed), and the association of strong
$H$ and $HeI$ emission with blue absorptions in $HeI\lambda 10830$ that are
indicative of a stellar wind. The $H$ and $HeI$ specific fluxes, either
at the line wings or at line center, are not fundamental issues in the
wind model, even though their observed values, as well as the absence of
blue $Pa\gamma$ and $HeI\lambda 5876$ absorptions, stipulate that the
emission arise from only a small fraction of the wind volume. The physical
conditions and filling factors deduced for the $H$ and $HeI$ emission
regions may indeed be brought about through the expansion
and cooling of high temperature, high pressure clumps. It does remain to be 
demonstrated, however, that there is a viable acceleration process that
produces a highly clumpy flow. Despite this uncertainty, the overall
positive comparision between model and observed line profiles and strengths,
together with earlier arguments against the accretion flow, lead us to 
decide squarely on the wind region as the site of the strong $H$ and
$HeI$ line emission.

The possibility of strong photoionization heating of the accretion flow
near the impact shock, however, means that $HeI$ line emission, possibly
strong, may occasionally arise from the accretion flow. In \S 7.2 it is pointed
out that statistically the $HeI\lambda 5876$ profiles show a definite
preference for a stronger blue wing, but 3 (DD, DF, \& DO Tau) clearly
have a red centroid (cf. Fig. 7 of BEK01). Such a red asymmetry can
be brought about if the emission arise primarily from the part of the
accretion flow with a $z$ velocity component towards the disk, specifically
the part at polar angle $\theta < 54.7^o$ in a dipolar trajectory, which
is the part close to the impact shock (BEK01). The three profiles are
much weaker at $v_{obs}\le -150~km~s^{-1}$ than at $v_{obs}\ge 150~km~s^{-1}$
and also have strong narrow components. Thus, based on the $HeI\lambda 5876$
profile morphologies, an origin of the helium emission in an accretion
flow is favored for those three objects.
A possible caveat is the uncertain extent to which azimuthal
asymmetries in flow geometry and physical conditions affect the profiles.
This can only be answered by time monitoring in both optical and $1\mu m$
spectral regions, so that the relative contributions from the radial wind
and accretion flow can be assessed through analysis of both $HeI\lambda
10830$ and $\lambda 5876$ profiles. The great majority of observed
$HeI$ profiles, however, show characteristics that favor a wind origin.

\subsection{Disk Boundary Layer and Accretion Flow as Sites of $CaII$
Line Emission}

In \S 7.3 we argue against the wind region as the site of the strong $CaII$
infrared triplet emission and mention that, in addition to the accretion
flow, the disk boundary layer is a possible site. In essence this boundary 
layer, where the accreting particles dissipate part of their rotational
energies before falling along the stellar field lines, is the base of
the accretion flow. Here we list
the reasons why the $CaII$ emission from this base may be significant.

1. The energy dissipated in the disk boundary layer is a significant 
fraction of the total gravitational potential energy released by the 
infalling gas. Most of this energy emerges in a photon continuum which
we identify as part of the observed continuum excess around $1\mu m$.
Like the situation in a stellar atmosphere, the chromospheric region
above the continuum formation zone in this boundary layer will produce
$CaII$ infrared triplet emission, but the line profile, instead of being
narrow, will be Doppler broadened by the strong rotational motion. There
appears to be a correlation between the $CaII\lambda 8498$ line
strength and the excess $1\mu m$ continuum flux (Fischer et al. 2010) 

2. The profiles of the strong $CaII\lambda 8498$ emitters (Muzerolle et al.
1998b, Edwards et al. 2010) appear fairly
symmetrical. To be sure the $Pa~16$ line in the red wing
needs to be subtracted, and the profile contains uncertain amounts of
contribution from the accretion flow and the hydrogen emission regions in the
wind flow. But if this profile characteristic, as well as the narrower
width in comparison with the $H$ and $HeI$ lines, holds up in more strong
$CaII$ emitters, it is consistent with rotational broadening. The same
broadening also produces a depression at line center, but the centrally
peaked $CaII$ emission from the accretion flow and from the hydrogen emission
regions in the wind flow may fill it up.

3. The continuum produced in the disk boundary layer, like that produced
at the accretion footpoints, increases with mass accretion rate, but unlike
the latter, also increases with decreasing distance of its location from 
the star. Thus there may only be a weak correlation between the optical/UV
and $1\mu m$ continuum excesses. Then, if the $CaII$ narrow and broad emission
arise from the accretion shocks and disk boundary layer respectively, they
will reflect a similar relation between their strengths. The 11 $CaII$ line
profiles shown in Figure 1d of Muzerolle et al. (1998b) show that the
weak broad $CaII$ emitters have small $r_{8600} s$ and distinct narrow
components, while the strong broad emitters have larger $r_{8600} s$ and
at best weak narrow components, although their appearances are rendered
less conspicuous by the presence of the strong broad emission and the
stronger underlying continuum. It does appear that the correlation
between the $CaII$ broad and narrow emission strengths is weak, but more
objects are needed for better statistics.

We are confident that the strong $CaII$ emission do not originate from the
wind region, and think that a good portion of it arises from the disk
boundary layer.

\section{Discussion}

In this section we discuss the implications of the findings described 
earlier in \S 8. They include the need of a clumpy wind, identifying the
source of ionizing photons, and determining the underlying cause of the
correlation between strong $CaII$ and $HeI$ line emission. Most of this
discussion is speculative in nature.

A clumpy wind has also been advanced by Mitskevich et al. (1993). Their 
motivation is to explain the blue absorption seen in $H\alpha$ as being
an intrinsic part of line formation in a stellar wind that accelerates
to a peak velocity and then decelerates towards zero. Our arguments for a
clumpy wind are not dependent on a specific velocity structure. One is
based on the lack of blue absorptions in $Pa\gamma$ and $HeI\lambda 5876$,
whose observed intensities are not strong enough to cover up underlying
absorptions of the continuum. So, whatever the velocity structure, the
optically thick emission regions can only screen a small fraction of the
stellar surface at each $v_{obs}$. The other is based on the need to reduce
the model intensities of all the $H$ and $HeI$ lines, 
calculated assuming a smooth flow with
physical conditions suitable to produce the line ratios, 
in order to match the observed values.

With regard to the accelerating and then decelerating wind advocated by
Mitskevich et al. (1993), we have several reservations for its general
applicability to CTTSs. First, the $Pa\beta$ and $Pa\gamma$ lines are also
quite optically thick ($\tau_{Pa\gamma}>20$, cf. Fig. 11) but, unlike 
$H\alpha$, they rarely show a blue absorption. Second, the maximum blue
velocity seen in $HeI\lambda 10830$ or $Pa\gamma$ is often $\ge 350~km~s^{-1}$
(EFHK06). If this is reached at $3R_\ast$, the locale adopted by 
Mitskevich et al. (1993), the gravitational pull of a CTTS of $1~M_\odot$
and $2R_\odot$ will only decelerate it to $240~km~s^{-1}$, not low enough
to explain most of the $H\alpha$ blue absorptions, whose maximum depths
usually occur at $v_{obs}>-100~km~s^{-1}$ (cf. Fig. 15 of BEK01). We think 
that the $H\alpha$ blue absorptions, other than the broad ones that 
originate in a radial wind, are caused by disk winds lying beyond the
hydrogen emission zone. They are fairly sharp and narrow, similar to those
produced by a disk wind scattering the stellar continuum (Kwan et al. 2007).
However, because the hydrogen emission zone is more extensive than the
stellar surface, a revised modelling is needed to ascertain the absorption
profile.

One possible mechanism for accelerating the gas and initiating a clumpy flow
is the occurrence of multiple coronal mass ejections. These ejections will
have to be much more energetic than those in the solar corona and occur at a
much higher frequency, since the wind density needs to be $\geq 5\times 10^8 
~cm^{-3}$ to produce a $\tau_{HeI\lambda 10830}$ of $\geq 1$, thereby
implicating a minimum mass loss rate of
$\sim 0.5\times 10^{-9}~M_\odot~yr^{-1}$.
The clumpy gas distribution can arise from the initial multiple ejection
centers and the shocked regions produced when the different ejectas intersect.

There can be several sources for the photons needed for helium ionization.
One, known already, is the shocked region at the footpoint of
the accretion stream. For an impact velocity of $300~km~s^{-1}$, the 
temperature in the post-shock region reaches $10^6~K$. Cooling by free free
emission will produce many photons more energetic than $24.6~eV$. A
fraction of this radiative luminosity will propagate towards the star and
be re-processed into optical and UV continuum seen as veiling, but a
fraction will propagate away from the star and may even escape absorption
by the infalling flow if the gases fill only partially the overall
accretion envelope (Fischer et al. 2008). For an accretion rate of
$3\times 10^{-9}~M_\odot~yr^{-1}$, if 1\% of the radiative luminosity
generated escape, the luminosity in photons with energies $\ge 24.6~eV$ will
be $\sim 10^{-4} L_\odot$. Ionizing photons can also originate from the
wind region, e.g., at the sites of coronal mass ejections and the filaments 
formed by intersecting ejectas, likely with comparable luminosities. As
mentioned before, however, the issue with ionizing photons is not so much
the availability of sources, but more the propagation distance from a 
given source. Both a clumpy accretion flow and a clumpy wind help to increase
the propagation length, but a definitive understanding will probably need
Monte Carlo simulation of the propagation of photons with energies both near
and far away from ionization thresholds, and of photons produced from
subsequent recombinations to the ground state, as well as consideration 
of multiple ionizations when the ejected electron has an energy exceeding the
$HI$ ionization threshold.

We have identified the disk/magnetosphere interface where accreting 
gases dissipate part of their rotational energies before falling along
stellar field lines as a source of energy for both the veiling continuum
near $1\mu m$ and the $CaII$ line emission, and there appears to be a
correlation between $r_Y$ and the strength of the $CaII$ infrared triplet
(Fischer et al. 2010). The correlation of strong $HeI$ line emission and
strong $CaII$ line emission is probably a consequence of the correlation
between $HeI$ emission and $r_Y$. BEK01 noted that when $HeI\lambda 5876$ has 
a signature indicative of a wind origin the NC is comparatively weak and
may be even absent, and suggested a scenerio in which high accretion rates
or weak stellar fields may cause the magnetosphere to be crushed
sufficiently that the disk extends almost to the star, so impact velocities
of the accreting matter will be smaller and less energy will be available
for the NC emission. At the same time a larger fraction of the rotational
energy of the accreting matter needs to be dissipated, possibly leading to
a larger $r_Y$. Also, the equatorial region of the star will be preferentially
torqued, and the resulting differential rotation with latitude may induce
stronger magnetic activities for field lines anchored at the polar region
and produce a greater mass ejection, leading to stronger $HeI$ line
emission.

The minimum coronal density we derive is $\sim 5\times 10^8~cm^{-3}$, and the 
corresponding mass loss rate is $\sim 0.5\times 10^{-9}~M_\odot~yr^{-1}$ in 
order to produce a $\tau_{HeI\lambda 10830}$ of 1. The density and mass loss
rate probably span a factor of $\ge 10$ in range. They are germane to
the investigation of the role of a stellar wind on the angular momentum
evolution of a CTTS (Matt \& Pudritz 2005) and the numerical simulation of 
magnetospheric accretion of matter in various field geometries
(Romanova et al. 2009).

We have applied our model calculations to CTTSs because of their extensive
information on line fluxes and profiles. These results on line optical depths,
line emissivity ratios, and specific line emissivities from local excitation
calculations are usable for the broad lines of other pre-main-sequence
stars because the individual star affects the calculations only through its
stellar and veiling continua. For $H$, $HeI$, and $OI$ excitations, the
effects of these continua via photoionization of and stimulated absorption
by excited states are not significant. The effects on photoionizing $CaII$
and $NaI$ are stronger. They can be gauged from comparing Figures 6 \& 7,
which show results for the locations of 4 and $2.5R_\ast$ respectively. 
The $CaII$ and
$NaI$ line optical depths in the $r=2.5R_\ast$ calculation are affected by a
smaller $2r/v$ factor in the $\tau$ expression and a larger photoionization
rate by a factor of 2.56 (cf. \S 5.1). Together the two factors affect little
the $\tau_{CaII\lambda 8498}$ contour, but produce a displacement of the
$\tau_{NaI\lambda 5892}$ contour by a factor of $\sim 2$ or less.

\section{Conclusion}

We first summarize the main results of this work and then comment on
future studies that may shed further light on the formation and origins
of the line emission in CTTSs. Our primary conclusions are:

1. $UV$ photoionization is needed to produce the $HeI\lambda 10830$
opacities in both the accretion flow and radial wind that generate the
observed red and broad blue absorptions respectively. It is also the
most probable excitation mechanism responsible for the narrow, sharp blue
absorptions and the central absorptions.

2. The strong $HeI$ and $H$ line emission originate primarily in a radial
outflow that is highly clumpy. The bulk of the wind volume is filled by
gas at a density $\sim 10^9~cm^{-3}$ and optically thick to $HeI\lambda
10830$ and $H\alpha$, but optically thin to $HeI\lambda 5876$, $Pa\gamma$,
and the $CaII$ infrared triplet. The optically thick $HeI\lambda 5876$
emission occur mostly in regions of density $\ge 10^{11}~cm^{-3}$ and
temperature $\ge 1.5\times 10^4~K$, while the optically thick $H\alpha$
and $Pa\gamma$ emission occur mostly in regions of density around 
$10^{11}~cm^{-3}$ and temperature between $8750$ and $1.25\times 10^4~K$. In 
producing the observed line flux at a given $v_{obs}$ the area covering factor 
of these emission clumps is sufficiently small to not incur significant
absorption of the stellar and veiling continua in either $HeI$ and $H$ lines.
$HeI$ emission, possibly strong, may occasionally arise from the accretion
flow close to the impact shock as a result of photoionization heating by
the $UV$ photons.

3. The strong $CaII$ line emission likely arise in both the magnetospheric
accretion flow and the disk boundary layer where the gases dissipate part
of their rotational energies before infalling along magnetic field lines.
The needed density and temperature are $\sim 10^{12}~cm^{-3}$ and
$\le 7500~K$ respectively. Weak $CaII$ line emission, on the other hand, 
can come from the clumps producing the $H$ emission in the wind.

We plan to follow up this work by presenting all the optical and $1\mu m$ 
spectral data procured simultaneously or near-simultaneously 
(Edwards et al. 2010), with an aim
to provide more comparisons with model results on the ratios and specific
fluxes of not only lines alluded to here but also others, such as
$H\beta$, $H\gamma$, $HeI\lambda 6678$, that are useful as consistency
checks. The larger sample will also convey information on the ranges of
variation in the line properties and related physical conditions. It will
also apprise of possible correlations among continuum veilings, line
ratios, and line fluxes that may elucidate the relationship between
accretion and stellar wind activities.

A future project that will be enlightening is time monitoring at optical
and $1\mu m$ spectral regions simultaneously over more than a rotational
period for several CTTSs. The simultaneous coverage of both spectral
regions is needed to include $HeI\lambda 10830$, the key indicator of
intervening kinematic structures through its absorption features, and the
many optical lines whose strengths and profiles delimit the requisite
physical conditions. The time sequence data can test the often presumption
of azimuthal symmetry for either the accretion flow or radial wind and 
its physical conditions, and provide detailed information for a realistic
modelling of both the accretion flow and radial wind geometric structures.

Further understanding of the origins and formation of spectral lines can
be gained from studying the highly ionized lines, notably $CIV\lambda 1549$,
and optical as well as $UV$ $FeII$ and $FeI$ lines. There are two potential
sites for the highly ionized lines, the stellar wind region and the part
of the accretion flow close to the impact shock. It is important to
determine their origin, through analysis of their profiles and excitations,
and check for consistency/conflict with the formation and origin of the
$HeI$ lines. $FeII$ and $FeI$ emission can be significant from both the
$H$ and $CaII$ emission regions. Their strengths and profiles, in comparison
with those of $H$ and $CaII$ lines, as well as the similarities/contrasts
between $FeII$ and $FeI$ lines, can provide additional information on the
excitation conditions in those regions.

Eventually confirmation of a clumpy stellar wind requires an understanding
of the energy generation and acceleration process. It also needs numerical
simulation of the ejectas and gas flows to check if the resulting 
density, temperature, and velocity distributions are consistent with
those inferred from the line emission strengths and profiles.

In conclusion, the frequent presence in $HeI\lambda 10830$ of absorption
features indicative of a radial flow, and our deduction of strong line
emission from this flow, if correct, indicate that a significant stellar
wind, in terms of mass loss and energy output rates, is an essential 
component of the star formation process. It is clearly in response to the
active accretion of matter onto the star. This dynamic action-reaction 
between accretion disk and star may ultimately determine the mass and
angular momentum of the emerging star.

We thank Suzan Edwards and Lynne Hillenbrand for use of several preliminary
results, and Suzan for many comments and suggestions that greatly improve
the presentation of the paper.  This work was partially supported by NASA grant NNG506GE47G issued through the Office of Space Science.

\appendix

\section{Atomic Parameters}

\subsection{HI}

We obtain the rate coefficients for collisional transitions between $HI$
levels from Anderson et al. (2002). Unfortunately they have included only
levels $n=1-5$. The $\Delta n=1$ transitions have the largest rates and we
fit their collisional rate coefficients, i.e. $C_{n+1,n}$, $1\leq n\leq 4$,
with the expression $C_{n+1,n}=C_{3,2}(n/2)^{\kappa}exp(\chi[ln(n/2)]^2)$,
where $\kappa$ and $\chi$ are the fit parameters, and extrapolate to
$5\leq n\leq 14$. To check the accuracy of the extrapolated rate coefficients,
we compare them to those obtained from the formulas given in Johnson (1972).
The differences are $\sim 30$\%, comparable to those for $n\leq 4$. The
$\Delta n=2$ collisional rate coefficients are the next strongest and,
for the same lower level, are a factor of $\sim 2.5$  or more smaller than
the $\Delta n=1$ ones, i.e. $C_{n+2,n}/C_{n+1,n}\leq 1/2.5$. They are
quite different from Johnson's values, being larger by a factor of between
$1.2$ and $1.8$ at $T=10^4~K$. We fit them, i.e. $C_{n+2,n}$, $1\leq n\leq 3$,
with the simple power-law expression $C_{n+2,n}=C_{4,2}(n/2)^\delta$,
where $\delta$ is the fit parameter, and the proviso that the extrapolated
$C_{n+2,n}$ for $4\leq n \leq 13$ is no more than a factor of $\sim 2$
larger than Johnson's corresponding rate coefficient. The $\Delta n=3$
collisional rate coefficients are a factor of $\sim 2.3$ or more smaller
than the $\Delta n=2$ ones, i.e. $C_{n+3,n}/C_{n+2,n}\leq 1/2.3$,
$1\leq n\leq 2$. They are larger than Johnson's values by a factor of
$\sim 2.2$ at $T=10^4~K$, so we again extrapolate them to higher values
of $n$ with a power-law fit, i.e. $C_{n+3,n}=C_{5,2}(n/2)^\epsilon$, and the
proviso that the extrapolated rate coefficients are no more than a factor
of $\sim 2.5$ larger than Johnson's. For level $n=6$ only, the upper
state of the $Pa\gamma$ transition, we also extrapolate the rate
coefficients of Anderson et al. (2002) to obtain $C_{6,1}$ and $C_{6,2}$. We
fit $C_{n,1}$, $2\leq n\leq 5$, with the expression $C_{n,1}=C_{2,1}[1/(n-1)]^
\zeta$, and $C_{n,2}$, $3\leq n \leq 5$, with the expression
$C_{n,2}=C_{3,2}[1/(n-2)]^\eta$, and extrapolate to $n=6$. It turns out that
in the regions responsible for the observed emission the electron densities
are quite high, $\geq 3\times 10^{10}~cm^{-3}$, so collisions dominate
the population exchange among the higher energy levels. The level
population then depend mostly on the scaling of the collisional rate
coefficients with $n$, which we hope to capture with our fitting procedure.

We calculate the rate coefficients for collisional ionization from the
formulas given in Johnson (1972) and determine the three-body recombination
coefficients by detailed balance. In addition to $UV$ photoionization of
the ground state, we include
photoionizations of the $n=2-4$ levels by the stellar and veiling continua.
The photoionization cross-sections are gathered from Allen (1973) and the
radiative recombination coefficients to the 15 levels from Seaton (1959). We
also include radiative absorptions of the stellar and veiling continua
at all permitted line transitions except the Lyman ones. In principle,
the probability of stimulated absorption is the same as the escape
probability calculated with the velocity gradient along the trajectory of the
incident continuum photon. It is therefore different from the escape
probability of a photon emitted by spontaneous emission, in general. We will
ignore this distinction here, since it does not affect the order of the lines
in their opacity magnitudes.

\subsection{HeI}

We adopt the helium atomic parameters and many of the collision strengths
from the Chianti data base (Young et al. 2003) which lists energy levels,
Einstein A rates, and collision strengths for transitions involving one
of the four lowest energy states. For the rest of the collision strengths,
we obtain them from Sawey and Berrington (1993), the basis of Chianti's
compilation. We determine the collisional ionization rate coefficient of
the ground state from the cross-sections measured by Montague,
Harrison, \& Smith (1984), and
that of the $2s~^3S$ level from the theoretical calculation of cross-sections
by Taylor, Kingston, \& Bell (1979), and gather the rest from the
compilation of Benjamin, Skillman, \& Smits (1999). We obtain the $2s~^3S$
and $2s~^1S$ photoionization cross-sections from Fernley, Taylor, \& Seaton
(1987), and the radiative recombination rate coefficients from Benjamin et al.
(1999).

The absence of collision strengths involving $n\geq 5$ levels is the reason
why our helium model atom has only $19$ levels. There is the concern that,
like the case of hydrogen, collisional excitation from the $19$ levels to
higher ones will likely lead to ionization, so limiting helium to $19$
levels will underestimate the helium ionization rate and overestimate the
helium level population. This problem may be somewhat less severe for
helium because the separations in energy of the angular momentum states
with the same energy quantum number $n$ lead to more radiative decay
channels, and the ionization of hydrogen by $\lambda 584$ photons provides
a steady drainage of the $2p~^1P$ population. Both features tend to dampen
the rapid build up of population into levels of higher $n$ as the
collisional excitations and line opacities increase with increasing density.
From the results of the calculations we find that under the physical
conditions responsible for the $\lambda 5876$ emission the $8~n=4$ levels
account for less than 20\% of the total helium ionization rate. Hopefully
the errors in helium line fluxes and line ratios are within 20\%.

\subsection{OI, CaII, NaI}

We obtain most of the Einstein A rates of the pertinent $OI$ transitions from
Hibbert et al. (1991), Bi$\acute{e}$mont \& Zeippen (1992), and
Carlsson \& Judge (1993). The $\lambda 11287$ Einstein A rate of
$1.1\times 10^7~s^{-1}$ is deduced from the experimental work of
Christensen \& Cunningham (1978). The $3s~^3S$ level has also a radiative
decay route to the $2p^4~^1D$ state that is not indicated in Figure 3.
Its Einstein A rate of $1.83\times 10^3~s^{-1}$ (Bi$\acute{e}$mont \&
Zeippen 1992) is included as an additional decay rate of $3s~^3S$ to the
ground level. We gather the collisional rate coefficients for the
various transitions from Barklem (2007), the radiative recombination
rate coefficients from P$\acute{e}$quignot, Petitjean, \& Boisson (1991),
and the $OI-HII$ and $OII-HI$ charge-exchange reaction rates from Field \&
Steigman (1971).

We obtain the Einstein A rates and collisional rate coefficients for all
relevant $CaII$ transitions from Burgess, Chidichimo, \& Tully (1995), the
photoionization cross-sections from Verner et al. (1996), and Shine \&
Linsky (1974), the collisional ionization rate coefficients from Arnaud \&
Rothenflug (1985), and Shine \& Linsky (1974), the total radiative
recombination rate coefficient from Shull \& Van Steenberg (1982), and
the direct radiative recombination rate coefficients to the three
levels from their photoionization cross-sections through detailed balance.

For the relevant atomic parameters of $NaI$, we obtain the
Einstein A rate from Sansonetti (2008), the $3s~^2S$ and $3p~^2P$
photoionization cross-sections from the experimental work of Hudson \&
Carter (1967) and Rothe (1969) respectively, the $3s~^2S$ collisional
ionization rate coefficient from Arnaud \& Rothenflug (1985), the
$3s~^2S\rightarrow 3p~^2P$ collisional excitation rate coefficient from
Clark et al. (1982), and the radiative recombination rate coefficient from
Verner \& Ferland (1996).

\clearpage
\begin{deluxetable}{ccc}
\tablecaption{Line Types and Designated Temperatures}
\tablehead{
\colhead{ Line Type } & 
\colhead{ $T(10^4~K)$ } &
\colhead{ Figures } 
}
\startdata
\object{short dash-long dash} & 0.5 & 5,8,9,10,11,12,15,16,$~~$,$~~~$,
19,20,21,22,23,24 \\
\object{short dash} & 0.75 & 5,8,9,10,11,12,15,16,17,18,19,20,21,22,23,24 \\
\object{long dash} & 0.875 & 5,8,9,$~~$,11,12,15,16,17,18,19,20,21,22,23,
24 \\
\object{solid} & 1.0 & 5,8,9,10,11,12,15,16,17,18,19,20,21,22,23,24 \\
\object{dotted} & 1.5 & 5,8,9,10,11,12,15,16,17,18,19,20,21,22,23,24 \\
\object{dot-short dash} & 2.0 & 5,8,9,10,11,12,15,16,17,18,19,20,21,22,23,24 \\
\object{dot-long dash} & 1.25 & 5,8,9,10,$~~~$,$~~~$,$~~~$,$~~~$,
$~~~$,$~~~$,$~~~$,$~~~$,$~~~$,22,23,$~~~$ \\
\object{$~~$} & 3.0 & $~~$,$~~$,$~~$,$~~$,11,$~~~$,$~~~$,
$~~$,$~~~$,$~~~$,19,20,21,$~~~$,$~~~$,$~~~$ \\
\object{$~~$} & 0.625 & $~~$,$~~$,$~~$,$~~$,$~~$,$~~~$,15,16,17,18, 
$~~~$,$~~~$,$~~~$,$~~$,$~~~$,24 \\
\enddata
\end{deluxetable}

\clearpage
\begin{figure*}
\plotone{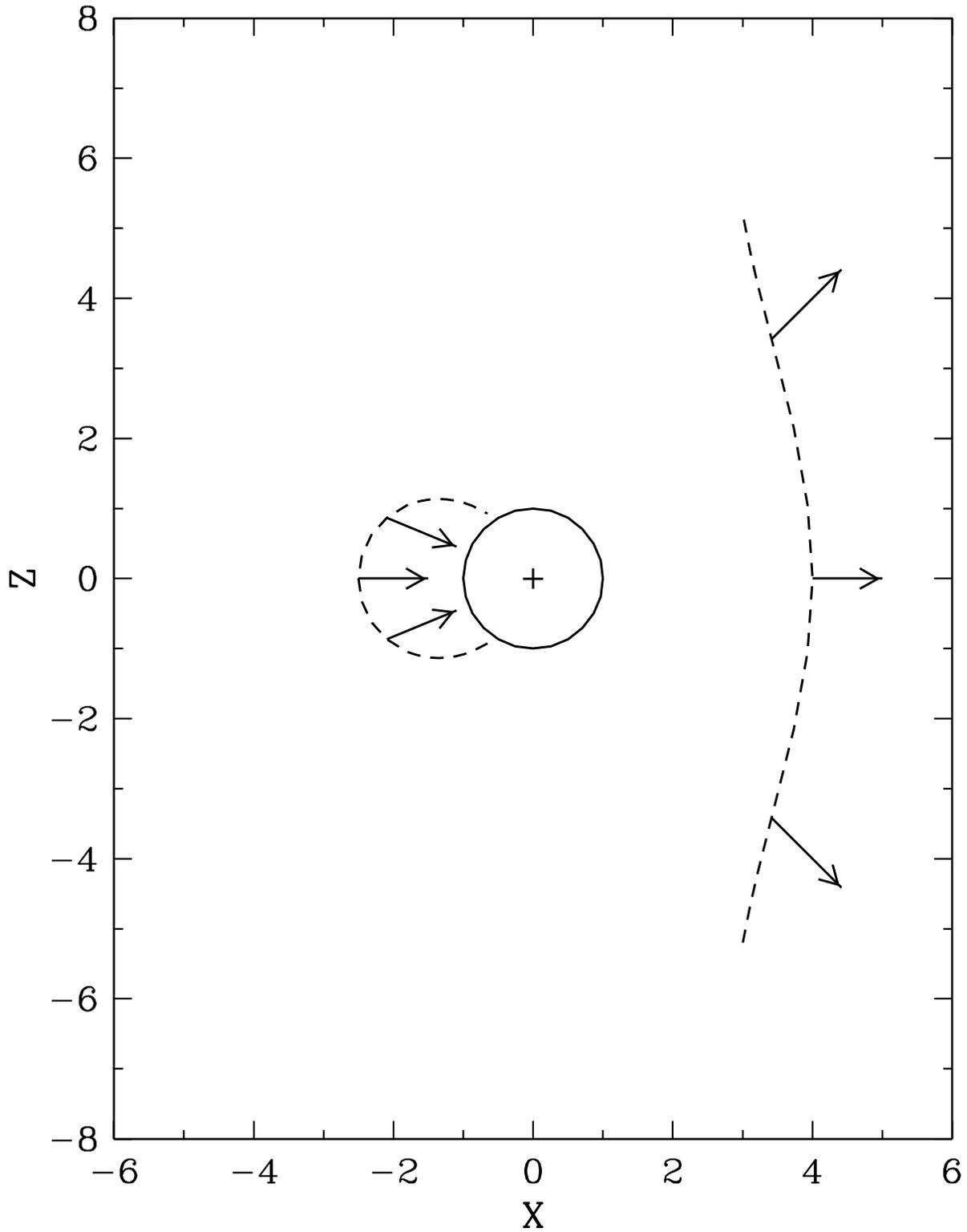}
\caption{Contour of $v_{obs}=-150~km~s^{-1}$ for a spherical wind (right
side) reaching $150~km~s^{-1}$ at $4R_\ast$
and radial infall (left side) reaching $150~km~s^{-1}$ at $2.5R_\ast$
, for a viewer at $x\rightarrow \infty$. The contour gives rise to a 
projected area of $27\pi R_\ast^2$ and $1.286\pi R_\ast^2$ for the wind
and infall flow respectively.}
\end{figure*}

\clearpage
\begin{figure*}
\plotone{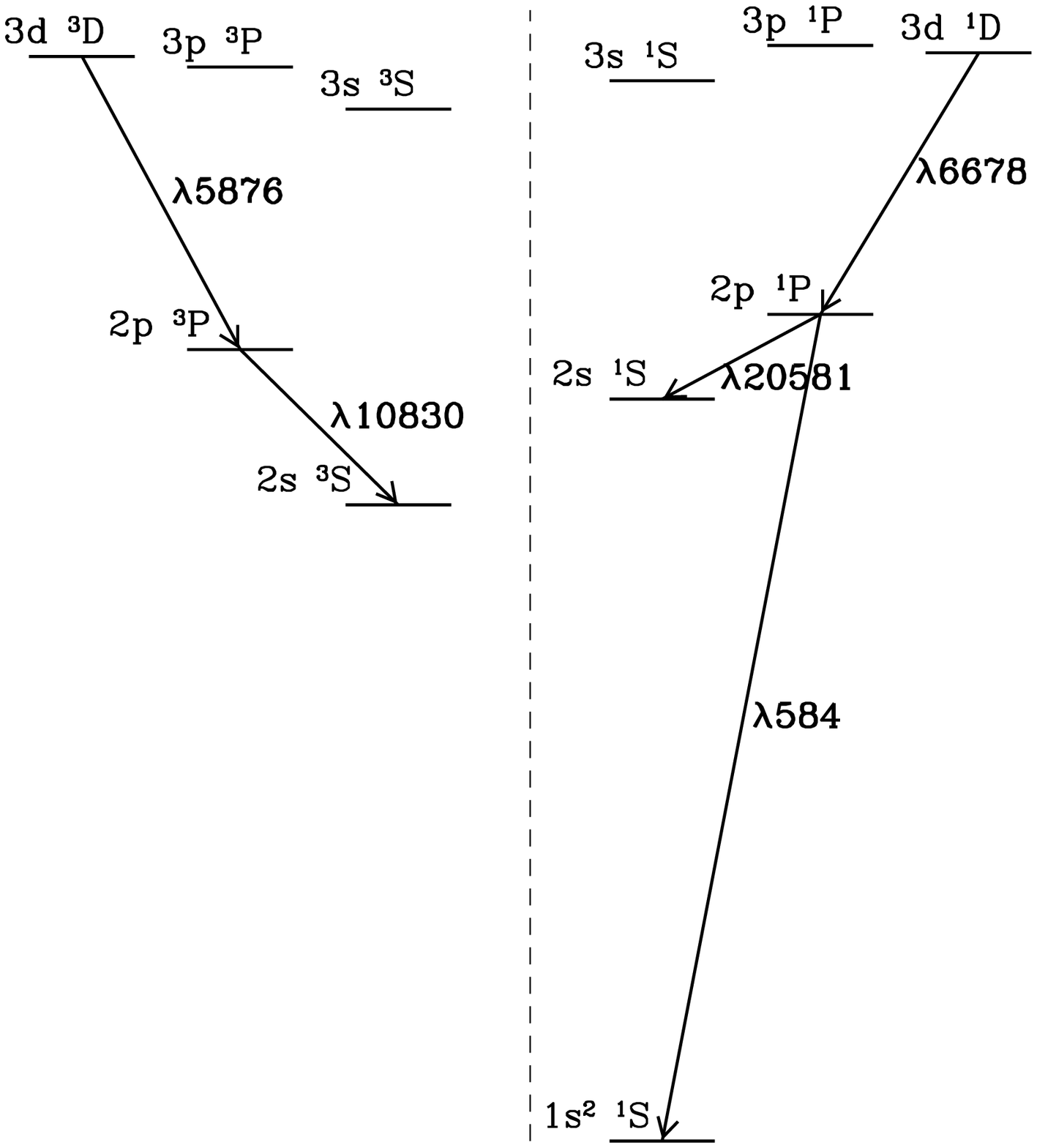}
\caption{Energy level diagram of $HeI$, not drawn to scale.}
\end{figure*}

\clearpage
\begin{figure*}
\plotone{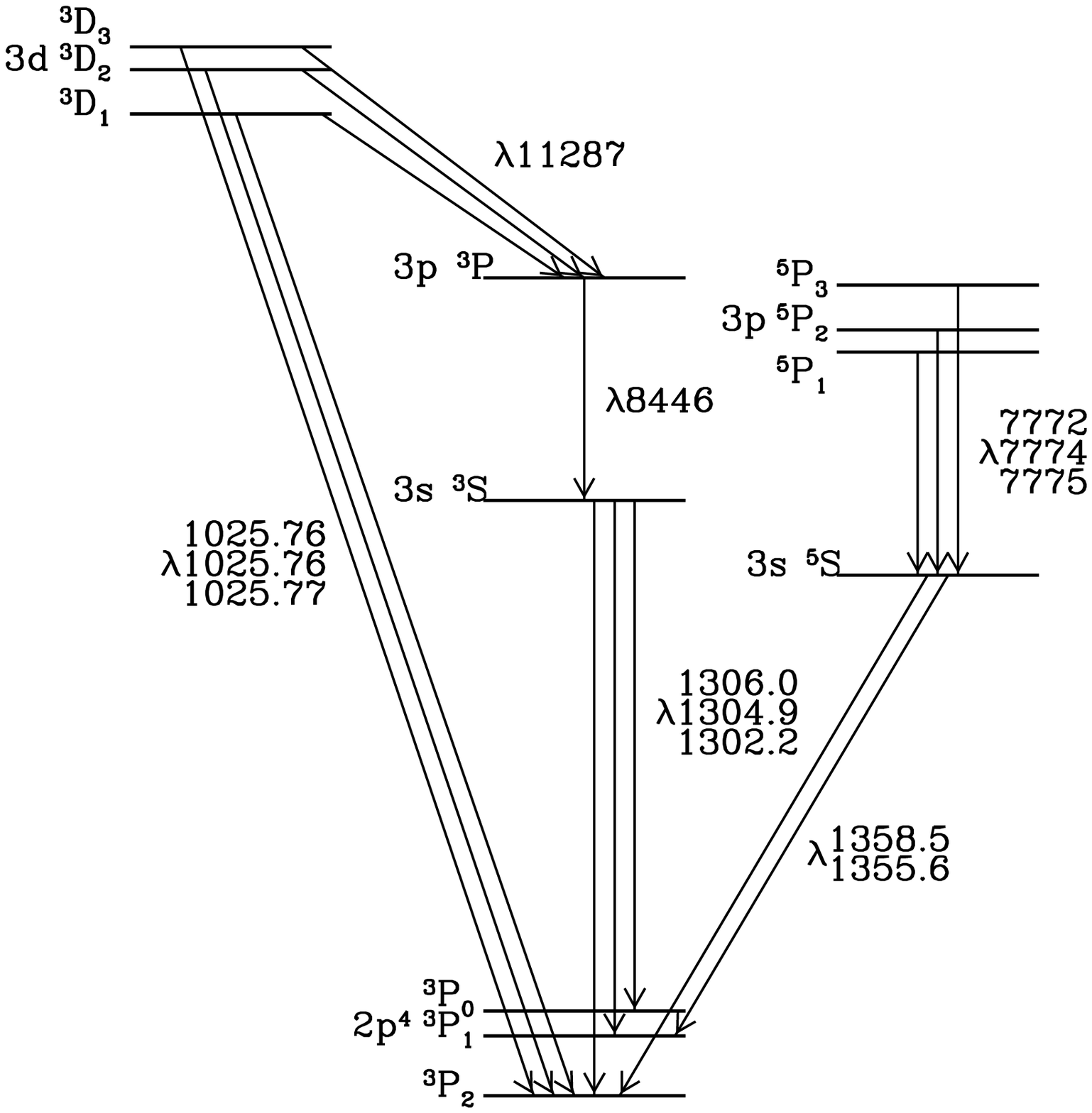}
\caption{Energy level diagram of $OI$, not drawn to scale.}
\end{figure*}

\clearpage
\begin{figure*}
\plotone{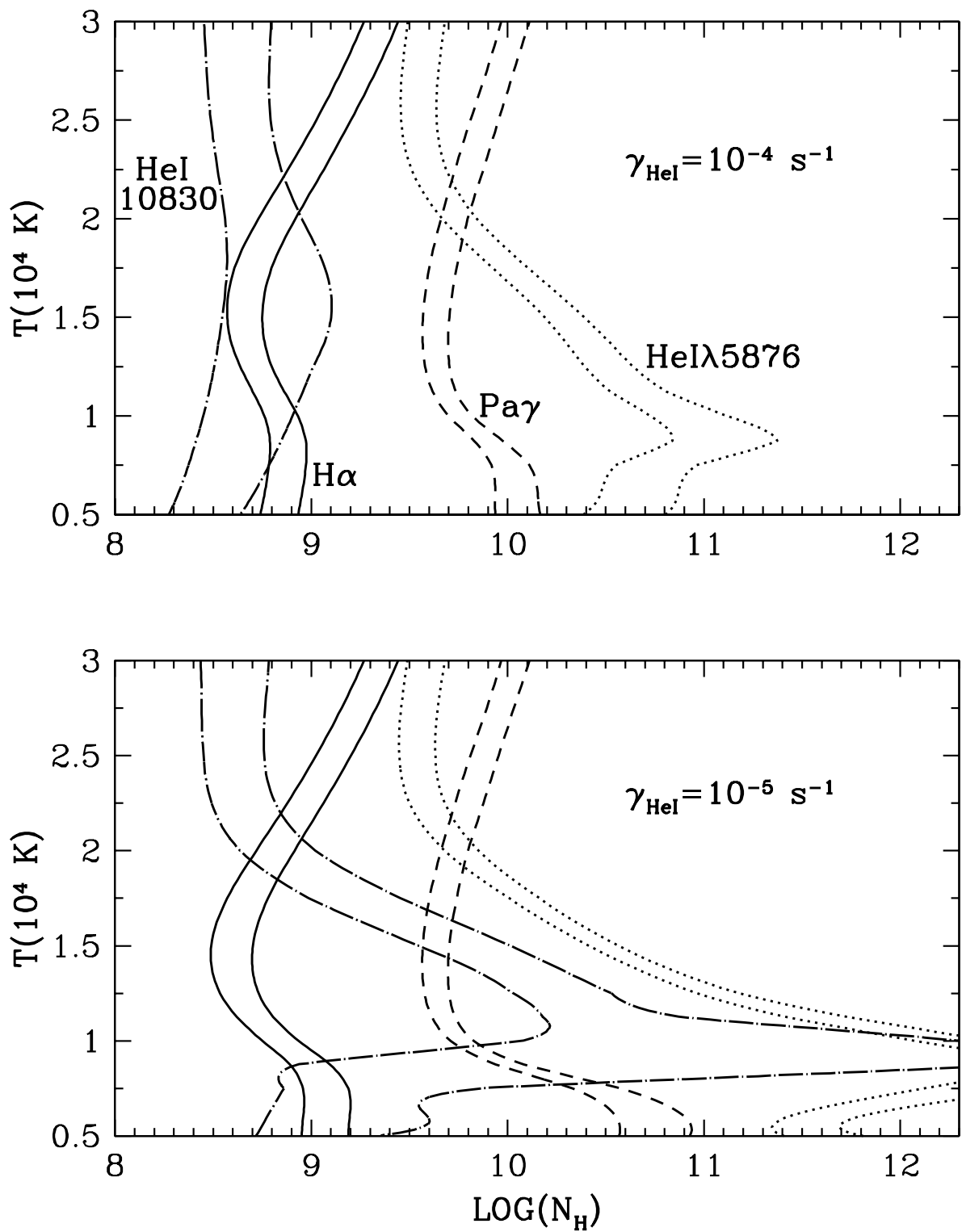}
\caption{Contour plots of line optical depths 
for $r=4R_\ast$ and $\gamma_{HeI}
=10^{-4}$ (top panel) and $10^{-5}~s^{-1}$ (bottom panel). $HeI\lambda
10830$ and $H\alpha$ have $\tau=1$ and 3.16 contours, and $HeI\lambda 5876$
and $Pa\gamma$ have $\tau=0.1$ and 0.316 contours.}
\end{figure*}

\clearpage
\begin{figure*}
\plotone{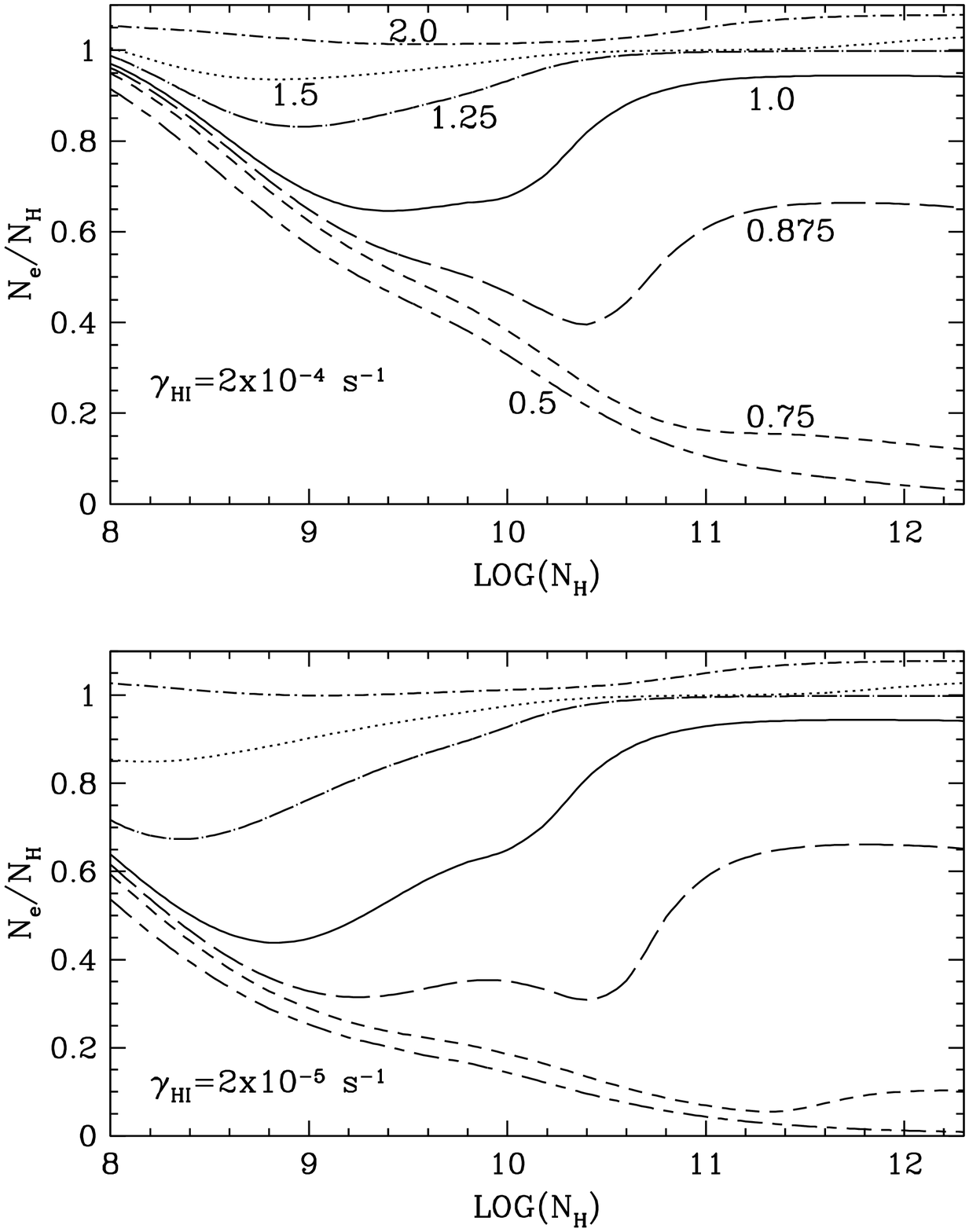}
\caption{Dependence of electron fraction $N_e/N_H$ on $N_H$ for $r=4R_\ast$,
$\gamma_{HI}=2\times 10^{-4}$ (top panel) and $2\times 10^{-5}~s^{-1}$
(bottom panel), and seven temperatures (in unit of $10^4~K$).}
\end{figure*}

\clearpage
\begin{figure*}
\plotone{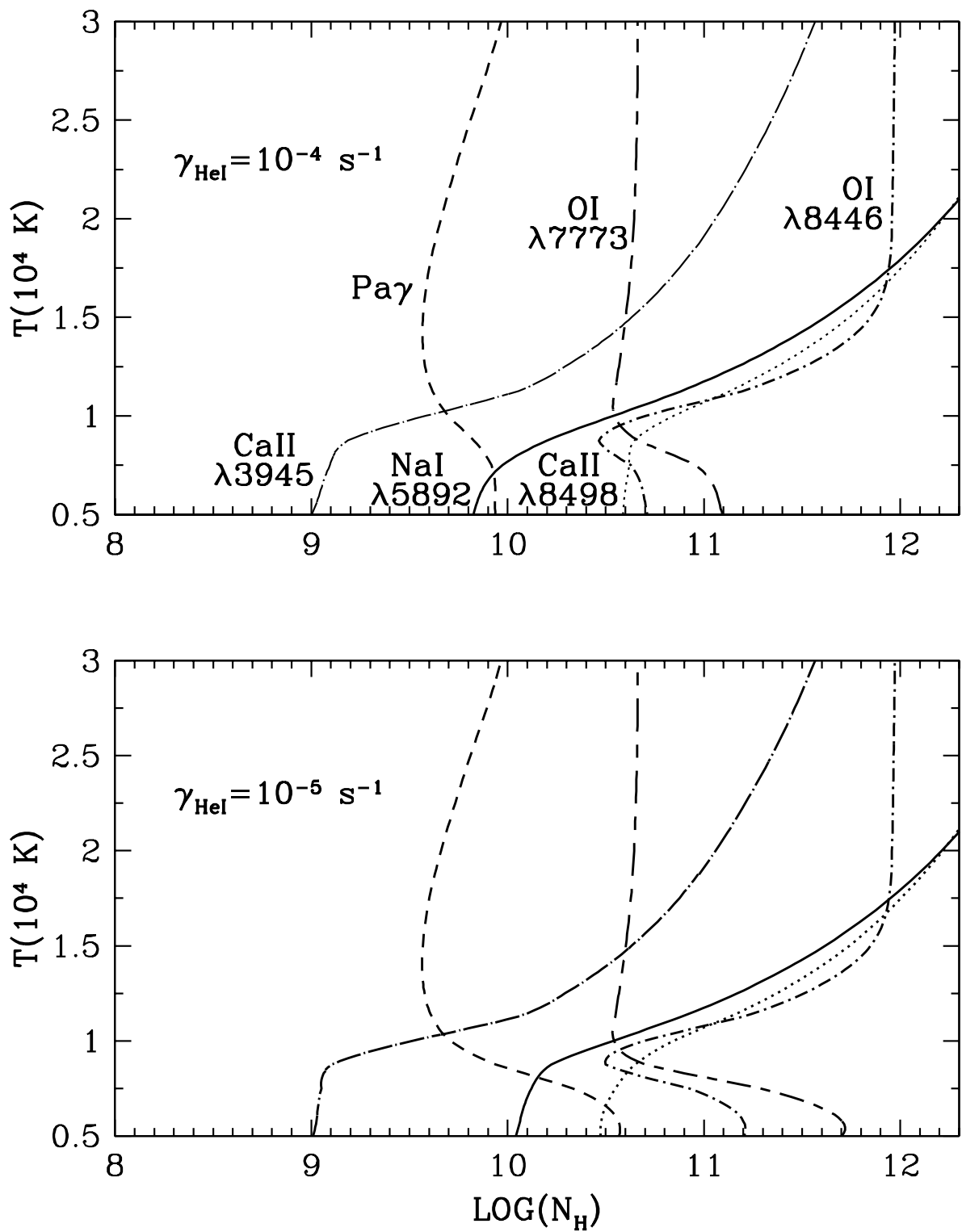}
\caption{Contour plots of line optical depths for $r=4R_\ast$ and
$\gamma_{HeI}=10^{-4}$ (top panel) and $10^{-5}~s^{-1}$ (bottom panel).
$CaII\lambda\lambda 3945, 8498$, and $NaI\lambda 5892$ have $\tau=1$
contours, and $Pa\gamma$, $OI\lambda\lambda 8446, 7773$ have $\tau=0.1$
contours.}
\end{figure*}

\clearpage
\begin{figure*}
\plotone{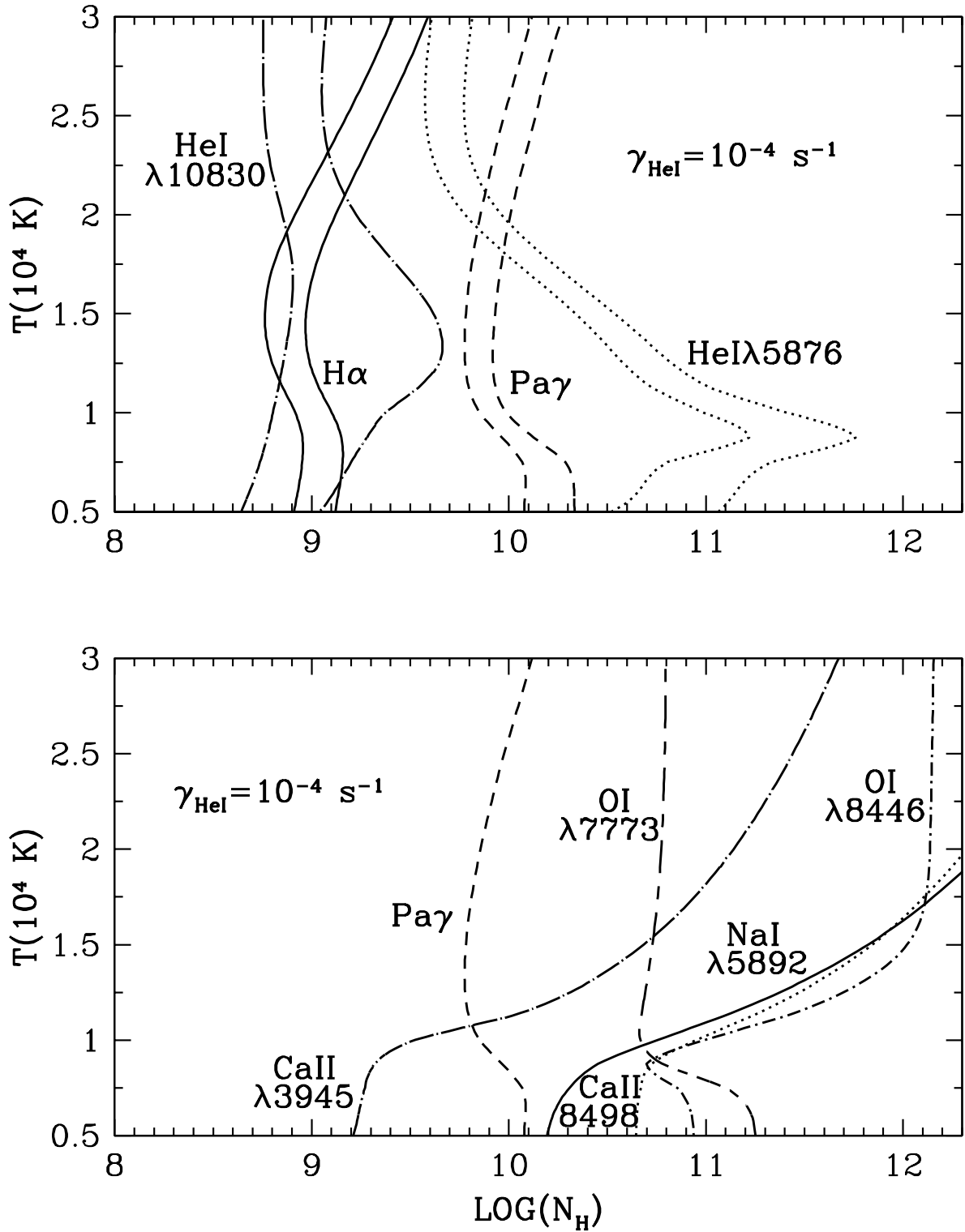}
\caption{Top panel same as top panel of Fig. 4 and
bottom panel same as top panel of Fig. 6 but for $r=2.5R_\ast$.}
\end{figure*}

\clearpage
\begin{figure*}
\plotone{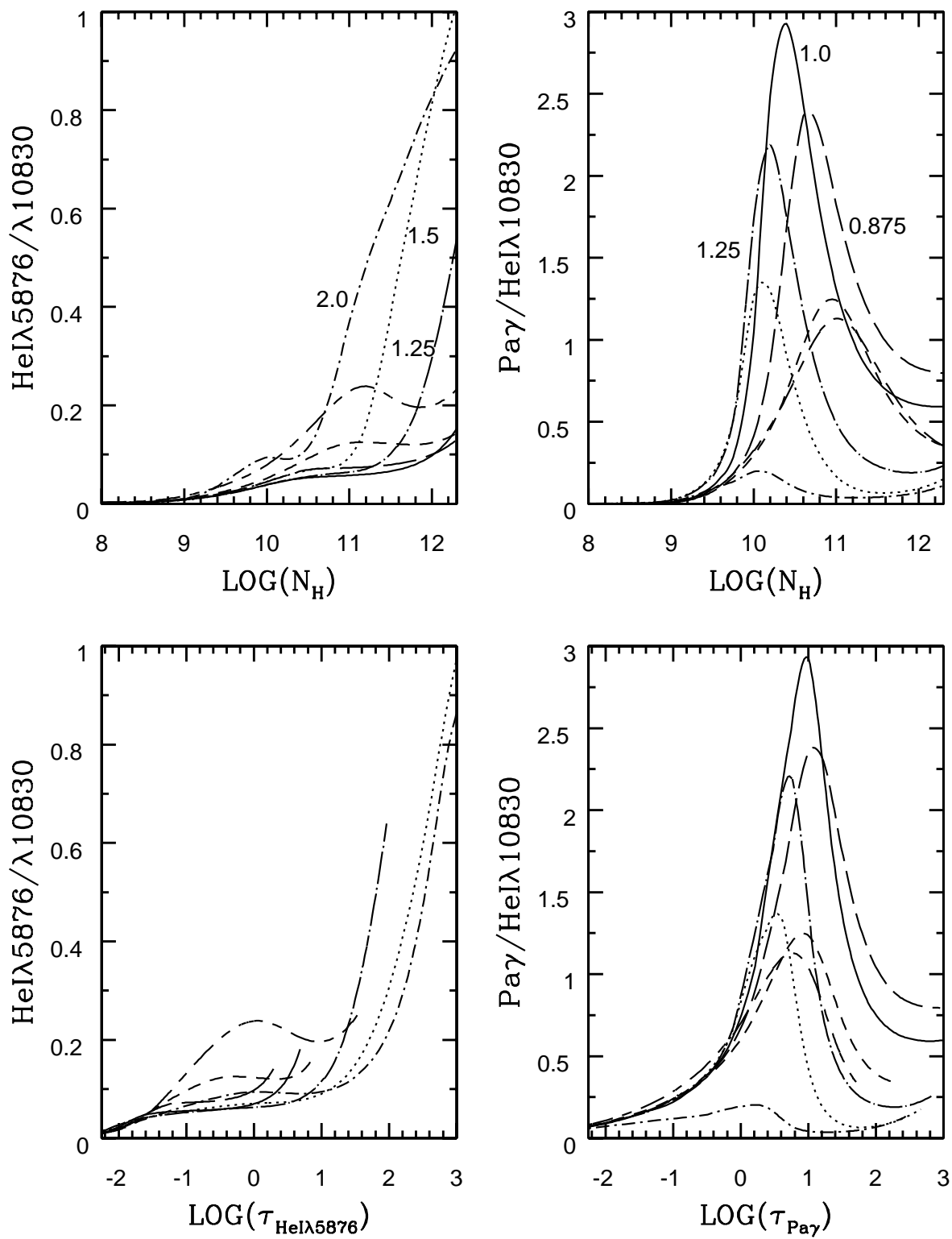}
\caption{Dependences of $HeI\lambda 5876/\lambda 10830$ and $Pa\gamma/
HeI\lambda 10830$ on density and line optical depth
for $r=4R_\ast$, $\gamma_{HeI}=10^{-4}~s^{-1}$, and
various temperatures. See Table 1 for the designations between line types
and temperatures.}
\end{figure*}

\clearpage
\begin{figure*}
\plotone{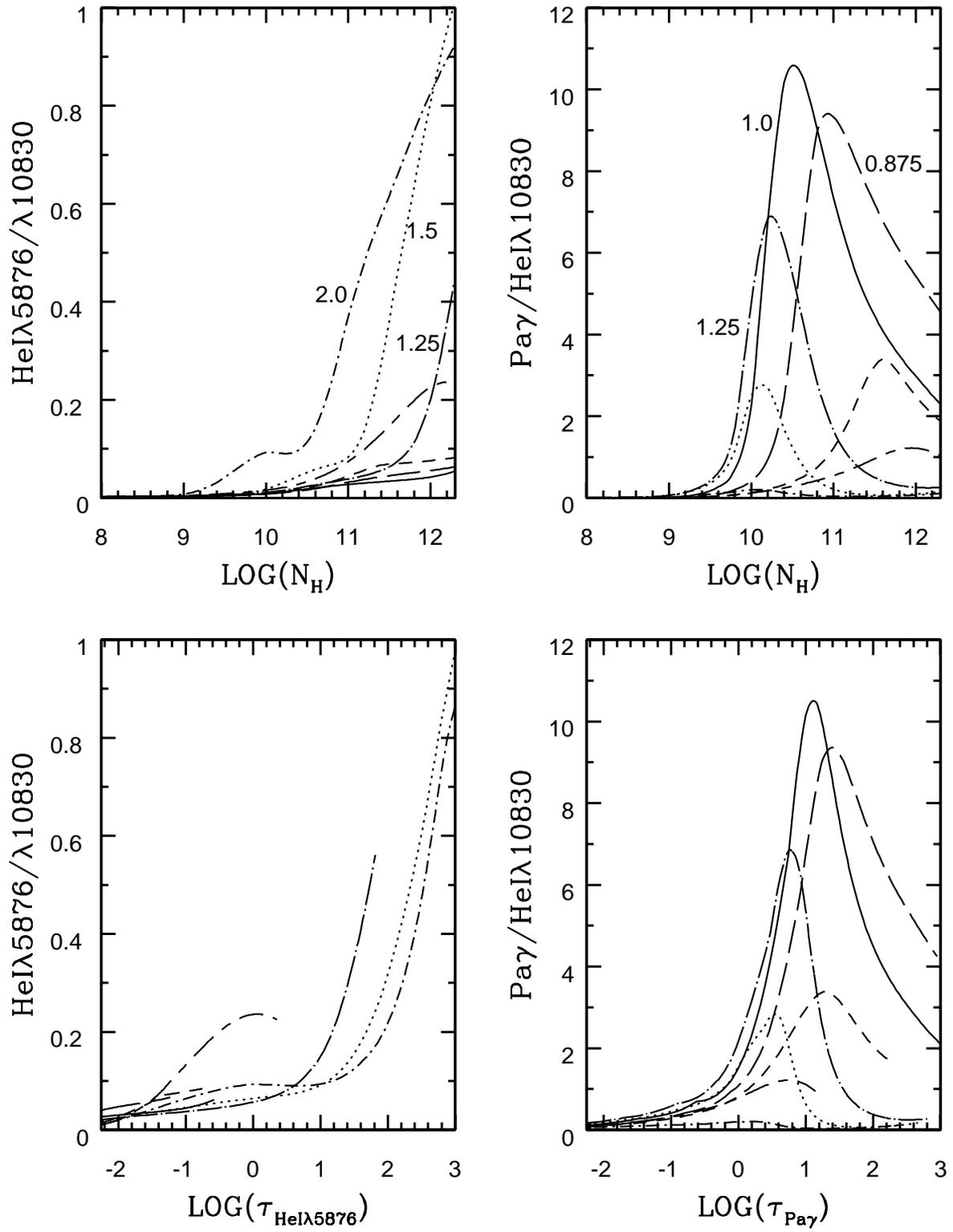}
\caption{Same as Fig. 8 but for $\gamma_{HeI}=10^{-5}~s^{-1}$.}
\end{figure*}

\clearpage
\begin{figure*}
\plotone{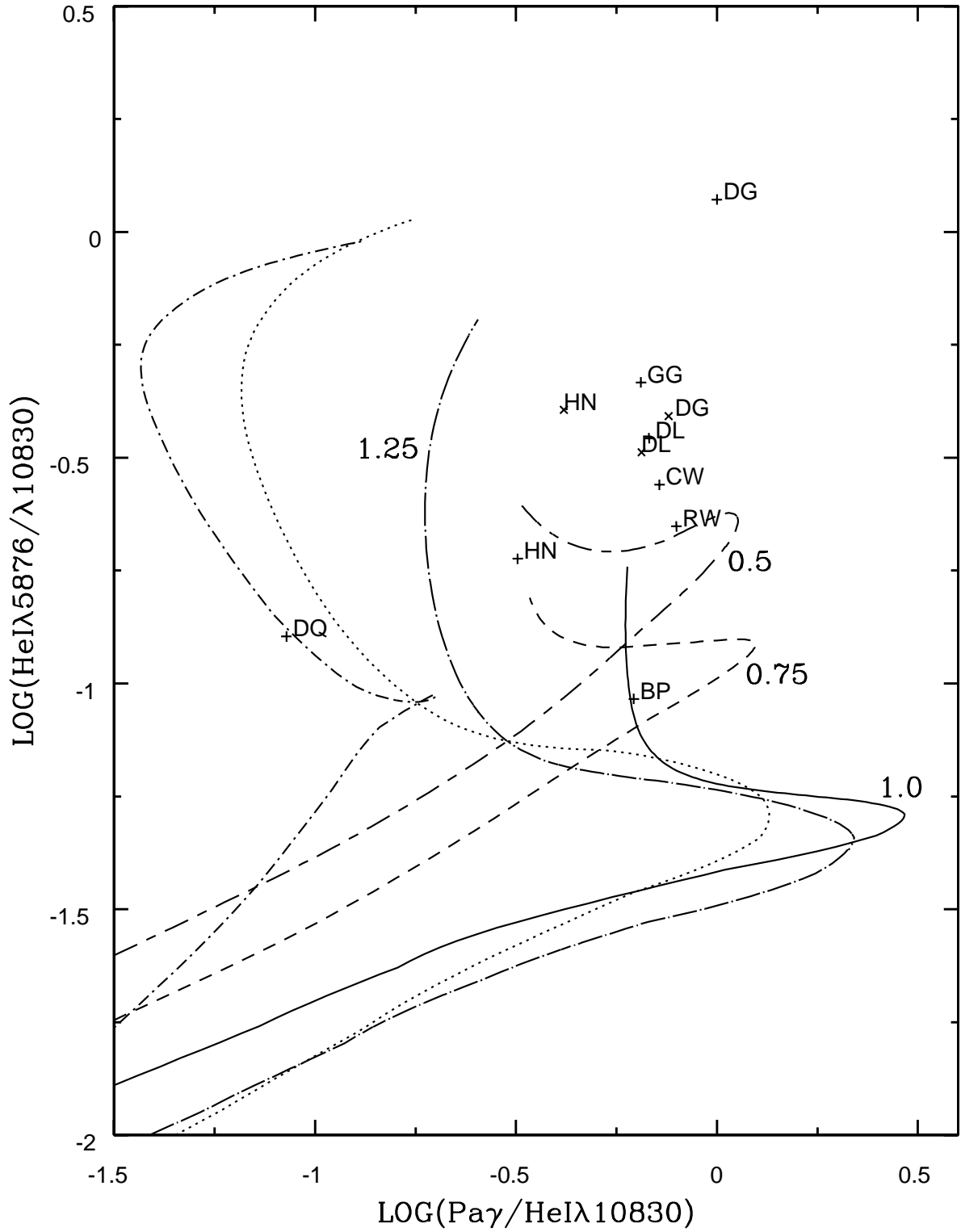}
\caption{Plot of log($HeI\lambda 5876/\lambda 10830$)
versus log($Pa\gamma/HeI\lambda 10830$)
for $r=4R_\ast$,
$\gamma_{HeI}=10^{-4}~s^{-1}$, and various temperatures(cf. Table 1).
Data points marked by +s are determined from BEK01 and EFHK06. Those marked
by xs are from Edwards et al. (2010).}
\end{figure*}

\clearpage
\begin{figure*}
\plotone{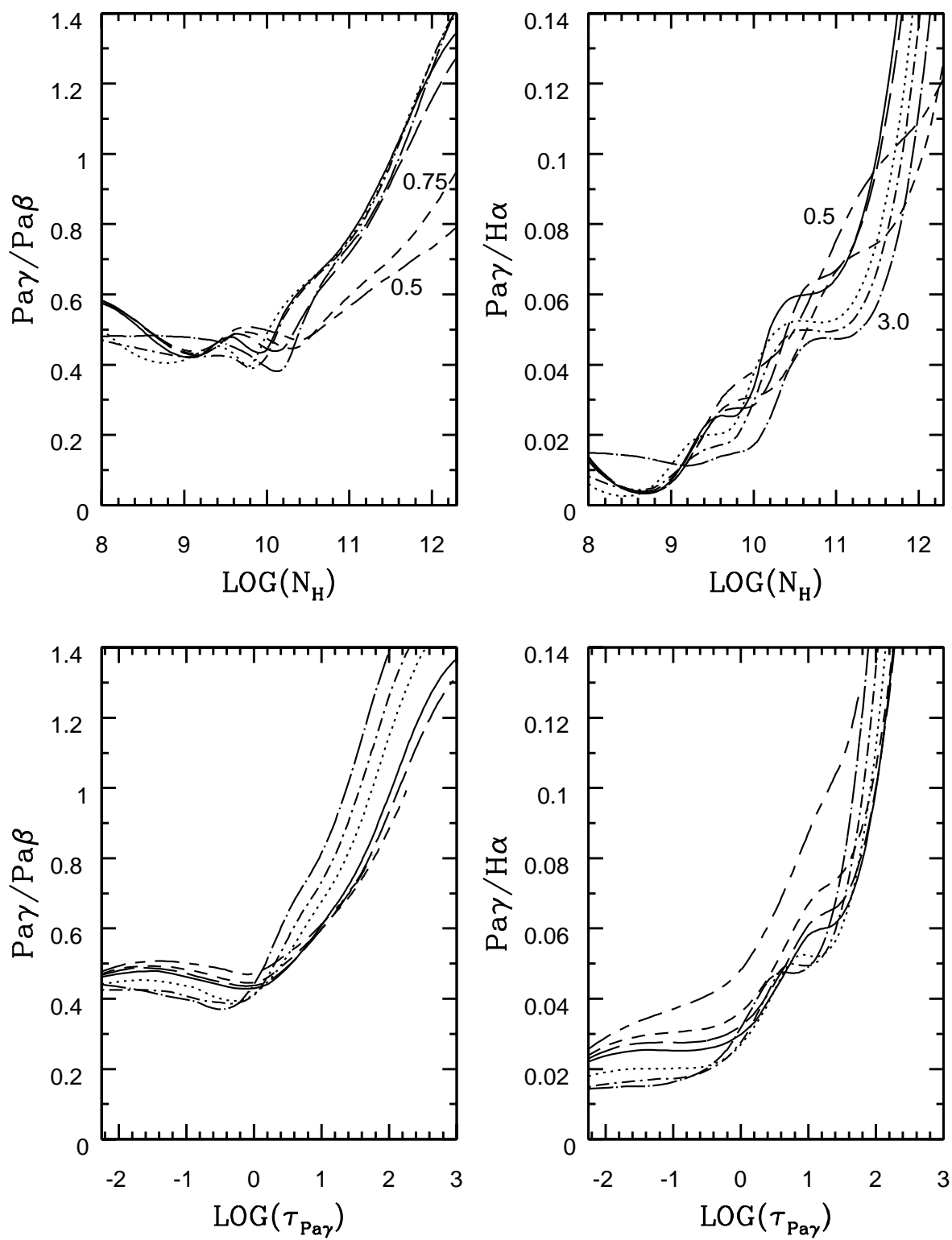}
\caption{Dependences of $Pa\gamma/Pa\beta$ and $Pa\gamma/H\alpha$ on density
and line optical depth for $r=4R_\ast$, $\gamma_{HI}=2\times 10^{-4}~s^{-1}$,
and various temperatures(cf. Table 1).}
\end{figure*}

\clearpage
\begin{figure*}
\plotone{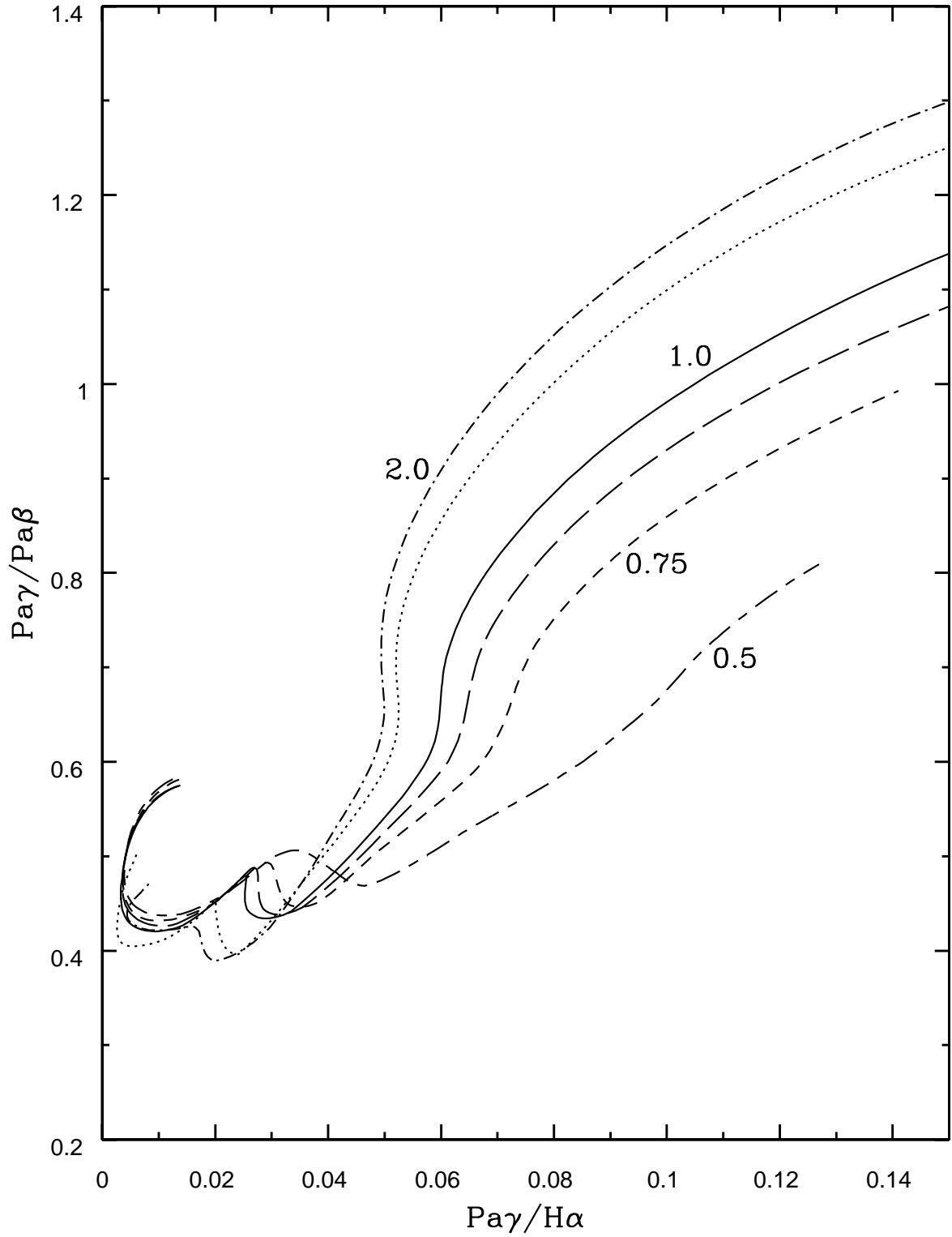}
\caption{Plot of $Pa\gamma/Pa\beta$ versus $Pa\gamma/H\alpha$ 
for $r=4R_\ast$, $\gamma_{HI}=2\times
10^{-4}~s^{-1}$, and various temperatures(cf. Table 1).} 
\end{figure*}

\clearpage
\begin{figure*}
\plotone{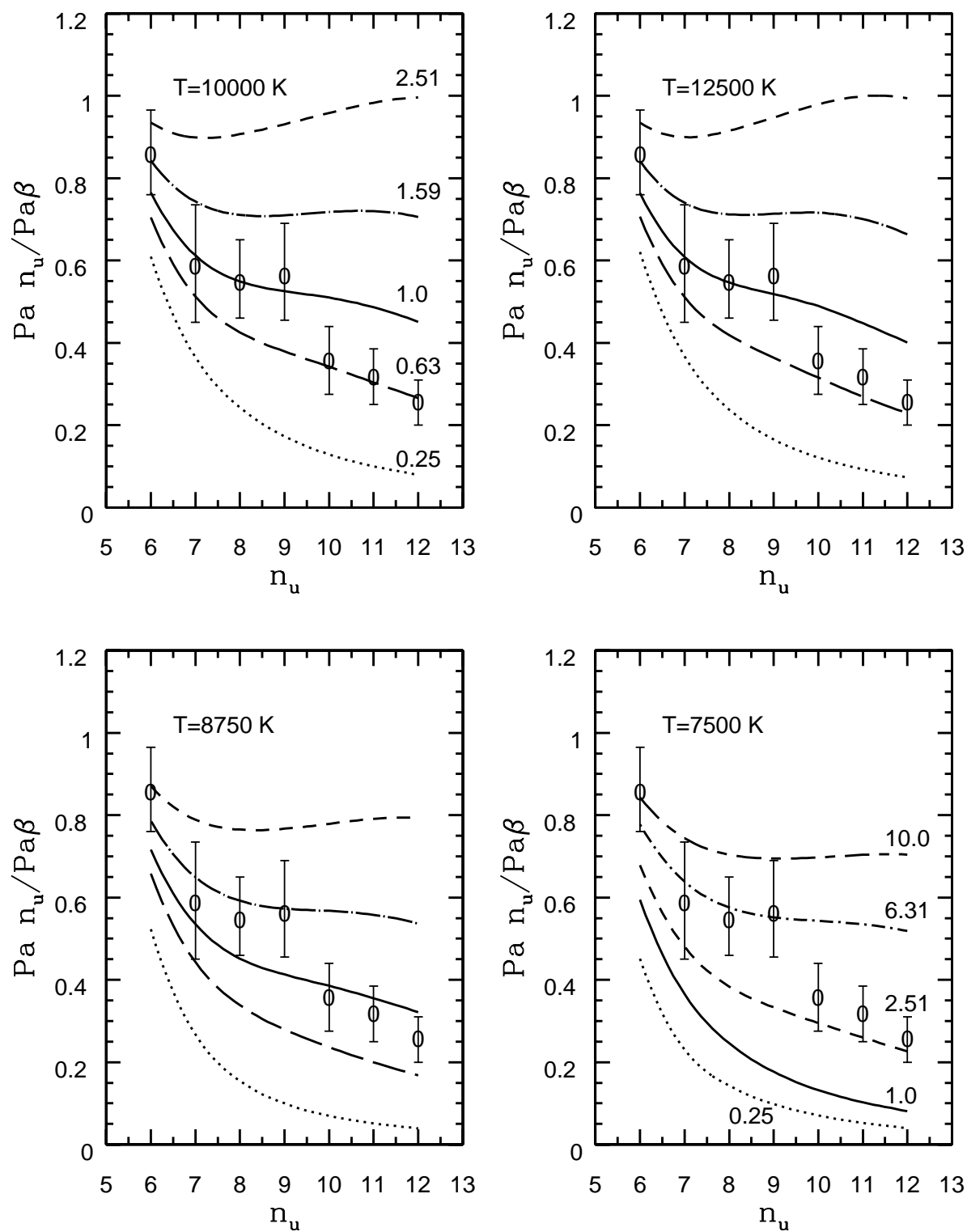}
\caption{Plot of $Pa~n_u/Pa\beta$ versus $n_u$ for four temperatures,
$r=4R_\ast$, $\gamma_{HI}=2\times 10^{-4}~s^{-1}$, and various values of
$N_H$ (in unit of $10^{11}~cm^{-3}$).
Data marked by open circles
are from Bary et al. (2008).}
\end{figure*}

\clearpage
\begin{figure*}
\plotone{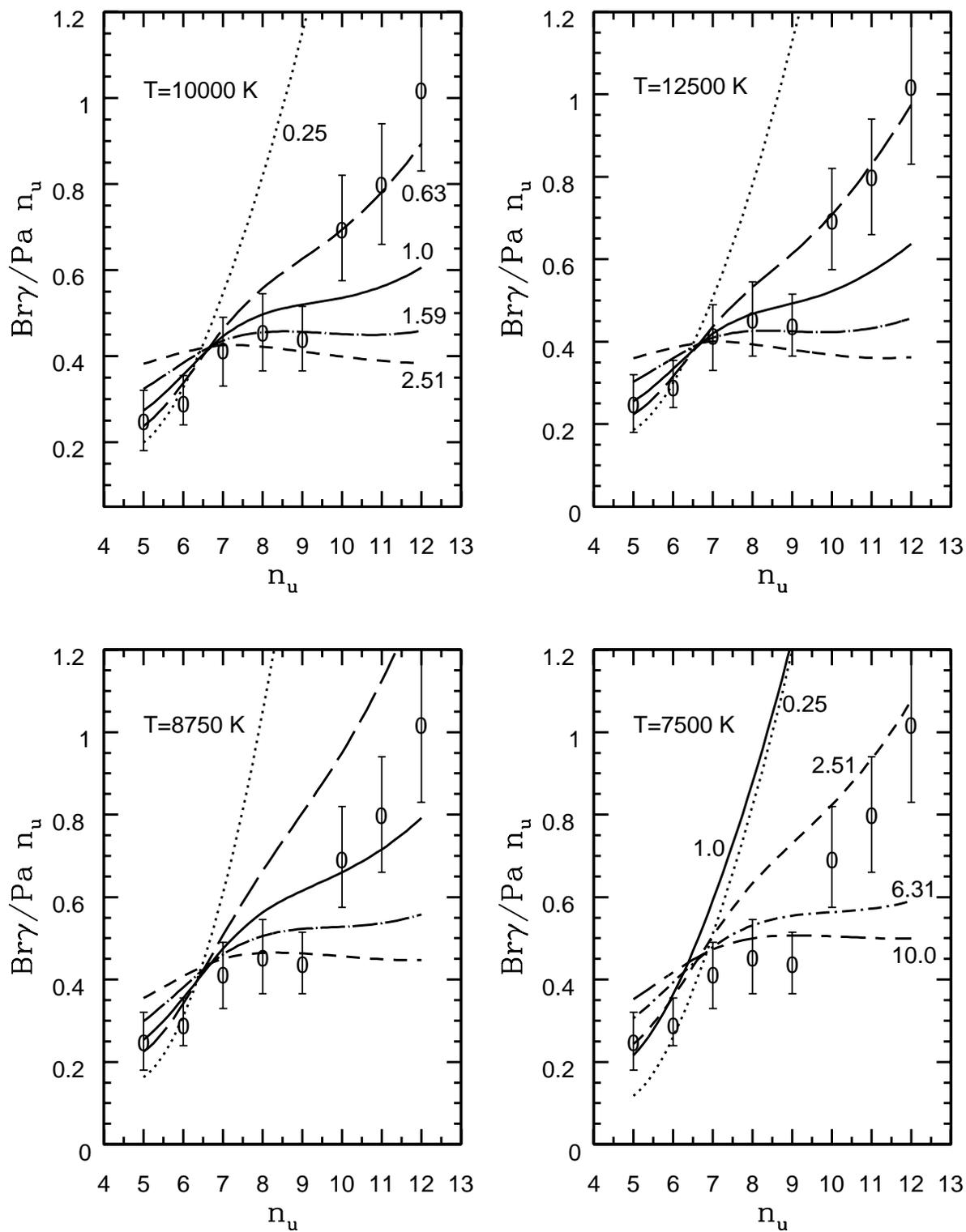}
\caption{Same as Fig. 13 but for $Br\gamma/Pa~n_u$ .}
\end{figure*}

\clearpage
\begin{figure*}
\plotone{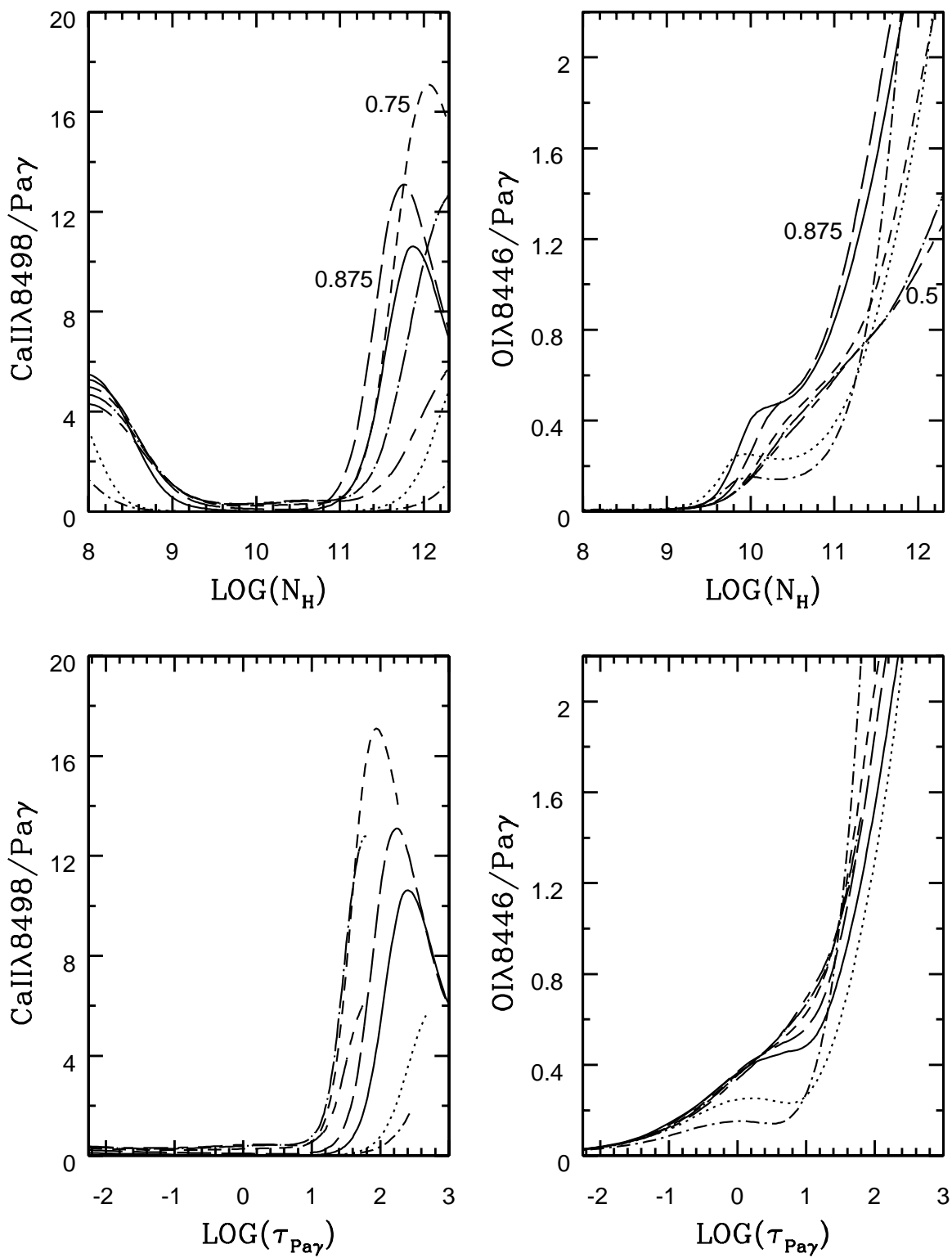}
\caption{Dependences of $CaII\lambda 8498/Pa\gamma$ and $OI\lambda 8446/
Pa\gamma$ on density and line optical depth for $r=4R_\ast$, $\gamma_{HI}=
2\times 10^{-4}~s^{-1}$, and various temperatures(cf. Table 1). The
large $CaII\lambda 8498/Pa\gamma$ values at $N_H<10^9~cm^{-3}$ are
not significant because they occur at $\tau_{Pa\gamma}\ll 1$.}
\end{figure*}

\clearpage
\begin{figure*}
\plotone{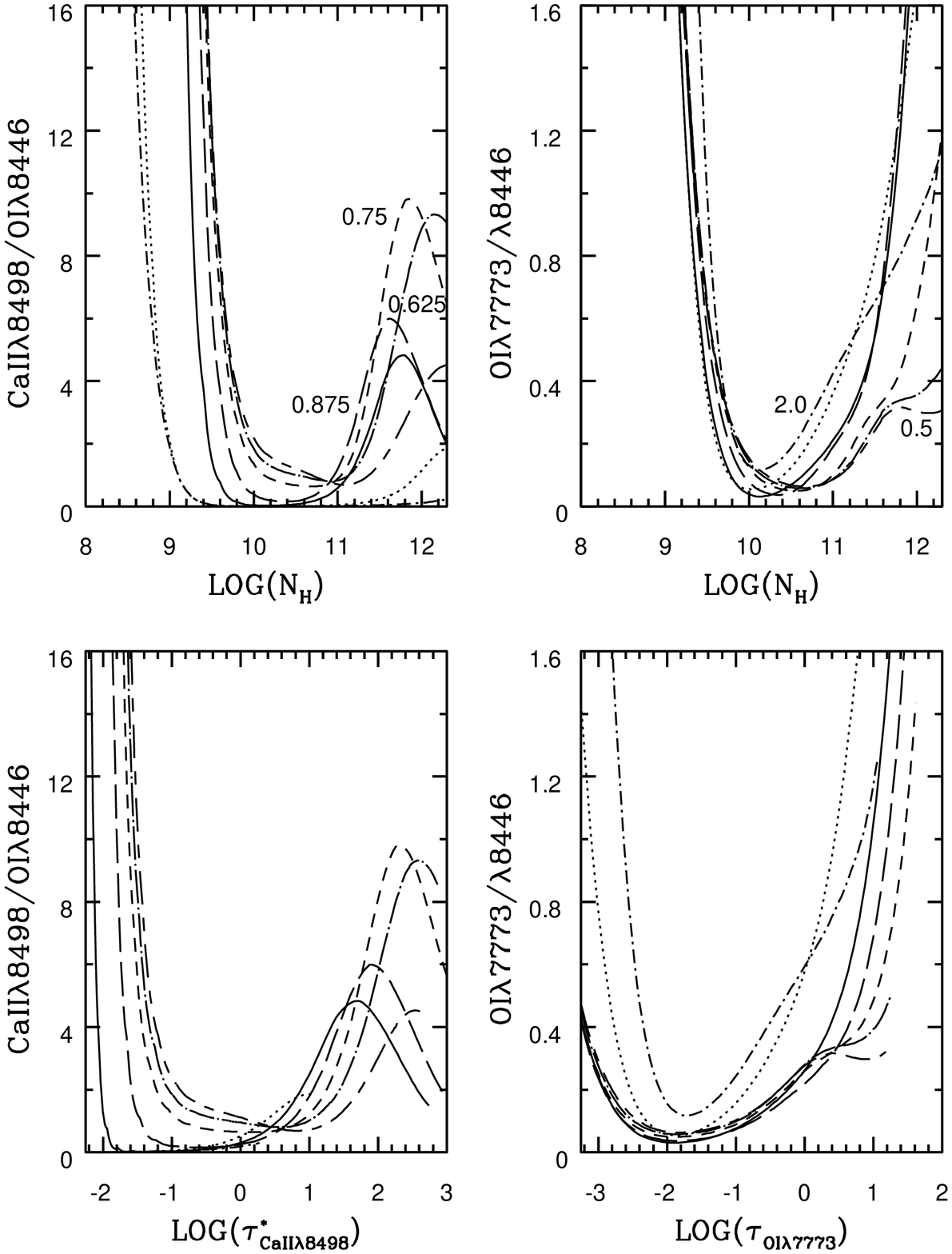}
\caption{Dependences of $CaII\lambda 8498/OI\lambda 8446$ and $OI\lambda 7773/
\lambda 8446$ on density and line optical depth for $r=4R_\ast$, $\gamma_{HI}=
2\times 10^{-4}~s^{-1}$, and various temperatures(cf. Table 1).
The large $CaII\lambda 8498/OI\lambda 8446$ values at $N_H<10^{10}~cm^{-3}$
are not significant because they occur at $\tau^*_{CaII\lambda 8498}\ll 1$.}
\end{figure*}

\clearpage
\begin{figure*}
\plotone{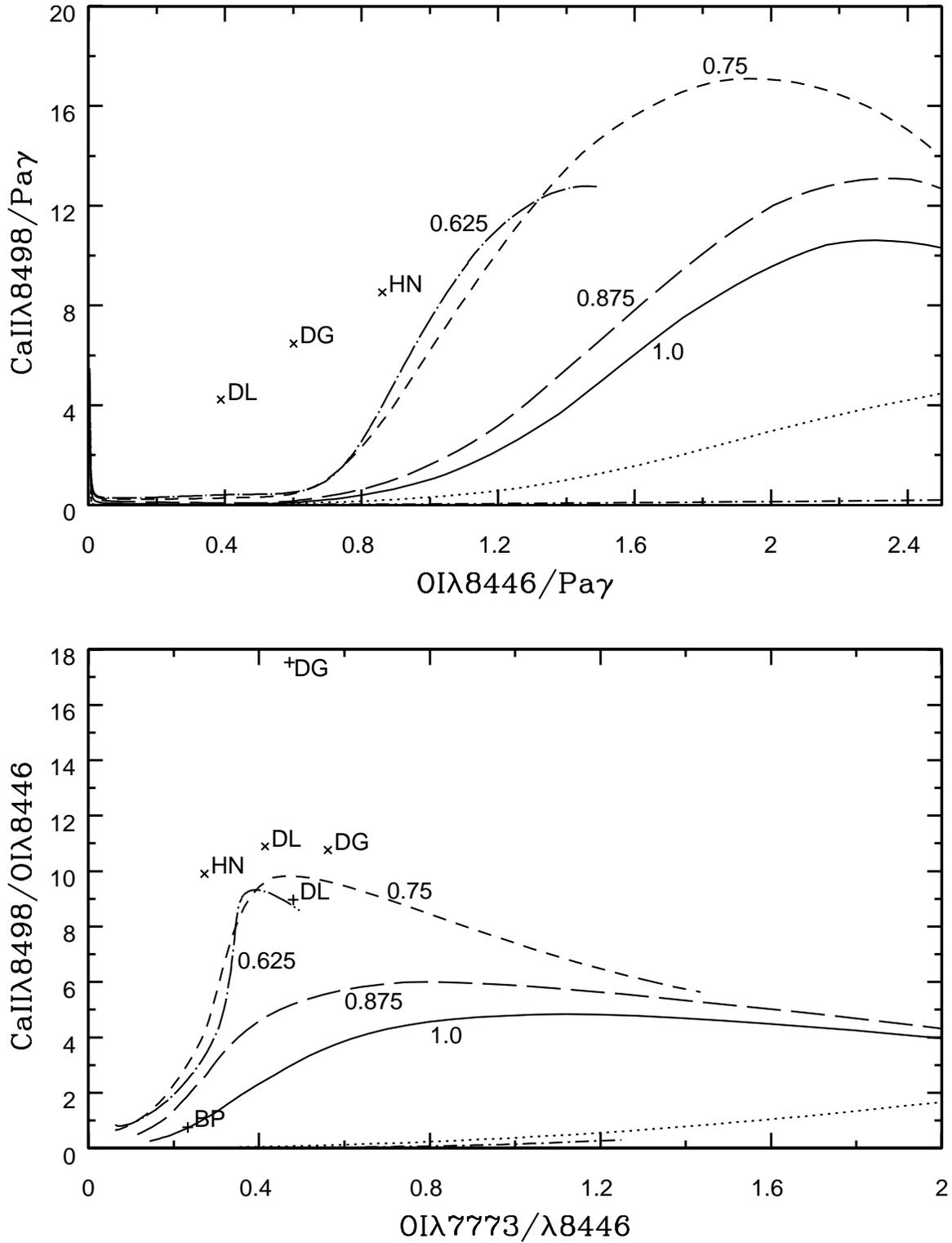}
\caption{Plot of $CaII\lambda 8498/Pa\gamma$ versus $OI\lambda 8446/Pa\gamma$
(top panel) and $CaII\lambda 8498/OI\lambda 8446$ versus $OI\lambda 7773/
\lambda 8446$ (bottom panel)
for $r=4R_\ast$, $\gamma_{HI}
=2\times 10^{-4}~s^{-1}$, and various temperatures(cf. Table 1). Data
points marked by xs and +s are from Edwards et al. (2010) and 
Muzerolle et al. (1998) respectively.}
\end{figure*}

\clearpage
\begin{figure*}
\plotone{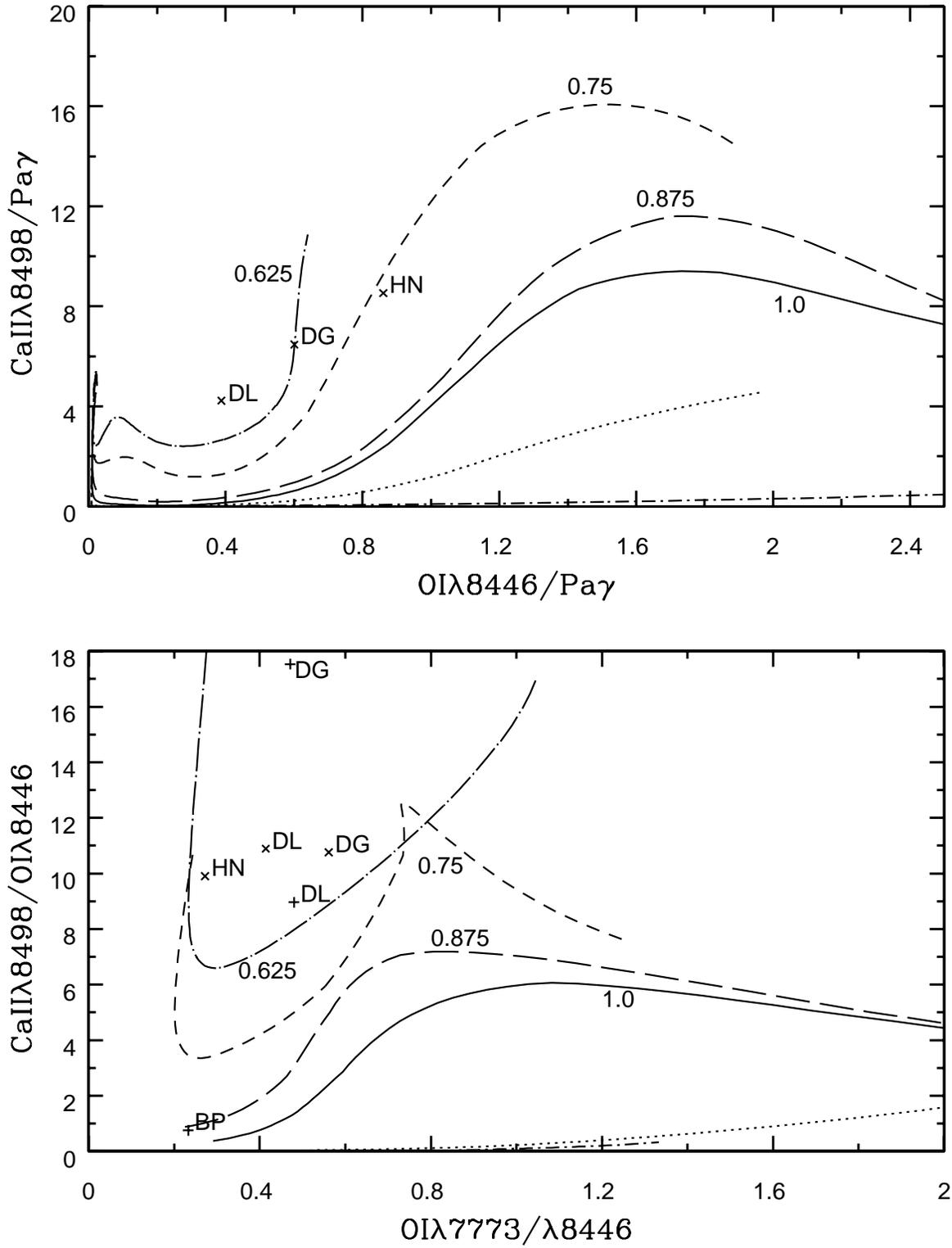}
\caption{Same as Fig. 17 but for $r=2.5R_{\ast}$, $\gamma_{HI}=2\times 10^{-5}~
s^{-1}$.}
\end{figure*}

\clearpage
\begin{figure*}
\plotone{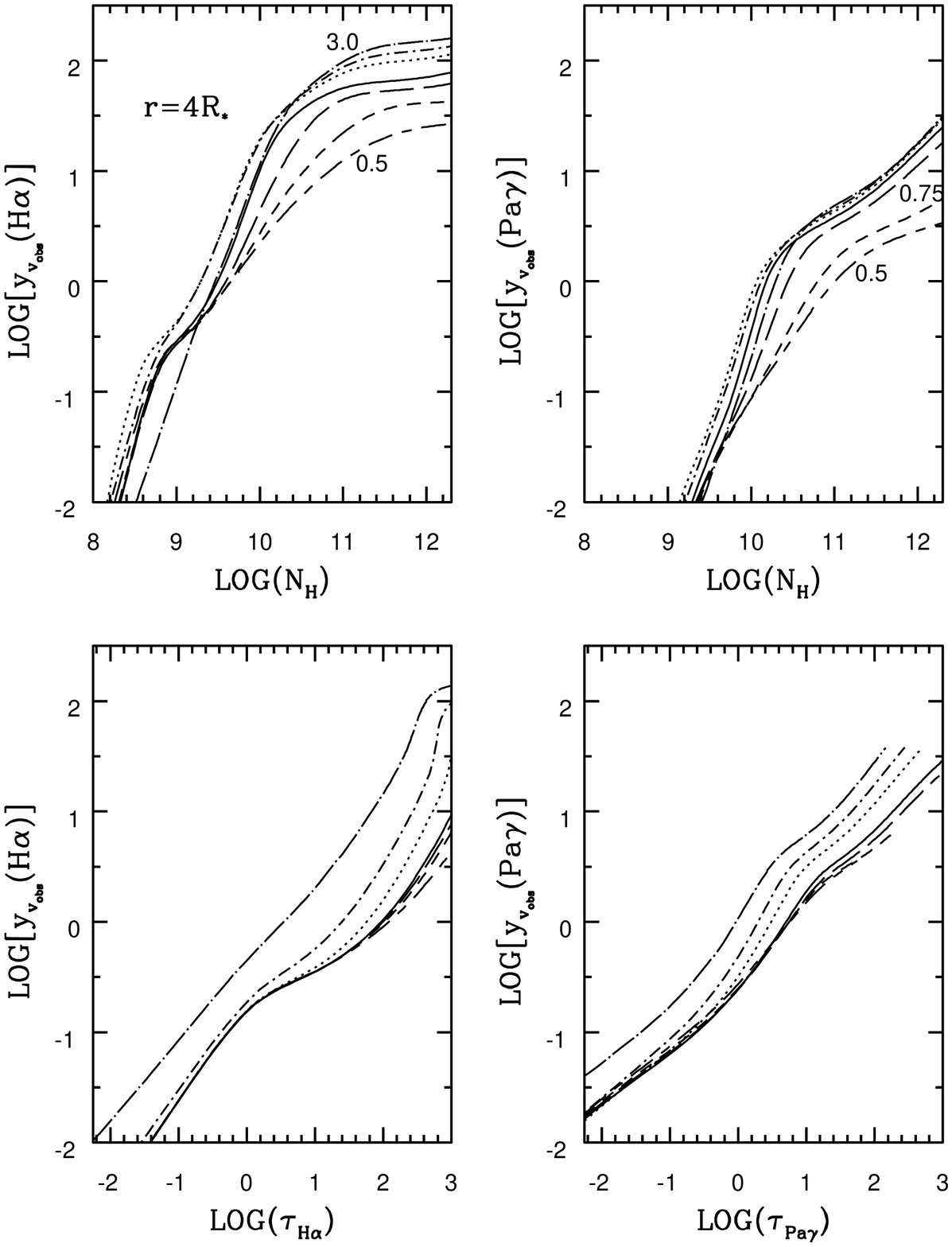}
\caption{Dependences of $H\alpha$ and $Pa\gamma$ specific flux (measured
relative to the local continuum), $y_{v_{obs}}$, on density
and line optical depth for $\gamma_{HI}=2\times 10^{-4}~s^{-1}$ and
various temperatures (cf. Table 1) in the wind model.}  
\end{figure*}

\clearpage
\begin{figure*}
\plotone{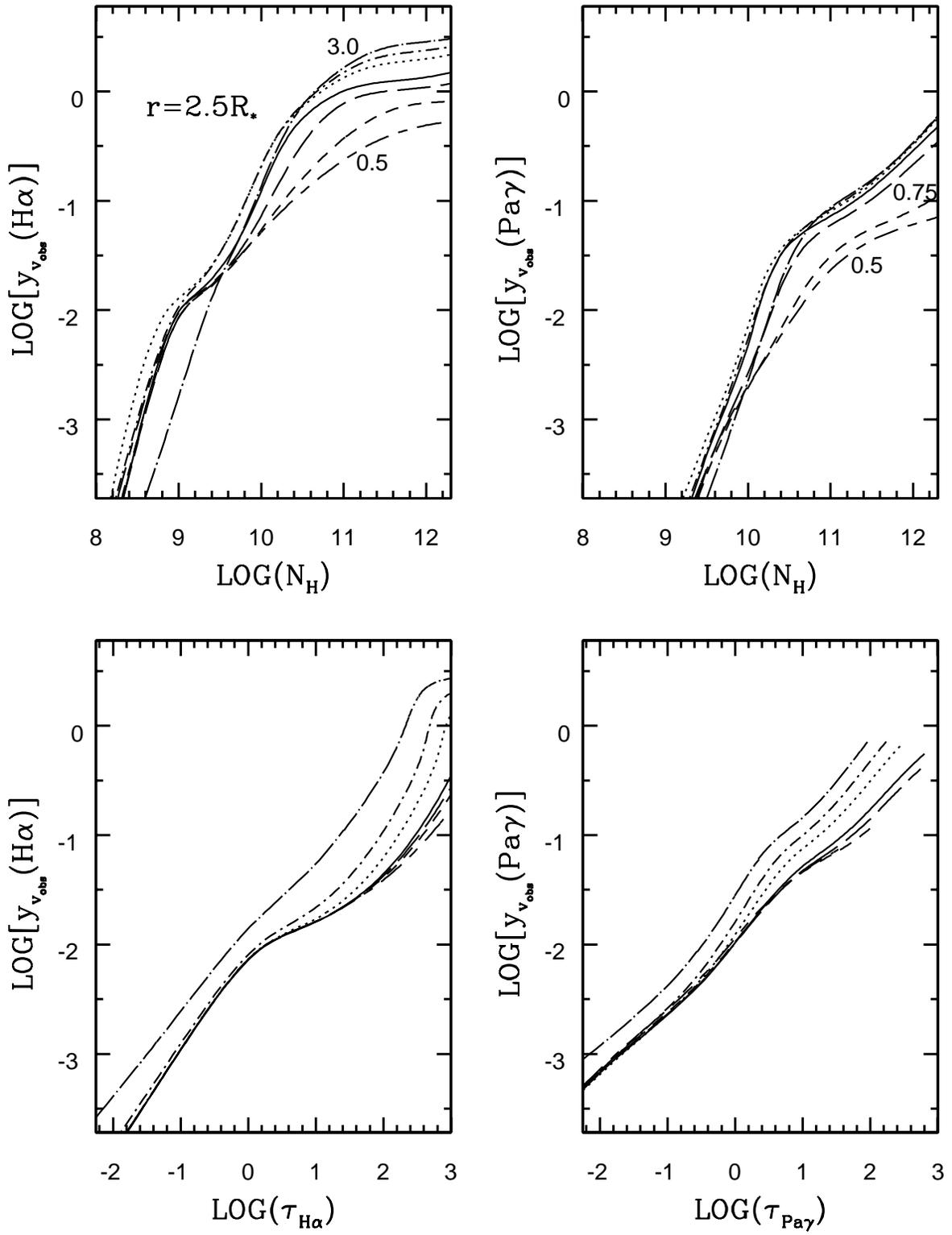}
\caption{Same as Fig. 19 but for the infall model.}
\end{figure*}

\clearpage
\begin{figure*}
\plotone{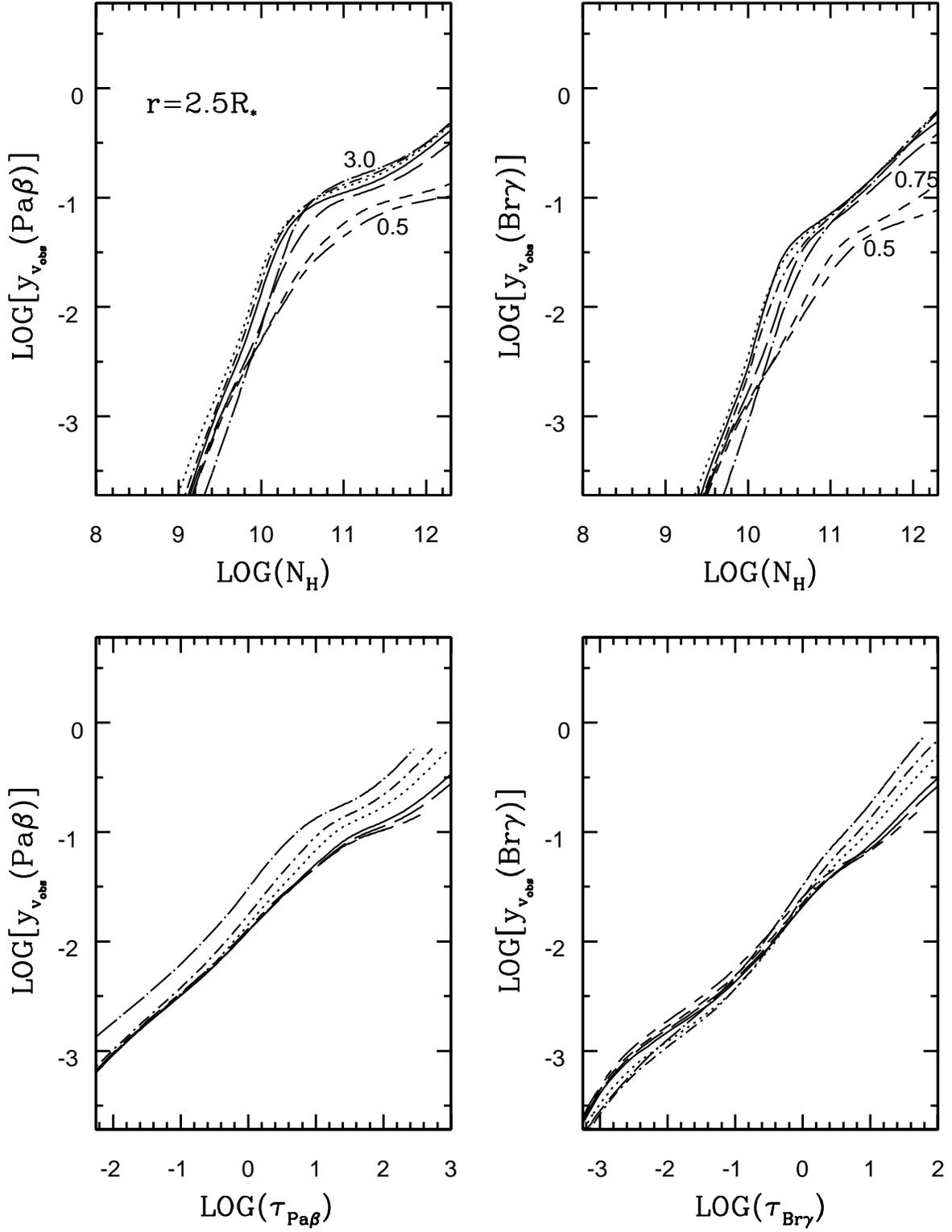}
\caption{Dependences of $Pa\beta$ and $Br\gamma$ specific flux (measured
relative to the local continuum), $y_{v_{obs}}$, on density
and line optical depth for $\gamma_{HI}=2\times 10^{-4}~s^{-1}$
and various temperatures (cf. Table 1) in the infall model.}   
\end{figure*}

\clearpage
\begin{figure*}
\plotone{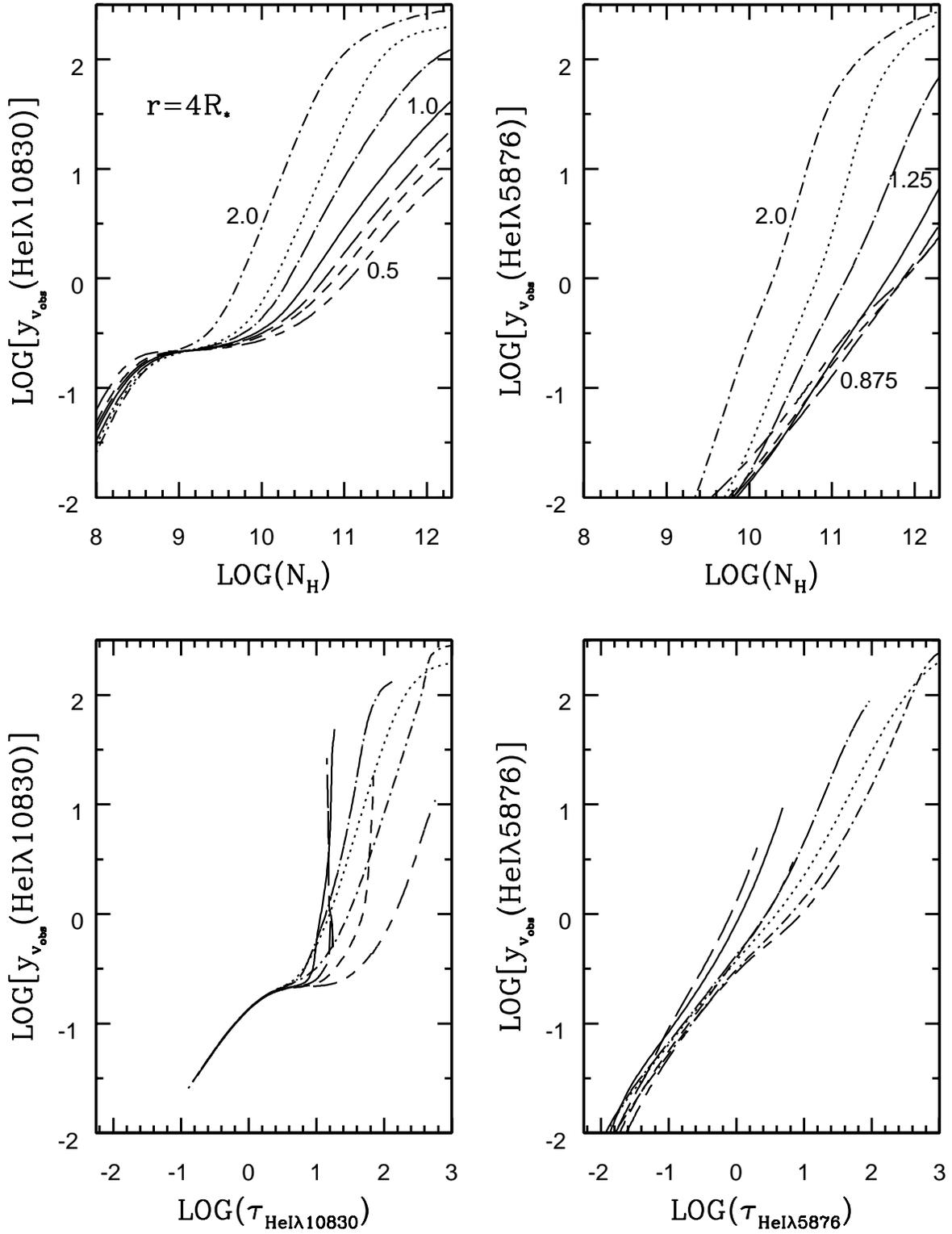}
\caption{Dependences of $HeI\lambda 10830$ and $HeI\lambda 5876$ specific
flux (measured relative to the local continuum),
$y_{v_{obs}}$, on density and line optical depth for $\gamma_{HeI}=
10^{-4}~s^{-1}$ and various temperatures (cf. Table 1) in the wind model.}  
\end{figure*}

\clearpage
\begin{figure*}
\plotone{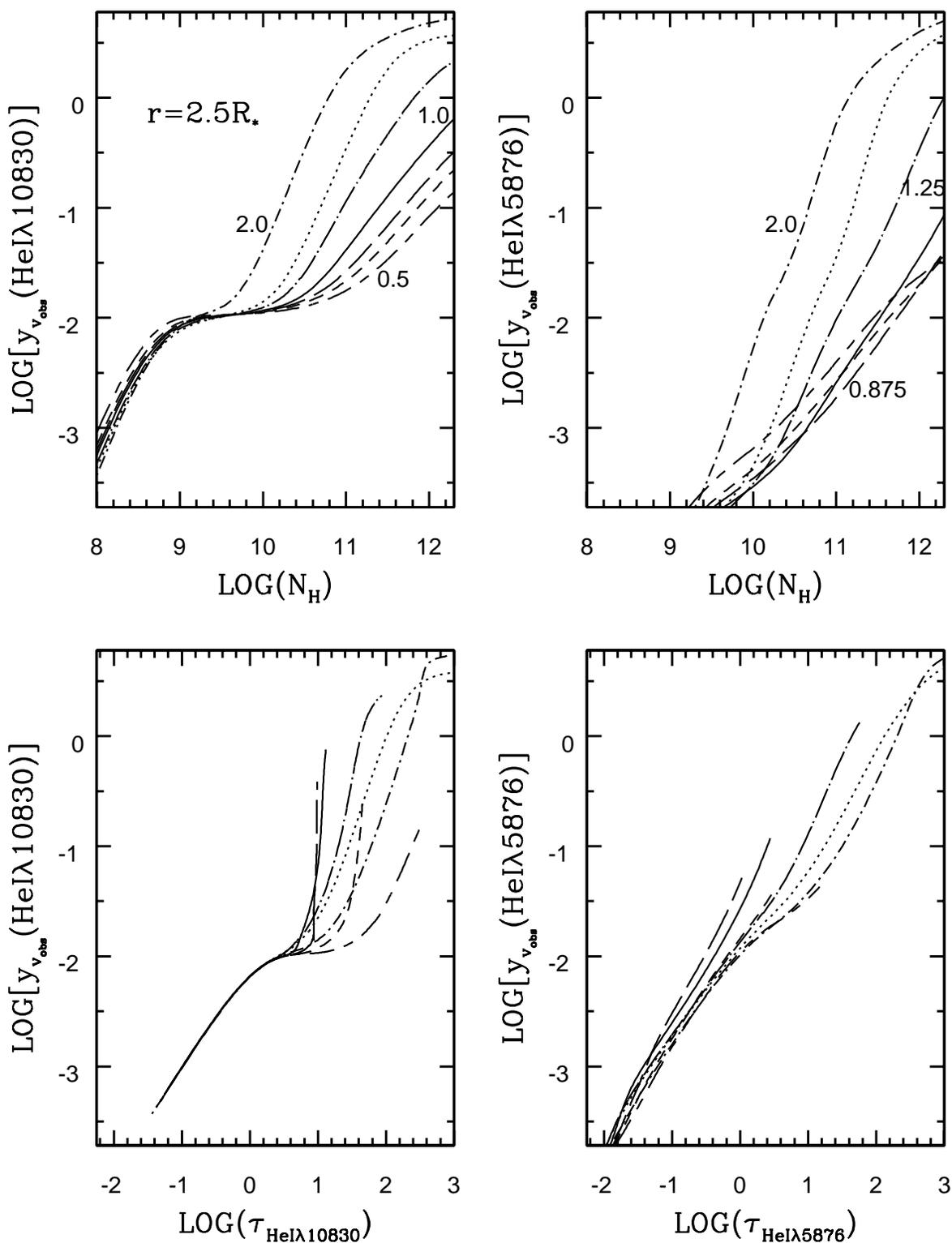}
\caption{Same as Fig. 22 but for the infall model.}
\end{figure*}

\clearpage
\begin{figure*}
\plotone{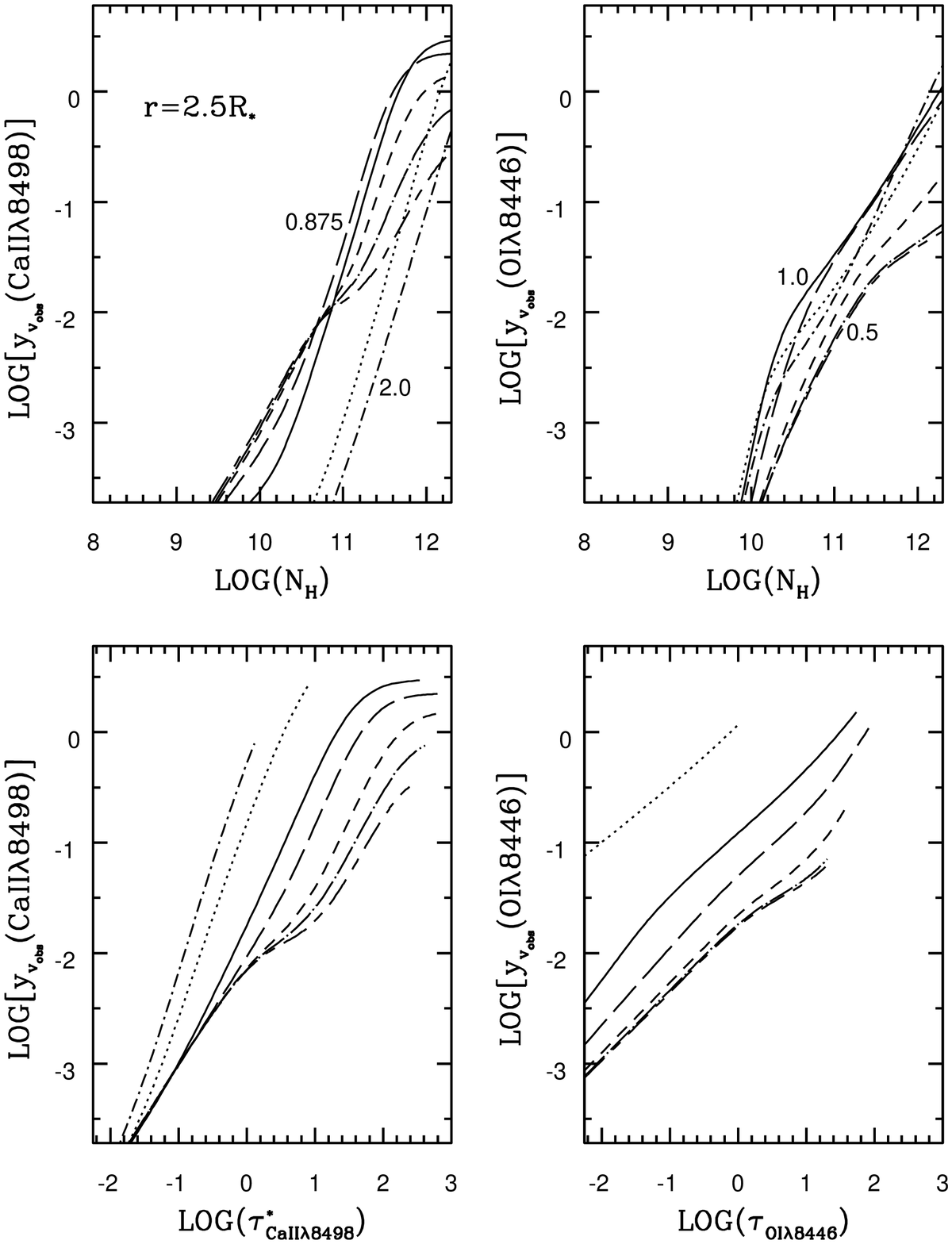}
\caption{Dependences of $CaII\lambda 8498$ and $OI\lambda 8446$ specific
flux (measured relative to the local continuum),
$y_{v_{obs}}$, on density and line optical depth for $\gamma_{HI}=2\times
10^{-4}~s^{-1}$ and various temperatures (cf. Table 1) in the infall model.} 
\end{figure*}

\end{document}